\documentclass{aa}  
\usepackage{graphicx}
\usepackage{txfonts}
\usepackage[flushleft]{threeparttable}
\usepackage{multirow}
\usepackage{booktabs}
\usepackage{xcolor}
\usepackage{tabularx}
\usepackage{booktabs}
\usepackage{amssymb}
\usepackage{array}

\usepackage{textcomp}
\usepackage{gensymb}
\usepackage{hyperref}
\usepackage[dvipsnames]{xcolor} 
\hypersetup{
    colorlinks=true,
    linkcolor=RoyalBlue,
    filecolor=magenta,      
    urlcolor=RoyalBlue, 
    citecolor=RoyalBlue,
    pdftitle={Overleaf Example},
    pdfpagemode=FullScreen,
    }

\newcommand\Msun{\; {M}_{\odot}}
\newcommand\Msunyr{\; {M}_{\odot}\, {\rm yr}^{-1}}

\newcommand\kms{\; {\rm km}\;{\rm s}^{-1}}

\newcommand\pc{\;{\rm pc}}

\newcommand\kpc{\;{\rm kpc}}

\newcommand\cm{\;{\rm cm}}
\newcommand\yr{\; {\rm yr}}
\newcommand\Myr{\;{\rm Myr}}

\newcommand\Mdot{\dot{M}}

\newcommand\Aunit{\Msun\yr^{-1}}

\newcommand\simgt{\lower.5ex\hbox{$\; \buildrel > \over \sim \;$}}
\newcommand\simlt{\lower.5ex\hbox{$\; \buildrel < \over \sim \;$}}
\newcommand{\RNum}[1]{\uppercase\expandafter{\romannumeral #1\relax}}

\newcommand\Htwo{\rm H_2}
\newcommand\HI{{\rm H{\sc I}}}
\newcommand\Hp{\rm H^{+}}

\graphicspath{{./}{figures/}}
\mathchardef\mhyphen="2D

\begin{document}
\title{Simulations of gas inflow in the Milky Way – I. Stellar-Feedback-Regulated Transport from the Central Molecular Zone to the Circumnuclear disk}

\newcommand{\CLAP}    {Como Lake centre for AstroPhysics (CLAP), DiSAT, Universit{\`a} dell’Insubria, via Valleggio 11, 22100 Como, Italy}
\newcommand{\ZAHITA}  {Universit\"{a}t Heidelberg, Zentrum f\"{u}r Astronomie, Institut f\"{u}r Theoretische Astrophysik, Albert-Ueberle-Straße 2, D-69120 Heidelberg, Germany}
\newcommand{\IWR}     {Universit\"{a}t Heidelberg, Interdisziplin\"{a}res Zentrum f\"{u}r Wissenschaftliches Rechnen, Im Neuenheimer Feld 225, 69120 Heidelberg, Germany}
\newcommand{\EPFL}    {Institute of Physics, Laboratory for Galaxy Evolution and Spectral Modelling, EPFL, Observatoire de Sauverny, Chemin Pegasi 51, 1290 Versoix, Switzerland}
\newcommand{\UConn}   {University of Connecticut, Department of Physics, 196A Auditorium Road, Unit 3046, Storrs, CT 06269}
\newcommand{\chalmers}{Department of Physics and Astronomy, Chalmers University of Technology, SE-412 96 Gothenburg, Sweden}
\newcommand{\ISSA}{Institute of Space Sciences {\&} Astronomy, University of Malta, Msida MSD 2080, Malta}
\newcommand{\UF}{Department of Astronomy, University of Florida, Gainesville, FL 32611 USA}
\newcommand{\UK}{Department of Physics and Astronomy, University of Kansas, 1251 Wescoe Hall Dr., Lawrence, KS 66045, USA}
\newcommand{\MPA}{Max-Planck-Institut für Astronomie, Königstuhl 17, D-69117, Heidelberg, Germany}
\newcommand{\LJMU}{Astrophysics Research Institute, Liverpool John Moores University, 146 Brownlow Hill, Liverpool, L3 5RF, UK}
\newcommand{\ESO}{European Southern Observatory (ESO), Karl-Schwarzschild-Straße 2, 85748 Garching, Germany}
\newcommand{\UT}{Institute of Astronomy, The University of Tokyo, Mitaka, Tokyo 181-0015, Japan}

\author{Zi-Xuan Feng    \inst{\ref{CLAP}, \ref{ZAHITA}  } \and
    Mattia~C.~Sormani   \inst{\ref{CLAP}                } \and
    Robin~G.~Tress      \inst{\ref{EPFL}                } \and 
    Simon~C.~O.~Glover  \inst{\ref{ZAHITA}              } \and 
    Ralf~S.~Klessen     \inst{\ref{ZAHITA}, \ref{IWR}   } \and
    Jonathan Petersson  \inst{\ref{EPFL}                } \and 
    Michaela Hirschmann \inst{\ref{EPFL}                } \and
    Ashley T. Barnes    \inst{\ref{ESO}                 } \and 
    Cara Battersby      \inst{\ref{UConn}               } \and 
    Marco Donati        \inst{\ref{CLAP}                } \and 
    Karl Fiteni         \inst{\ref{CLAP}, \ref{ISSA}    } \and 
    Jonathan D. Henshaw \inst{\ref{MPA}                 } \and 
    Adam Ginsburg       \inst{\ref{UF}                  } \and 
    Savannah Gramze     \inst{\ref{UF}                  } \and 
    Xingchen Li         \inst{\ref{CLAP}                } \and 
    Dani~R.~Lipman      \inst{\ref{UConn}               } \and 
    Steven N. Longmore  \inst{\ref{LJMU}                } \and
    Elisabeth Mills     \inst{\ref{UK}                  } \and 
    Maya~A.~Petkova     \inst{\ref{chalmers}            } \and 
    Yoshiaki Sofue      \inst{\ref{UT}                  } \and
    Arianna Vasini      \inst{\ref{CLAP}                }
}

\institute{
            \CLAP       \label{CLAP}        \and 
            \ZAHITA     \label{ZAHITA}      \and
            \EPFL       \label{EPFL}        \and 
            \IWR        \label{IWR}         \and
            \ESO        \label{ESO}         \and 
            \UConn      \label{UConn}       \and
            \ISSA       \label{ISSA}        \and
            \MPA        \label{MPA}         \and
            \UF         \label{UF}          \and 
            \LJMU       \label{LJMU}        \and 
            \UK         \label{UK}          \and 
            \chalmers   \label{chalmers}    \and
            \UT         \label{UT}          
}

\abstract
    {We perform hydrodynamical simulations with radially varying resolution to study the effects of stellar feedback on the radial inflow of gas from the Central Molecular Zone (CMZ, $R\sim200\pc$) to the Circumnuclear Disk (CND, $R\sim 5\pc$) of the Milky Way. The simulations include a realistic Milky Way barred gravitational potential, a cooling function coupled to a non-equilibrium chemical network, gas self-gravity, star formation, supernova feedback, and radiation feedback from massive stars computed via on-the-fly radiative transfer. Our main findings are as follows: 
    1) Stellar feedback drives a radial inflow that decreases monotonically with decreasing Galactocentric radius.  The time-averaged inflow rate in our fiducial SNRad simulation, which includes both supernova and radiation feedback, declines from $\langle \dot{M} \rangle\sim5\times10^{-3}\Msun\yr^{-1}$ at $R\sim 100\pc$ to $\langle\dot{M}\rangle\sim 10^{-6}\Msun\yr^{-1}$ at $R\sim1\pc$. 
    2) The total inflow rate can be broken down into two components driven by two distinct mechanisms. First, feedback-driven turbulence redistributes the angular momentum of gas clouds, producing a smooth (secular) transport of mass inward, similar to a Shakura-Sunyaev viscous accretion disk. This component contributes inflow rates that vary from $\dot{M}\sim5\times10^{-4}\Msun\yr^{-1}$ at $R\sim100\pc$, to $\dot{M}\sim3\times10^{-5}\Msun\yr^{-1}$ at $R\sim10\pc$, to $\dot{M}\sim10^{-7}\Msun\yr^{-1}$ at $R\sim1\pc$. Second, episodic inflow events can transiently increase the inflow rate by several orders of magnitude, reaching $\dot{M}\sim 10^{-3}\Msun\yr^{-1}$ over timescales of $\Delta t \sim 3\mhyphen5\Myr$ at $R=10\pc$.
   3) The stellar feedback model significantly affects the episodic inflow but has little 
impact on the smooth component. Simulations including radiation feedback produce substantially more episodic events than those with supernova feedback alone.}
 
\titlerunning{Simulations of Stellar-Feedback-Regulated Transport from the CMZ to the CND in the Milky Way}


\maketitle

\section{Introduction}
\label{sec:introduction}

It is now established that most (perhaps all) massive galaxies host supermassive black holes (SMBHs) at their centre, whose properties correlate with properties of their host galaxy such as bulge velocity dispersion and mass \citep{mag_etal_98, fer_mer_00, geb_etal_00, hop_etal_07, hop_etal_07a, all_ric_07, kor_etal_11, kor_ho_13,Zhang2024}. Gas accretion onto these black holes fuels Active Galactic Nuclei (AGN), and their release of energy in the form of radiation, winds, or jets of plasma (collectively known as ``feedback'') can shape the formation and evolution of their host galaxies \citep[e.g.][]{Tumlinson2017,Weinberger2018,Oppenheimer2020}. It remains a crucial open question how the SMBHs are fed, i.e. how gas is transported from kpc (or larger) scales down to the sphere of influence of the SMBHs \citep[e.g.][]{Storchi-Bergmann2019}. 

For the gas to migrate inwards, it must lose angular momentum. The key question is how this transfer occurs. Recent simulations that ``zoom-in'' from cosmological to sub-pc scales, as well as from galaxy scales to sub-pc scales in isolated systems, made considerable progress in answering this question \citep[e.g.][]{hop_etal_24b,guo_etal_24,shi_etal_25}. These works have shown that there is no single universal mechanism that operates in all galaxy types. For example, \citet{hop_etal_24b} have shown that for out-of-equilibrium galaxies produced by galaxy interactions, gravitational torques created by non-axisymmetries in gas and stars are the dominant physical process that transports gas from galactic ($>100$~pc) down to $<1$~pc scales. \citet{guo_etal_24} have shown that in massive elliptical galaxies, magnetic fields can increase the accretion rate by a factor of $\sim 10$ compared to the non-magnetised case. \citet{shi_etal_25} found that in dwarf galaxies, gravitational torques are dominant in transporting the gas down to the accretion disk immediately surrounding the central black hole, similar to \citet{hop_etal_24b}. However, it is not clear how these results can be applied to more quiescent massive disk galaxies, such as the Milky Way, or to nearby Seyfert galaxies such as NGC 1097, NGC 1365 or NGC 1433.

The Milky Way (MW) is a typical barred massive spiral galaxy. Its centre \citep[distance $R_0 = 8275\pm9_{\rm stat}\pm33_{\rm sys}$ pc; ][]{gra_etal_21} is a hundred times closer to us than that of the next comparable galaxy (Andromeda) and can be studied in much greater detail than any other galactic centre. It is the only galactic nucleus in which we can observe the interstellar medium down to the scale of forming stars and the kinematics of individual stars. Although the MW central SMBH (Sgr A*) is currently inactive, it shares many similarities with what we see around nearby AGNs (Seyfert galaxies): the presence of a molecular ring at $R \sim200$ pc from the centre (known as the Central Molecular Zone or CMZ, \citealt{mor_ser_96,hen_etal_23}), the presence of a compact molecular disk at few pc from the centre (known as the circum-nuclear disk or CND, e.g.\ \citealt{gen_etal_85, chr_etal_05,Lau2013}) and the presence of energetic outflows that can be attributed to past ejection events associated with Sgr A*. Indeed, X-ray echoes suggest that Sgr A* was a factor of $\sim 10^5$ more luminous a few hundred years ago \citep[e.g.][]{Ryu2013,Chuard2018}, while energetic outflows suggest that it might have been a factor of $10^{8\mhyphen 9}$ more luminous a few Myr ago \citep[e.g.][]{Su2010,Ponti2019,Heywood2019,Predehl2020}. Thus, studying the mechanisms by which gas is transported in the centre of the MW could give us important insights on the feeding of SMBHs.

The radial inflow in the present-day MW from Galactic (kpc) scales down to the sphere of influence of Sgr A* ($R\lesssim 1\pc$, defined as the region where Sgr A* dominates the gravitational potential) can be schematically thought of as a sequence of steps. The first step is performed by the Galactic bar, that channels the gas from the Galactic disk ($R\gtrsim 3$ kpc) down to the CMZ ring at a rate of $ \sim 1 \Msunyr$ \citep[e.g.][]{sor_bar_19,hat_etal_21,Su2024}. Although there are some uncertainties on the inflow rate number, mostly stemming from uncertainties in the CO-to-H$_2$ conversion factor \citep{Gramze2023}, the basic physical process that removes the angular momentum from the gas is well understood: it is the gravitational torque of the Galactic (stellar) bar.

The next step in the inflow process is the transport of gas from the CMZ down to the CND ($1.5 \lesssim R \lesssim 5\pc$). As the closest large reservoir of molecular gas to Sgr~A*, the CND is a clumpy, inhomogeneous, and kinematically disturbed concentration of molecular gas with a total mass of $\sim 10^4\Msun$ \citep{req_etal_12}. This second step is much less understood than the first one, but simulations have shown that gas transport on these scales is not primarily driven by the Galactic bar \citep{tre_etal_20a,tre_etal_24}. Several possible mechanisms have been proposed, including 1) A secondary bar could in principle repeat on a smaller scale the inflow process of the main bar \citep{shl_etal_89}. However, it remains uncertain whether the MW is a double barred galaxy (see Section~5.2.1 in \citealt{sch_etal_25} and \citealt{fit_etal_26}). The a priori probability is $\sim 50\%$, as approximately this fraction of barred galaxies with total mass comparable to the MW host a secondary bar \citep{Erwin2024}. 2) Stellar feedback associated with the intense star formation activity in the CMZ may drive turbulence that redistributes angular momentum among gas parcels. In particular, \citet{tre_etal_20a} compared two simulations of gas flow in a Milky Way barred potential that differ only by the inclusion of gas self-gravity and supernova feedback. They found that the radial inflow rate \emph{from} the CMZ inwards is zero in the simulation without supernova feedback (the gas simply piles up in the CMZ ring-like structure), while it is $\sim 10^{-3} \Msunyr$ in the simulation with supernova feedback (at this rate the CND would take only $\sim 10 \Myr$ to build up). 3) Magneto-hydrodynamic instabilities may also generate transport of mass and angular momentum, in a way similar to standard accretion disk models \citep{bal_haw_98}. In particular, \citet{tre_etal_24} found using non self-gravitating simulations that the magneto-rotational instability (MRI) can cause an in-plane inflow of matter from the CMZ gas ring towards the central few parsecs of $\sim 0.01\mhyphen0.1\Msunyr$ that is absent in unmagnetised simulations.

The simulations of \citet{tre_etal_20a} and \citet{tre_etal_24} have several important limitations: 1) The adopted gravitational potential was overly simplified, with several distinct components (nuclear stellar disk, nuclear star cluster) clumped together in a simple power law. This washes out any important radial variations in the rotation curve and in the shear. Moreover, the gravitational potential of Sgr A* was not included, although it can influence the morphology and formation of the CND \citep{map_tra_16, tra_etal_18}. (2) The resolution of $20\Msun/{\rm cell}$ was insufficient to reliably keep track of the gas flows down to the innermost pc. (3) \citet{tre_etal_20a} only included supernova feedback, and neglected early feedback such as photoionisation from massive stars, which can affect the inflow by halting accretion and growth of sink particles before the first supernova explosions occur at $t \sim 4 $ Myr. \citet{tre_etal_24} did not include star formation at all, and so the combined effect of stellar feedback and magnetic fields \citep[see also][]{moo_etal_23} could not be tested.

In this paper, we perform new high-resolution simulations ($M_{\rm cell} \lesssim 1\Msun$ within central $50\pc$) of gas flow in the CMZ to study the effects of stellar feedback, in particular supernova and photoionisation feedback, on the radial inflow to the central $1\pc$. We include a realistic MW barred potential, gas self-gravity, star formation and associated feedback, and a cooling function coupled to non-equilibrium chemical networks. We do not include magnetic fields, but plan to include them in future studies. We aim to address the following questions:
\begin{itemize}
    \item To what extent can stellar feedback drive radial inflow from the CMZ to scales of $R \lesssim 1$~pc?
    \item What is the relative importance of supernova feedback vs early feedback (photoionisation from massive stars)?
    \item How does the inflow depend on radius?
    \item How does the inflow depend on time? Is it episodic or steady?
\end{itemize}
This paper is organised as follows. In Sect.~\ref{sec:numerical_setting} we introduce the numerical setting. In Sect.~\ref{sec:results} we present the general ISM and star properties in simulations, and the overall gas flow patterns. In Sect.~\ref{sec:discussion} we mainly discuss our results and compare with existing literature. We summarize in Sect.~\ref{sec:conclusion}. 

\section{Numerical scheme}
\label{sec:numerical_setting}

We use the moving-mesh code AREPO \citep{spring_10, pak_etal_16, wei_etal_20} to model the gas flow in the innermost $5\kpc$ of the Milky Way (this includes the entire bar region). The hydrodynamical equations we solve are: 

\begin{align}
\label{eq:flux_equation}
& \frac{\partial \rho}{\partial t} + \nabla\cdot(\rho \bf{v}) = 0, \\
\label{eq:Euler_equation}
& \frac{\partial (\rho \bf{v})}{\partial t} + \nabla\cdot(\rho \mathbf{v}\mathbf{v} + P\mathbf{I}) = -\rho\nabla\Phi, \\
\label{eq:energy_equation}
& \frac{\partial (\rho e)}{\partial t} + \nabla\cdot[(\rho e + P)\mathbf{v}] = \rho\dfrac{\partial\Phi}{\partial t} - \mathcal{L}, 
\end{align}
where $\rho$ is the gas density, $\mathbf{v}$ is the velocity, $P$ is the thermal pressure, and $\mathbf{I}$ is the identity matrix. The energy per unit mass is defined as $e = e_{\rm therm} + \Phi + \mathbf{v}^2/2$, with $e_{\rm therm}$ is the thermal energy per unit mass. We use the equation of state for ideal gas, $P = (\gamma -1)\rho e_{\rm therm}$, where $\gamma = 5/3$. $\mathcal{L}$ is the net cooling (heating) rate per unit volume (this includes the processes described in Sect.~\ref{sec:chemistry}). The term $\Phi$ is defined as $\Phi = \Phi_{\rm ext} + \Phi_{\rm sg}$, where the first term on the right-hand-side is the externally imposed gravitational potential, while the second is the self-gravity of gas, central sink (Sect.~\ref{sec: SMBH_sink_particle}), and $N$-body star particles. 

\begin{figure*}[ht!]
\includegraphics[width=\textwidth]{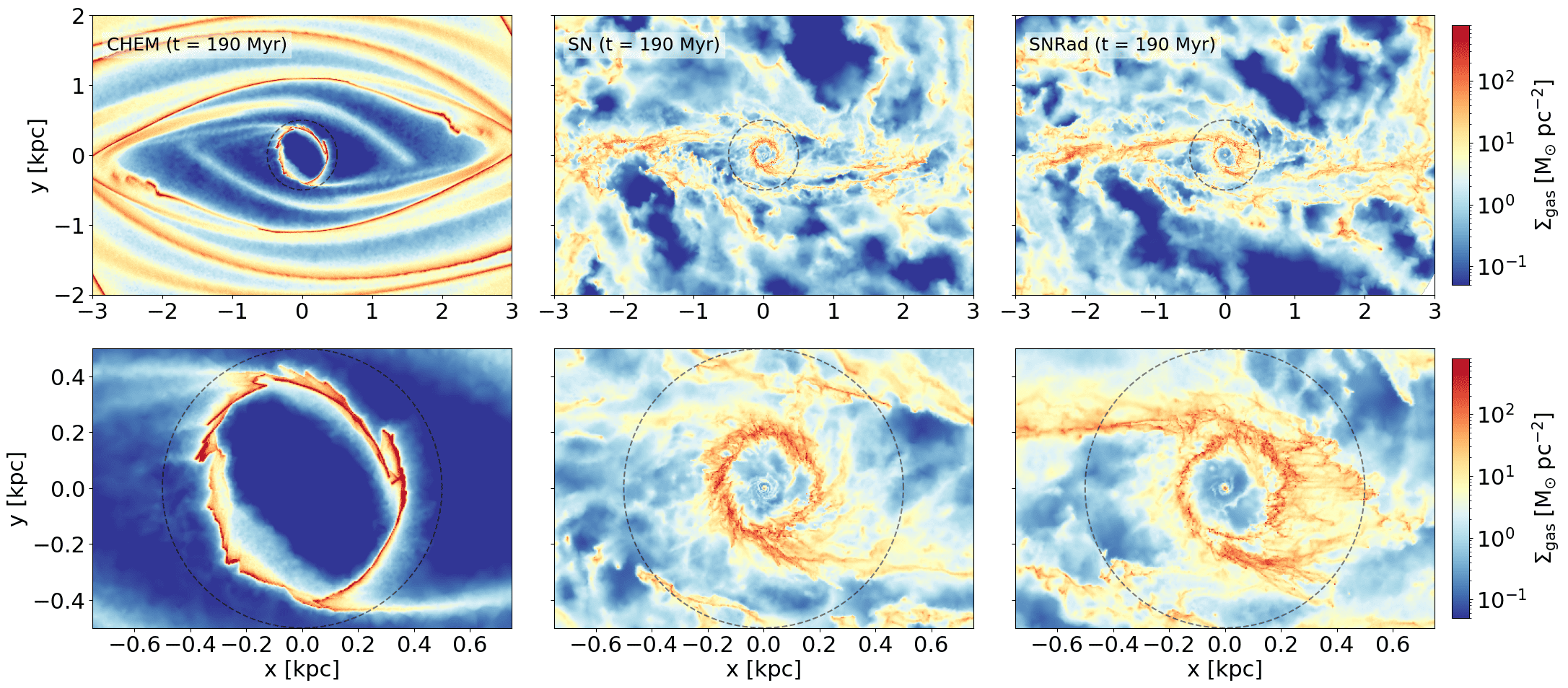}
\centering
\caption{\textit{Top:} Gas surface density in models CHEM (left), SN (middle), and SNRad (right). Black dashed circles mark $R = 0.5\kpc$. \textit{Bottom:} Zoom-in views of the CMZ.}
\label{fig:model_overview}
\end{figure*}

We run three simulations with different amounts of physics, as summarised in Table~\ref{tab:list_simulations}.
All models share the same gravitational potential, initial conditions, and use similar resolution settings, they differ only in the star-formation and feedback implementations. The CHEM simulation does not have star formation and does not include the gas self-gravity, while the SN and SNRad models include star formation and the gas self-gravity. The overall morphology in the bar and the CMZ region for these three models are presented in Fig.~\ref{fig:model_overview}. In the following we use $(R,\theta,z)$ to denote Galactocentric cylindrical coordinates, and $(r,\theta,\phi)$ for Galactocentric spherical coordinates.

\begin{table*}
\caption{Summary of simulations analysed in this paper.}
\label{tab:sim_summary}
\centering
\begin{tabularx}{\textwidth}{>{\centering\arraybackslash}X
                            >{\centering\arraybackslash}X
                            >{\centering\arraybackslash}X
                            >{\centering\arraybackslash}X
                            >{\centering\arraybackslash}X
                            >{\centering\arraybackslash}X
                            >{\centering\arraybackslash}X}
\toprule
Simulation & BH accretion & chemical network & self-gravity & star formation & supernova feedback & radiation feedback \\
\midrule
CHEM  & \checkmark & \checkmark &  &  &  &  \\
SN    & \checkmark & \checkmark & \checkmark & \checkmark & \checkmark &  \\
SNRad & \checkmark & \checkmark & \checkmark & \checkmark & \checkmark & \checkmark \\
\bottomrule
\end{tabularx}
\label{tab:list_simulations}
\end{table*}

\subsection{Gravitational potential} \label{sec:gravpot}

We use the external Milky Way barred potential from \citet{hun_etal_24}, which is publicly available in the software package {\sc Agama} \citep{vasili_19}. This is fine-tuned to reproduce the properties of the Milky Way, particularly in the central regions. It is composed by the sum of several components as follows:
\begin{equation}
    \Phi_{\rm ext} = \Phi_{\rm SgrA*} + \Phi_{\rm NSC} + \Phi_{\rm NSD} + \Phi_{\rm bar} + \Phi_{\rm disk} + \Delta \Phi_{\rm spiral} + \Phi_{\rm halo} \,,
\end{equation}
where
\begin{itemize}

\item $\Phi_{\rm SgrA*}$ is the potential of the SMBH at the centre of the Galaxy. In the first part of the simulation ($t=0\mhyphen150\Myr$) it is modelled with a Plummer sphere with softening of $0.1\pc$ and a fixed mass of $4.154\times10^6\Msun$ \citep{gra_etal_19}. At $t>150\Myr$ it is modelled with a sink particle with initial mass $4\times10^6\Msun$, which can later accrete in time. See Sect.~\ref{sec: SMBH_sink_particle} for more details on the SMBH sink particle. This component dominates the potential at $R \lesssim 1$pc.

\item $\Phi_{\rm NSC}$ is the potential of the nuclear star cluster (NSC) from the dynamical model of \citet{cha_etal_15}, which is fitted to the star counts and kinematics of several thousands of stars from \citet{fri_etal_16}. This component dominates the potential at $1 \lesssim R\lesssim 30\pc$.

\item $\Phi_{\rm NSD}$ is the potential of the nuclear stellar disk from the Jeans model 3 of \citet{sor_etal_20a}. This component dominates the potential at $30 \lesssim R\lesssim 300\pc$.

\item $\Phi_{\rm bar}$ is the potential of the bar/bulge from the made-to-measure model of \citet{por_etal_17} as parameterized by \citet{sor_etal_22}. It includes an X-shaped bulge and a long-bar component. We adopt a bar pattern speed of $\Omega_{\rm b} = 37.5\;\rm km\;s^{-1}\kpc^{-1}$. This component dominates the potential at $0.3 \lesssim R\lesssim 3\kpc$.

\item $\Phi_{\rm disk}$ is a modified version of the stellar disk potential from \citet{mcmill_17}, which comprises both thin and thick disk components.  It dominates the potential at $3 \lesssim R\lesssim 10\kpc$.

\item $ \Delta \Phi_{\rm spiral}$ is a rigidly rotating logarithmic spiral potential perturbation modified from \citet{li_etal_22} as detailed in \citet{hun_etal_24}. It consists of two pairs of $m=2$ spiral arms rotating with a pattern speed of $\Omega_{\rm spiral} = 22.5\;\rm km\;s^{-1}\kpc^{-1}$.

\item $\Phi_{\rm halo}$ is a dark matter halo that follows a spherical \citet{einast_69} profile with a total mass of $1.1\times10^{12}\Msun$. The mass distribution is constrained from the circular velocities of the Galaxy and the dynamical modelling of satellite galaxies and stellar streams \citep{cau_etal_20, cor_vas_22, vas_etal_21, kop_etal_23}.

\end{itemize}

\noindent The rotation curve of the potential is shown in Fig.~1 of \citet{hun_etal_24}. We refer the reader to this paper for further details on the mass model and the resulting gravitational potential.

\subsection{Resolution settings}
\label{sec:resolution}

\begin{figure}[ht!]
\includegraphics[width=\columnwidth]{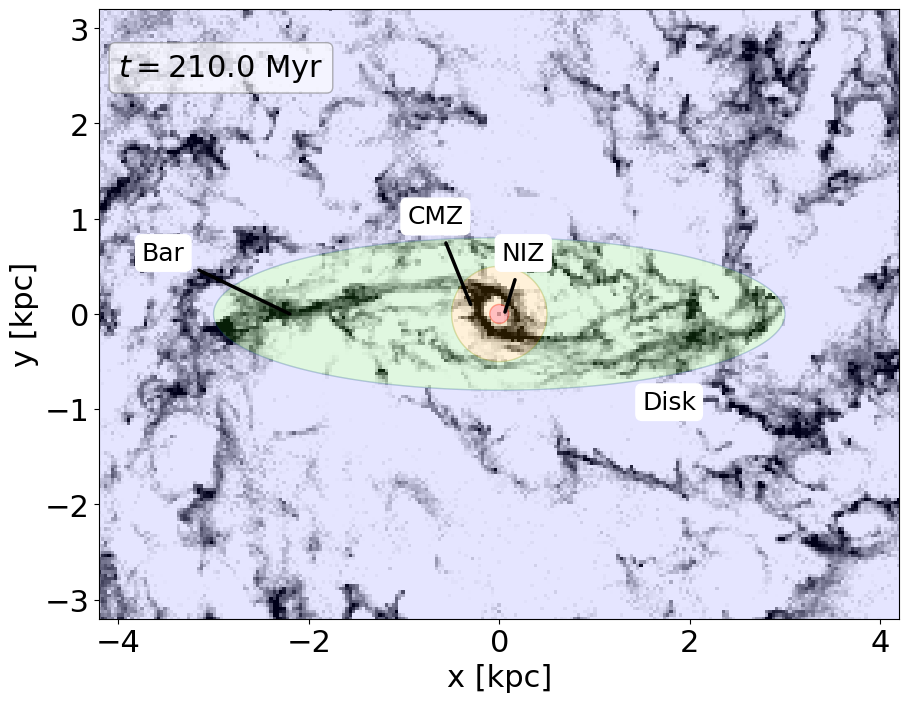}
\centering
\caption{The four spatial regions (Disk, Bar, CMZ, NIZ) that we use to define our resolution settings (see Sect.~\ref{sec:resolution}).}  
\label{fig:region_scheme}
\end{figure}

\begin{figure}[ht!]
\includegraphics[width=\columnwidth]{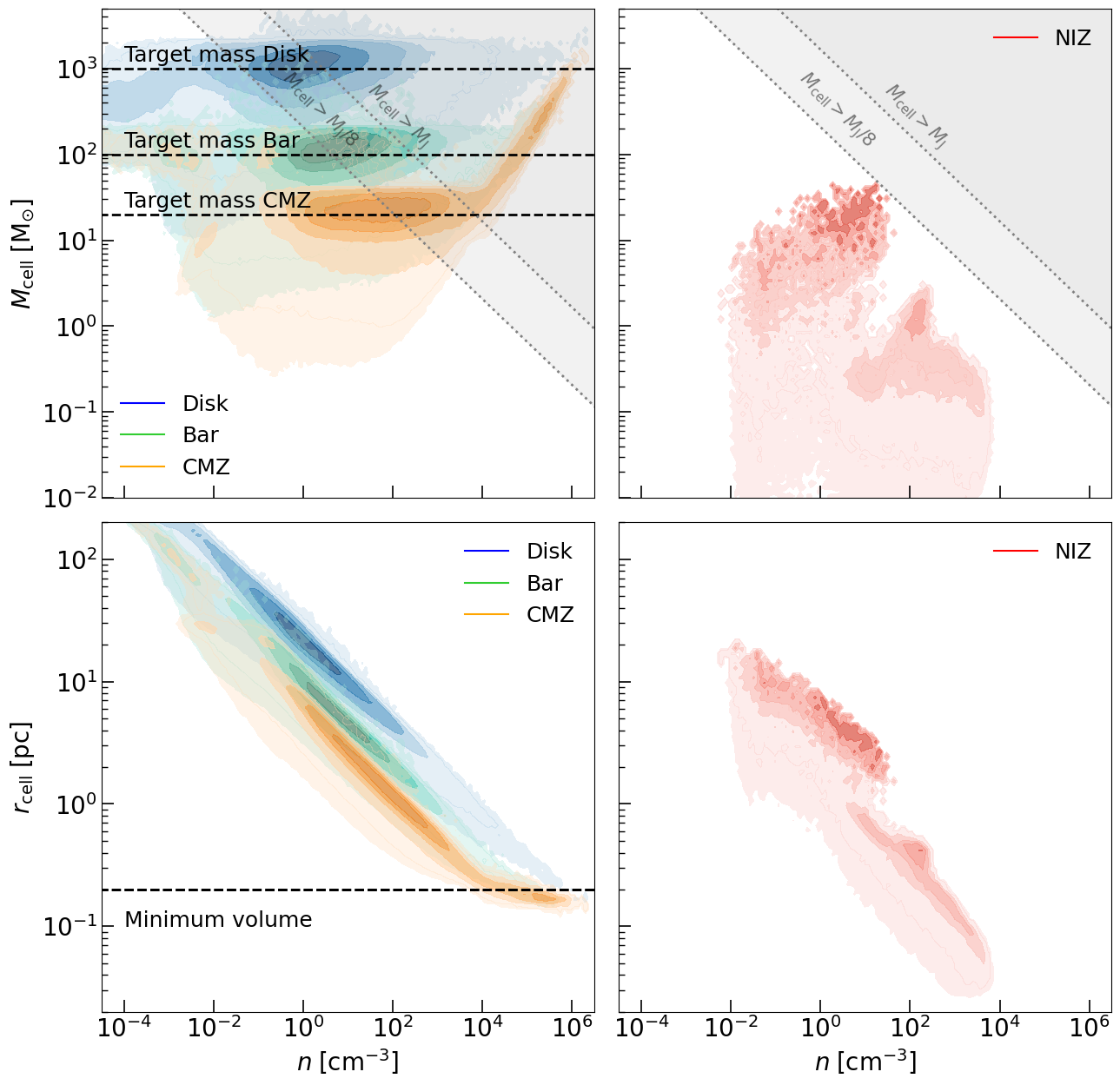}
\centering
\caption{Mass and spatial resolution in the SNRad simulation at $t = 190 \Myr$ as a function of gas density. Colours denote the different regions defined in Sect.~\ref{sec:resolution}. For each cell, $r_{\rm cell}$ is the radius of a sphere with the same volume. The dashed line in the top panel indicates the cell-volume limit ($r_{\rm cell} = 0.2\pc$) for regions outside $r=100\pc$. Horizontal dashed lines in the bottom panel represent the target masses for disk, bar, and CMZ regions at $500$, $100$, and $20\Msun$, respectively. The dotted lines in the bottom panel mark the star-formation thresholds at $M_{\rm cell} = M_{\rm J}$ and $M_{\rm cell} = M_{\rm J}/8$ (see Sect.~\ref{sec:starparticles})}  
\label{fig:resolution_diagrams}
\end{figure}

\begin{figure}[ht!]
\includegraphics[width=\columnwidth]{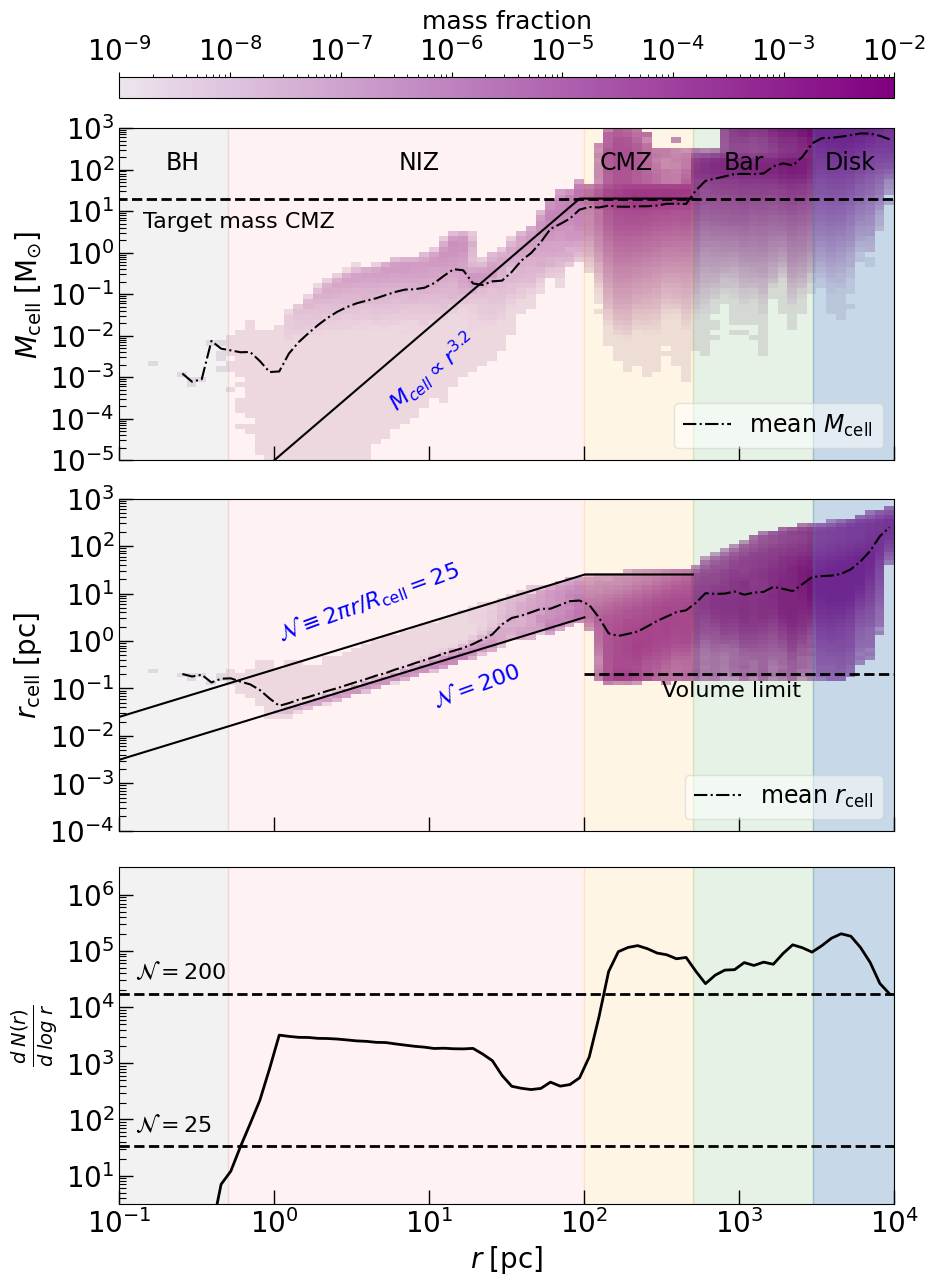}
\centering
\caption{Mass and spatial resolution as a function of Galactocentric spherical radius in the  SNRad simulation at $t = 190 \Myr$. The colormap at the top indicates mass fraction per bin. Bins are logarithmically spaced in both radius and mass. We have 120 bins in the $r$ range of $10^{-1}\mhyphen10^4\; {\rm pc}$. The ratio between the radius of consecutive bins is $10^{5/120}\sim1.2$. The vertical shaded regions approximately indicate the regions defined in Sect.~\ref{sec:resolution} using the same color coding as Fig.~\ref{fig:resolution_diagrams}. \textit{Top:} gas cell mass $M_{\rm cell}$ versus spherical radius $r$. The dot-dashed curve is the averaged cell mass per radial bin. The dashed line indicates the target mass for the CMZ region. The black solid line indicates $M_{\rm cell} = 40000\,(r/{\rm kpc})^{3.2} \Msun$. \textit{Middle:} spatial resolution. The dashed line denotes the $r_{\rm cell} > 0.2\pc$ volume limit outside $R=100\pc$. Inside $100\pc$, the oblique solid lines indicate radius-dependent volume limits set by $\mathcal{N}=25$ and $\mathcal{N}=200$ (see Sect.~\ref{sec:resolution}). 
\textit{Bottom:} number of gas cells per logarithmic radius bin. A flat trend indicates a constant fractional resolution $r_{\rm cell}/r$. Horizontal dashed lines mark $\mathcal{N}=25$ and $\mathcal{N}=200$.}
\label{fig:resolution_criteria}
\end{figure}

To follow the gas inflow down to $R=1 \pc$, we define custom schemes that prescribe the target mass resolution as a function of radius and density as detailed below. AREPO refines and derefines cells according to this scheme to achieve the target resolution within a tolerance factor $\sim 2$. Figure~\ref{fig:region_scheme} defines four spatial regions to which different target masses are imposed. Figures~\ref{fig:resolution_diagrams} and \ref{fig:resolution_criteria} summarise the mass and spatial resolution of a typical snapshot as functions of density and radius. We divide the simulation domain into the following spatial regions (see Fig.~\ref{fig:region_scheme}):
\begin{itemize}

\item {\bf Nuclear Inflow Zone (NIZ)}. This is the region inside spherical radius $r<100$ pc. Inside this region we require the target mass resolution to be $M_{\rm cell}=40000\, (r/{\rm kpc})^{3.2} \Msun $, and we set an upper limit of $M_{\rm cell} = 20\Msun$. This choice ensures that the number of cells per logarithmic radial bin remains approximately constant (bottom panel in Fig.~\ref{fig:resolution_criteria}). 
For the SN simulation only, we impose an additional lower limit of $M_{\rm cell} = 10^{-2}\Msun$ during a strong inflow episode at $t\sim 173\Myr$. This is to prevent excessive refinement and numerical noise.
\item {\bf CMZ}. This is the region satisfying $r>100\pc$ (spherical radius) and $R<500\pc$ (cylindrical radius). We use a uniform target mass resolution $M_{\rm cell} = 20\Msun$ for this region. 
\item {\bf Bar}.  This is an elongated region comprised between $R=500\pc$ and an ellipse with semi-major and semi-minor of $3\kpc$ and $0.8\kpc$ respectively that is aligned and corotating with the Galactic bar. We use a target mass resolution of $M_{\rm cell} = 100\Msun$.
\item {\bf Disk}. This is defined as those space outside the bar region. We set $M_{\rm cell} = 1000\Msun$ for this region.
\end{itemize}

In addition to the above mass resolution requirements, we also enforce volume limits (see the middle panel of Fig.~\ref{fig:resolution_criteria}): 
\begin{itemize}
    \item {\bf NIZ}. We require that $25 <\mathcal{N}\equiv 2\pi r/r_{\rm cell}<200$, where $r$ is the Galactocentric radius of the cell and $r_{\rm cell}$ is the radius of a sphere with volume equal to the volume of the cell. This quantity measures the number of azimuthal resolution elements along a circle of radius $r$. The upper limit prevents cells from becoming too large in the innermost few pc, while the lower limit helps controlling the computational cost when massive clouds enter the CND and the timestep can fall to $\sim 1$\,yr while preserving the ability to resolve the CND-forming streams.
    \item {\bf CMZ}. We impose $ 0.2 < r_{\rm cell} < 30 \pc$.
    \item {\bf Bar and Disk}. We impose $ 0.2 \pc < r_{\rm cell}$.
\end{itemize}

\subsection{Non-equilibrium chemistry} \label{sec:chemistry}
We keep track of the chemical composition of the gas using the same non-equilibrium chemical network as \citet{Tress2025}. We briefly summarise the key properties here and refer readers to \citet{sor_etal_18}, \citet{tre_etal_24,Tress2025} and references therein for more details.

The network is an updated version of the N97 network from \citet{glo_cla_12}, based on the work of \citet{nel_lan_97} and \citet{glo_mac_07, glo_mac_07a}. It tracks the hydrogen and carbon chemistry using a simplified set of equations aimed at capturing the behaviour of integrated quantities such as the mean CO fraction. In particular, it solves for the non-equilibrium abundances of $\{\mathrm{H},\,\mathrm{H}_2,\,\mathrm{H}^+,\,\mathrm{C}^+,\,\mathrm{O},\,\mathrm{CO},\,\mathrm{e}^{-}\}$ in each computational cell.

Radiative heating and cooling is based on a cooling function that includes the effect of the non-equilibrium abundances calculated using the chemical network. This cooling function includes radiative cooling from fine–structure lines (chiefly $\mathrm{C}^+$ and $\mathrm{O}$), molecular rotational and vibrational lines (CO, $\mathrm{H}_2$), permitted transitions of H and He (Lyman-$\alpha$ etc.), optical and UV transitions of metals, and gas–dust energy exchange. It includes heating due to the photoelectric effect, $\Htwo$ photodissociation and cosmic ray ionisation (see Table~2 in \citealt{sor_etal_18} for a full list of the processes included). In addition, in the SNRad simulation we also account for photoionisation heating within HII regions. The rates of these heating and cooling processes are calculated based on the local state of the gas (density $\rho$, temperature of gas and dust $T_{\rm gas}$ and $T_{\rm dust}$, abundances $\{x_i\}$), on the local values of the attenuated non-ionising interstellar radiation field (ISRF, see Sect.~\ref{sec:ISRF}), on the value of the cosmic ray ionisation rate (CRIR, see Sect.~\ref{sec:ISRF}), and (if present) on the stellar feedback processes described in Section~\ref{sec:feedback} (in particular, energy injection by SN in both the SN and SNRad simulations, and ionising radiation from massive stars in the SNRad simulation).

\textbf{We assume the ISM to be at solar metallicity. The C and O abundances used in the cooling and chemistry calculations are fixed gas-phase values adopted from the local WNM measurements of \citet{sem_etal_00}\footnote{These values differ from undepleted solar abundances, for which all metals are assumed to remain in the gas phase.}. We do not include the mechanism of metallicity deposition from young stars for simplicity. To avoid numerical instability, we enforce a temperature floor of $T_{\rm floor} = 20$~K, as in \citet{tre_etal_20}.}

\subsection{Star formation} \label{sec:starparticles}

In galaxy-scale simulations such as those presented here, it would be extremely computationally intensive to resolve the build-up of individual stars and star clusters. Thus we have to resort to sub-grid models that replace high density regions with star particles.

We model star formation with a stochastic star-particle method \citep[e.g.][]{katz_92, kat_etal_96, sti_etal_06, hu_etal_17, smi_etal_21, gol_etal_25}. In particular, we use the same implementation as described by \citet{gol_etal_25} and \citet{Tress2025}. We first flag all regions that satisfy all the following criteria:
\begin{itemize}
 \item \textit{Jeans-mass criterion}. We flag the cell if the Jeans mass, computed from the local gas density and temperature, cannot be resolved by at least eight cells, i.e.\ $M_{\rm cell}>M_{\rm J}/8$, according to the \citet{tru_etal_97} and \citet{fed_etal_10} criterion.
 \item \textit{Negative velocity divergence} ($\nabla \cdot \mathbf{v}<0$). We require that star particles only form in converging flows. This criterion helps to distinguish between truly collapsing gas and dense non-collapsing gas.
 \item \textit{Local potential minimum}. We require that the star particle only forms in a local minimum of the gravitational potential.
\end{itemize}

Gas cells satisfying all the above criteria are stochastically converted into star particles at a rate ${\rm SFR} = \epsilon_{\rm SF} {M_{\rm cell}}/{t_{\rm ff}}$, where $t_{\rm ff}$ is the local free-fall time and $\epsilon_{\rm SF}=0.1$ is the star formation efficiency per free-fall time. If a cell reaches $M_{\rm cell}=M_{\rm J}$ without having formed stars, we increase the star formation efficiency to $\epsilon_{\rm SF} =1.0$ and stop further refinement since it would lead to spurious results. Fig.~\ref{fig:resolution_diagrams} shows the spatial and mass resolutions as functions of the number density, with colours indicating the different regions defined in Sect.~\ref{sec:resolution}.

Each star particle represents a stellar population with a typical total mass of $\sim 10 \mhyphen 1000\Msun$. We then ``populate'' these particles with stars with masses in the range $1\mhyphen 1000\Msun$ drawn from a \citet{kroupa_01} initial mass function (IMF). The stellar content of the particle is then used to determine the number of massive stars that explode as supernovae (Sect.~\ref{sec:supernovae}) and that act as sources of ionising radiation (Sect.~\ref{sec:radiation_transfer_sweep}). Note that in this approach, the mass of the star particle coincides with the mass of the stellar content of the particle only on average. Although in most cases the star-particle mass is comparable to that of its stellar content, it can happen that the stellar content of the star particle exceeds the mass of the star particle \citep{sor_etal_17}. This has a negligible effect on the overall feedback budget.

\subsection{Stellar feedback} \label{sec:feedback}
\subsubsection{Supernovae} \label{sec:supernovae}

The treatment of supernova feedback follows \citet{tre_etal_20} and \citet{Tress2025}. Star particles that contain stars with masses over $8\Msun$ produce supernova at the end of those stars' lives, with the lifetimes determined according to Table 25.6 in \citet{maeder_08}. Note that in this work we only include core-collapse supernovae that are produced by young massive stars, and unlike \citet{Tress2025} we do not include type Ia supernovae. This choice allows us to isolate the effect of feedback from young stellar populations. We leave a systematic exploration of the effect of Type Ia supernovae on gas inflow an interesting direction for future studies.

Supernova explosions inject either energy or momentum into the nearest neighbours of the injection cell depending on the local resolution:
\begin{itemize}

    \item \textit{Energy injection.} If the Sedov–Taylor phase is resolved, we inject $10^{51}{\rm erg}$ of thermal energy isotropically into nearest neighbours of the injection cell within the supernova remnant radius at the end of the Sedov–Taylor phase, with an upper limit of $100\pc$ on the injection radius. This energy input fully ionises the surrounding gas. The code then self-consistently evolves the system into a blast wave, converting part of the injected energy into momentum of the ISM.
    
    \item \textit{Momentum injection.} If the adiabatic Sedov–Taylor phase cannot be resolved well, we inject the supernova energy directly in kinetic form \citep{kim_ost_15} isotropically into the injection region. The injected momentum is set as the terminal momentum at the end of the Sedov-Taylor phase,    \[
    p_{\rm ST} = 2.6 \times 10^5 \left(\dfrac{\bar{n}}{\mathrm{cm}^{-3}}\right)^{-2/17}\Msun\kms.
    \]
    Here, $\bar{n}$ denotes the mean number density within the injection region. Gas cells within the injection region are heated to at least $10000\;K$.

\end{itemize}
The energy injection dominates the SNe explosion events in the CMZ region, and $< 6\%$ SNe are unresolved and use momentum injection.

In the pre-processing phase only (see Sect.~\ref{sec:simulation_phases}), the supernova returns the stellar mass into the surrounding cells. For each star particle, it stores a list of $N_{SN}$ scheduled SN events together with their explosion times. The re-injected mass $\Delta M$ is assigned so that each explosion returns a fraction $1/N_{SN}$ of the stellar mass represented by the star particle. This mass is then deposited into the injection region, which has a total volume $V_{\rm inj}$, by applying a uniform density increment $\Delta \rho = \Delta M/V_{\rm inj}$. The cell masses, corresponding momenta, and total energies are then updated consistently.

\subsubsection{Ionising radiation}
\label{sec:radiation_transfer_sweep}
This is only used in the SNRad simulation. We give a brief account here and refer to \citet{Tress2025} for further details. Star particles that contain OB stars act as sources of ionising radiation. We assume that each star is a black body with a radiation flux determined by their radius and surface temperature according to the Zero Age Main Sequence values of \citet{eks_etal_12}. For simplicity, we assume that the radiation from each star remains unchanged during its lifetime.

We use the SWEEP module \citep{pet_etal_23} implemented in AREPO to perform the radiation transfer (RT) calculations. SWEEP is a discrete-ordinates RT solver that assumes steady state with infinite speed of light, improving upon our earlier SPRAI algorithm \cite[][]{Jaura2018, Jaura2020}. RT equations are discretized as functions of time, coordinates, photon frequency and angular direction. It computes intensities by performing transport “sweeps” across the whole AREPO Voronoi mesh. These intensities are then passed to the chemical network (Sect.~\ref{sec:chemistry}) to compute the temperature and ionisation state of each gas cell. Sweeps are only performed on hydrodynamic synchronization steps. To capture rapid source/opacity variations and maintain tight RT--chemistry coupling, the numerical synchronization timestep should remain shorter than $\Delta t \sim 4\; \rm kyr$. At our adopted resolution, the actual hydrodynamical timesteps are typically $\lesssim 100\yr$ in both the SN and SNRad simulations, which is well below the required upper limit. We adopt a single frequency bin for ionising photons with energies above $13.6\; eV$. Using multiple ionising bins would improve the accuracy of the thermal and ionisation structure of HII regionss, although they do not largely affect the gas dynamics. Here we choose to use a single ionising bin for simplicity, however it would be interesting to include more frequency bins for better accuracy in ISM thermal states in the future. The computational cost of SWEEP scales with the number of cells, angular bins, and frequency groups, rather than with the number of ionising sources. Therefore it is well suited to many-source regimes, yielding substantially improved efficiency in radiation-transfer calculations with large numbers of sources.

In the current implementation, SWEEP does not fully treat line-driven self-shielding against photodissociating radiation in the Lyman-Werner bands. We therefore apply SWEEP only to ionising radiation in this work, while the non-ionising diffuse radiation field is treated separately.

\subsection{Non-ionising diffuse radiation field} \label{sec:ISRF}

All the simulations include a background diffuse non-ionising (photons with energy less than $13.6\;{\rm eV}$) interstellar radiation field ($\rm ISRF$) set at the standard value $G_{0}$ obtained from Solar-neighbourhood observations \citep{draine_78}. This is locally attenuated using the $\rm TreeCol$ algorithm \citep{cla_etal_12} based on the local amount of gas (within 30 pc of each computational cell) to account for dust extinction and $\Htwo$ self-shielding. Note that this ISRF only contributes to the spectrum of photons with energy less than $13.6\;{\rm eV}$. This is complementary to the radiation transfer method SWEEP described in Sect.~\ref{sec:radiation_transfer_sweep}, which only accounts for the ionising component (photons with energy above $13.6\;{\rm eV}$). Therefore, the ionising component is absent in the CHEM and SN simulations, while it is treated using SWEEP in the SNRad simulation. The CRIR of atomic hydrogen is fixed at $\zeta_{\rm H} = 3\times10^{-17}\;s^{-1}$  \citep{gol_lan_78} in all simulations. 

Although the strength of our assumed ISRF and CRIR are an underestimate for the CMZ \citep{cla_etal_13, oka_etal_19}, this is unlikely to cause major differences to the inflow rates. Since the radial inflow in our simulations is driven mainly by stellar-feedback-induced turbulence, the inflow rate is not expected to be sensitive to the adopted ISRF and CRIR. In addition, \citet{sor_etal_18} and \citet{tre_etal_20} have explored effects of ISRF and CRIR on large-scale gas dynamics within the inner Galaxy, and found only minor differences in mass fractions of gas phases and gas dynamics. We therefore adopt the same settings for consistency with previous work \citep{sor_etal_18, tre_etal_20, sor_etal_20, tre_etal_24}.

\subsection{Supermassive black hole}
\label{sec: SMBH_sink_particle}
We model the Milky Way SMBH as an accreting sink particle, using the implementation described by \citet{pet_etal_25} for the Noctua simulations in the arepoNoctua branch of AREPO. This approach allows the SMBH to grow by accreting mass that falls within the accretion radius of the central sink particle, and it prevents continuous mass build-up in the very centre that can lead to spurious star formation and stellar feedback. Although not included in the present work, this framework can be extended to study gas accretion onto Sgr~A* and the associated AGN activity.

The central sink particle has an initial mass of $4\times10^6\Msun$, slightly smaller than the observed value of $\sim4.1\times10^6\Msun$ \citep{gra_etal_19} to allow for the black hole mass to grow during the simulation. In our simulations, the accretion rate onto the SMBH is less than $10^{-5}\Msun\yr^{-1}$, which gives very minor accretion ($\delta M < 1000 \Msun$) over Phase II of $\sim 90\Myr$ (see Sect.~\ref{sec:simulation_phases}) compared to the its initial mass. Therefore we do not expect the feedback from SMBH to have much effect on the gas dynamics.

Since our simulations aim to follow gas inflow down to the innermost 1 pc, rather than the detailed dynamics within this region, we set the accretion radius to $0.5\pc$. The sink particle has a softening length of $0.5\pc$, equal to its accretion radius. This number is comparable the local gas-cell size of $\sim0.2\pc$. We set the mass resolution to $M_{\rm cell}=20\Msun$ within the central $1\pc$\footnote{The region at $r=0.05\mhyphen1\pc$ acts as a transition zone where cells increase their radius self-consistently to match the SMBH sink accretion radius. This is compatible with the zoom-in resolution scheme in Sect.~\ref{sec:resolution}, which is designed to trace inflows down to $r=1\pc$.}. This choice ensures that the sink accretion region is resolved by a few tens of gas cells, corresponding to an average gas-cell radius of $0.1\mhyphen 0.2\pc$.

\subsection{Initial condition} \label{sec:ic}
We adopt an initial gas density profile given by
\begin{equation}
\label{eq:rho_profile_current}
\rho_0(R,z) =  f(R)\times\rho_{\rm M17}(R,z)
\end{equation}
where
\begin{equation}
f(R) = 3-\frac{2}{1+\exp\left(-\frac{R+b}{x_2}\right)}
\end{equation}
with $b=-1.8\kpc$ and $x_2=0.5\kpc$, and
\begin{equation}
\rho_{\rm M17}(R,z) =
\frac{\Sigma_1\exp{\left(-\frac{R_{\rm m1}}{R}-\frac{R}{R_{\rm d1}}\right)}}{2z_1\cosh^2(z/z_{1})} 
+ 
\frac{\Sigma_2\exp\left(-\frac{R_{\rm m2}}{R}-\frac{R}{R_{\rm d2}}\right)}{2z_2\cosh^2(z/z_{2})}
\end{equation}
is the gas density profile from \citet{mcmill_17}, with $\Sigma_{1} = 53.1 \;M_{\odot}\;\pc^{-2}$, $R_{\rm m1} = 4$, $R_{\rm d1} = 7$, $z_1=170\pc$, $\Sigma_{2} = 2180\;M_{\odot}\;\pc^{-2}$, $R_{\rm m2} = 12$, $R_{\rm d2} = 1.5$, and $z_2=90\pc$. 
The function $f(R)$ increases the mass in the innermost 3 kpc compared to the profile from \citet{mcmill_17}. We truncate the initial profile \eqref{eq:rho_profile_current} at $5\kpc$ as we are interested in the inner regions only. This profile yields an initial total gas mass of $1.26\times10^9\Msun$ within $R=5\kpc$.

The gas velocities are initialised to follow circular motion at each height $z$ above the plane, with the centrifugal force balancing gravity:

\begin{equation}
    v_0(R,z) = \left[R\dfrac{\partial\Phi_{\rm ext}(R,z)}{\partial{R}}\right]^{\frac{1}{2}} \,.
\end{equation}

We adopt the adiabatic equation of state for all three models. The initial temperature is set to be $10000\;K$.

\subsection{Simulation phases}
\label{sec:simulation_phases}
The simulations are evolved through a series of preparatory phases. These allow the system to self-adjust to the subgrid physics, avoid violent star formation cycles, reach quasi-equilibrium, remove transients, and reduce computational costs. The phases are as follows:
\begin{itemize}
\item \textit{Pre-processing phase} ($-120<t<0$ Myr). We start with the gas disk of Sect.~\ref{sec:ic} and evolve it in an axisymmetrised version of the gravitational potential of Sect.~\ref{sec:gravpot} (we only keep the monopole component) for $120$ Myr. Star formation is enabled with a shortened stellar lifetime (one-tenth of the fiducial value) to build turbulence and allow the disk to reach a quasi-equilibrium state. For both SN and SNRad simulations, only supernova feedback is active in this phase. We adopt a relatively low mass resolution of $500\Msun$. Only for this phase, we allow supernova feedback to re-eject stellar mass into the surrounding gas to keep the total gas mass nearly unchanged. 
\item \textit{Phase I} ($0<t<150$ Myr). We start with the axisymmetrised external potential and linearly turn on the non-axisymmetric component over $t=150\Myr$. This is customary in simulations of gas flow in barred potentials and avoids transients \citep[e.g.][]{athana_92,tre_etal_20a}. The SMBH sink particle is introduced at the beginning of this phase. We use the resolution settings described in Sect.~\ref{sec:resolution}, except for the inner $100\pc$ where we use a constant mass resolution $M_{\rm cell} = 20 M_{\odot}$. During this phase, we linearly increase the star-particle lifetime from one-tenth of its fiducial value to the fiducial value during the first 100$\Myr$. For the SNRad simulation, we switch on radiation feedback at $t=100\Myr$.
\item \textit{Phase II} ($150<t<240$ Myr). In this phase the bar and spiral potentials are fully turned on, and the resolution is as described in Sect.~\ref{sec:resolution}. At the beginning of the phase we reset the mass of the SMBH to its initial value of $4\times10^6\Msun$ \footnote{The mass of the SMBH sink increases marginally by around few hundreds of solar mass during Phase II, which has a negligible effect on the gravitational potential.}.

For the SN simulation without radiation feedback, we adopt higher resolution within the NIZ only after $t=160$ Myr, since radial inflow is negligible at earlier times. This is the production phase that will be used for the actual inflow analysis.

\end{itemize}

\section{Results}
\label{sec:results}

\subsection{General gas and stellar properties} \label{sec:CMZgeneral}

\subsubsection{Gas morphology}  
\label{sec:morphology_multiscale}

\begin{figure*}[h!]
\includegraphics[width=\textwidth]{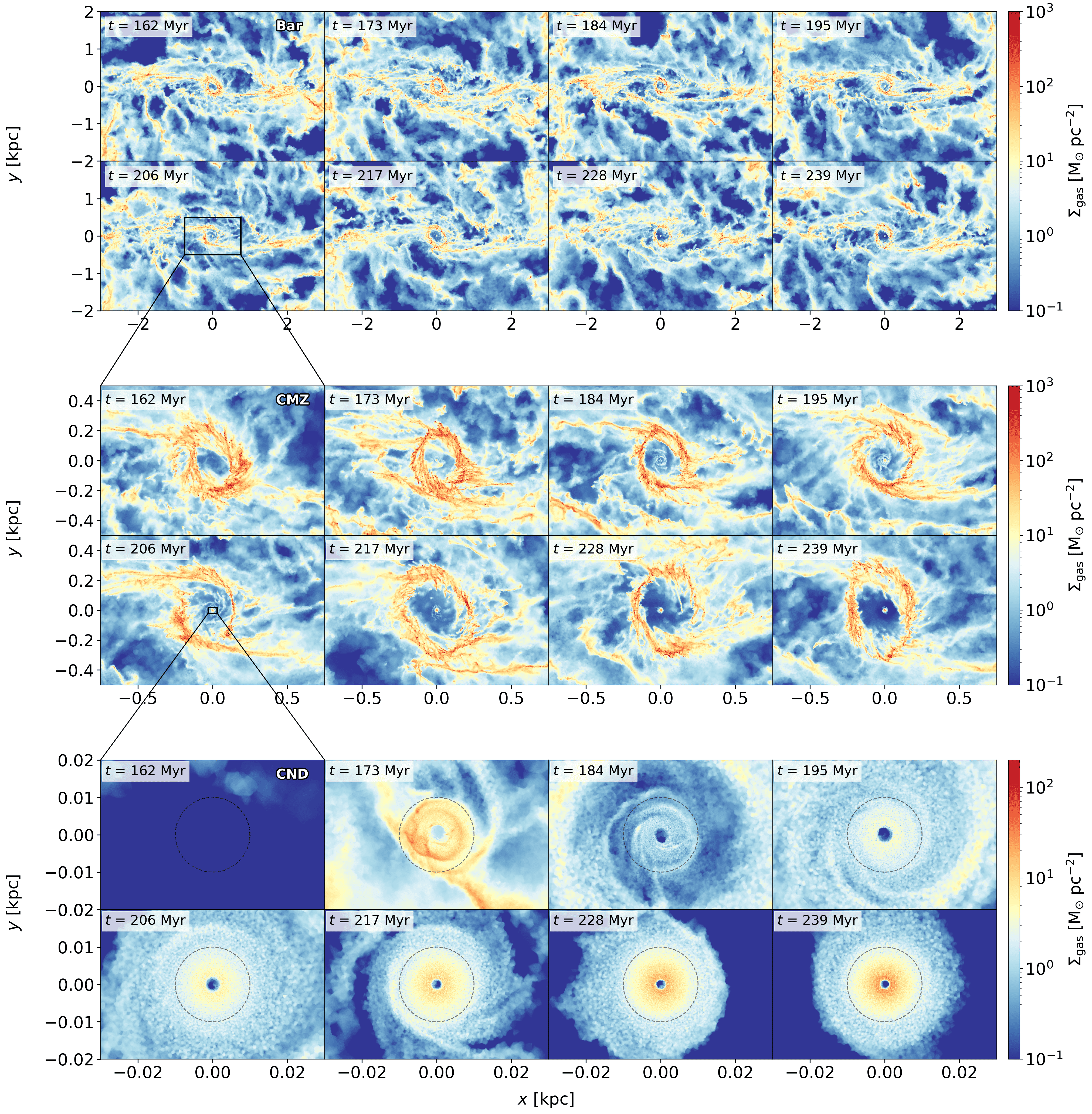}
\centering
\caption{Evolution of gas surface density in the SN simulation. Panels from top to bottom progressively zoom into the central regions. The bar major axis is aligned with the $x$-axis in all panels and rotates counterclockwise. The dashed circles in the bottom panel indicate $R=10$ pc.}  
\label{fig:morphology_multiscale_SN.png}
\end{figure*}

\begin{figure*}[h!]
\includegraphics[width=\textwidth]{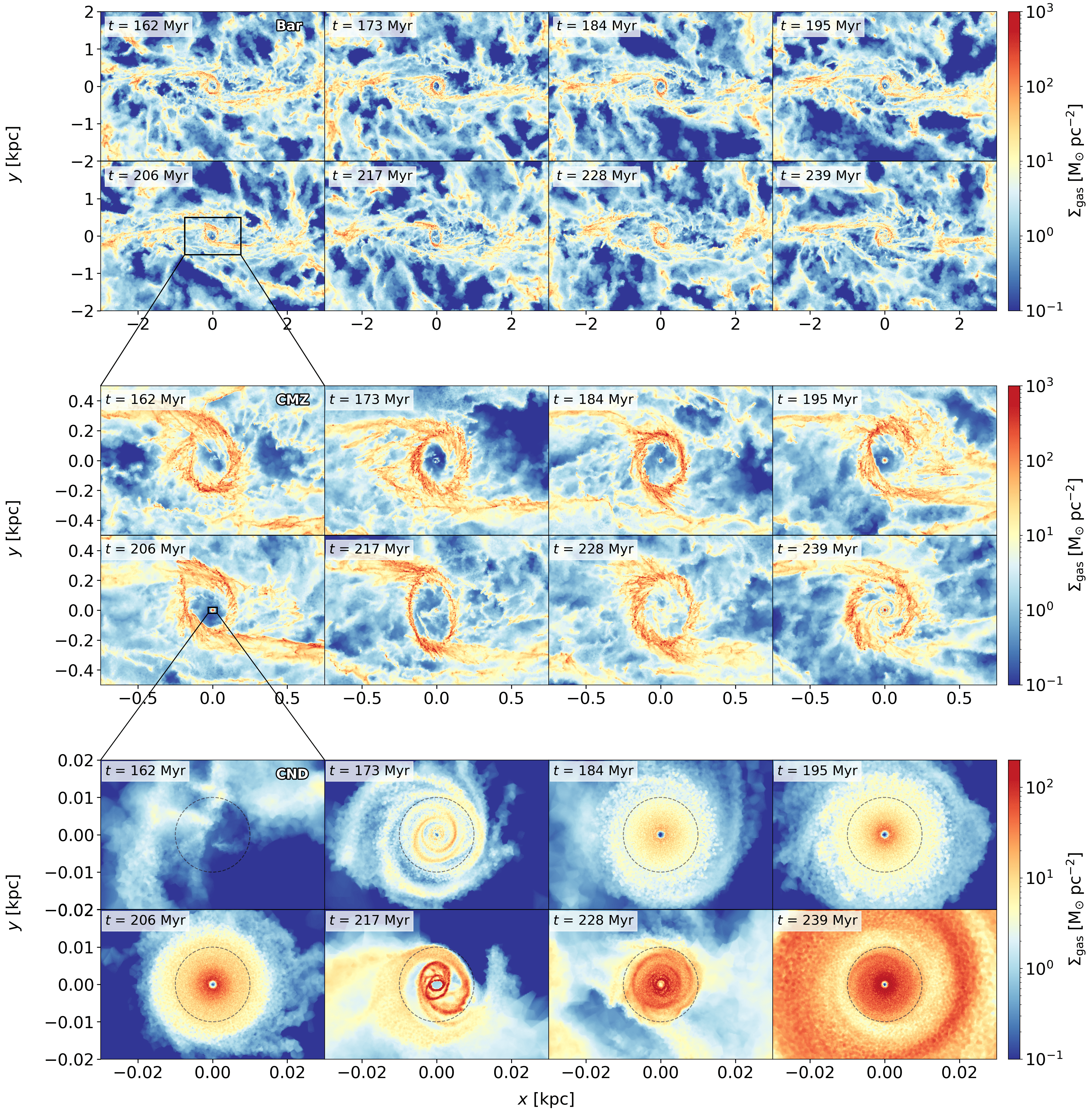}
\centering
\caption{Same as Fig.~\ref{fig:morphology_multiscale_SN.png} but for the SNRad simulation.}
\label{fig:morphology_multiscale_SNRad.png}
\end{figure*}

In this section we present the overall gas morphology in the SN and SNRad simulations with self-gravity, star formation and stellar feedback. In Figs.~\ref{fig:morphology_multiscale_SN.png} and~\ref{fig:morphology_multiscale_SNRad.png} we show the time evolution of the gas morphology at three spatial scales corresponding to three distinct structures: the bar on the largest scales (top panels), the CMZ ring on intermediate scales (middle panels), and the CND on the smallest scales (bottom panels) \footnote{The central hole with a radius of $\sim0.05pc$ inside the CND is produced by the SMBH sink. In our test simulations with a smaller SMBH accretion radius of $\sim 0.005\pc$, gas flows inwards to the innermost $0.1\pc$.}. 

On the largest scale (top panels), the simulations exhibit the characteristic bar-driven gas flow pattern discussed in numerous previous studies \citep[e.g.][]{athana_92, kim_etal_12, tre_etal_20a}. The stellar bar efficiently channels gas inward along the bar lanes. At intermediate scales (middle panels), the inflowing gas settles into a dense CMZ ring at $R \simeq 200\pc$. The CMZ ring has relatively high surface densities of $\Sigma_{\rm gas} \sim 50 \Msun\pc^{-2}$ compared to the bar and disk region (typically $\Sigma_{\rm gas} \sim 5 \Msun\pc^{-2}$), and is the site of intense star formation.

Stellar feedback drives turbulence and displaces gas in the vicinity of star-forming regions, redistributing the angular momentum and enabling a fraction of the material to migrate inward from the CMZ. Gas accumulates over time in the central $100\pc$ (the NIZ). On the smallest scale (bottom panels), where the gravitational potential is dominated by the NSC and the SMBH, inflowing clouds accumulate into a disk-like structure with a characteristic size of $\sim 2 \mhyphen10,\pc$ (see the dashed circle at $R = 10\pc$ in the bottom panels). We identify this structure as the CND.

The main differences between the SN and SNRad simulations are as follows (compare Figs.~\ref{fig:morphology_multiscale_SN.png} and \ref{fig:morphology_multiscale_SNRad.png}). In the SNRad run, radiation feedback increases the ionised gas fraction around star-forming regions and produces a smoother overall gas morphology relative to the SN simulation. The SNRad simulation produces a more massive CND compared to the SN simulation, primarily because of episodic radial inflow events (see Sect.~\ref{sec:episodic_radial_inflow}). Such events are present but less frequent in the SN simulation.

The CND remains more massive in the SNRad run throughout the simulation. Over $t=160\mhyphen240\Myr$, the time-averaged inflow rate at $R=10\pc$ in the SNRad run is $\dot{M}_{\rm 10pc}^{\rm SNRad}\sim2\times10^{-4}\Msun\yr^{-1}$, about one order of magnitude higher than in the SN run ($\dot{M}_{\rm 10pc}^{\rm SN}\sim2\times10^{-5}\Msun\yr^{-1}$). At the end of the simulation, the CND mass in the SNRad run becomes comparable to the observed CND mass of $M_{\rm CND}\sim 10^4\Msun$ in the Milky Way. By contrast, the SN run gives $\sim2\times10^3\Msun$ at the same time.

\begin{figure*}[ht!]
\includegraphics[width=\textwidth]{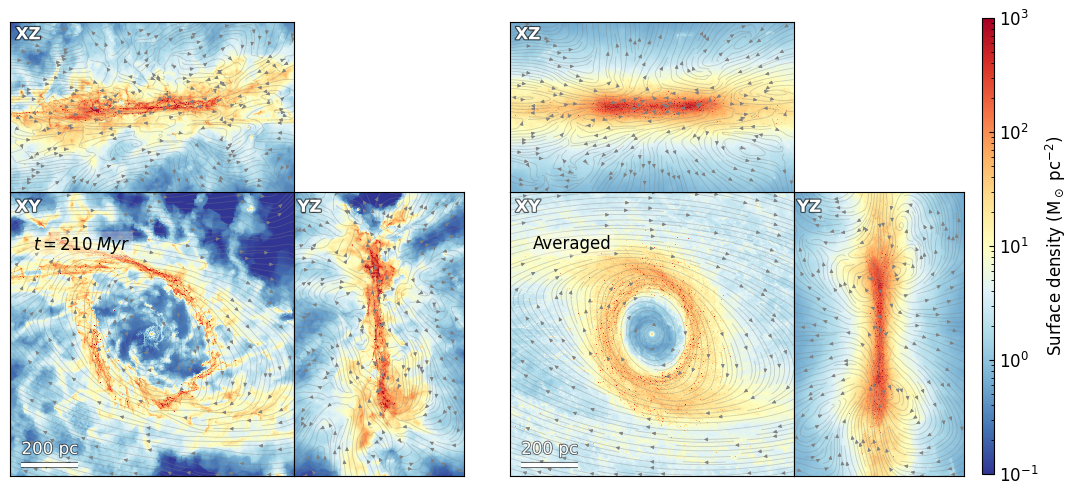}
\centering
\caption{\textit{Left:} Instantaneous gas streamlines superimposed on the CMZ surface density for the SN simulation at $t=210\Myr$. Projected velocities are mass-weighted integrating along the line of sight. Projections onto the ${\rm XY}$, ${\rm XY}$, and ${\rm XZ}$ planes highlight gas motions and the CMZ structure. \textit{Right:} Time-averaged streamlines and surface density over $t=180\mhyphen 230\Myr$ using 50 snapshots ($\Delta t=1\Myr$).}
\label{fig:streamline_SN.png}
\end{figure*}

\begin{figure*}[ht!]
\includegraphics[width=\textwidth]{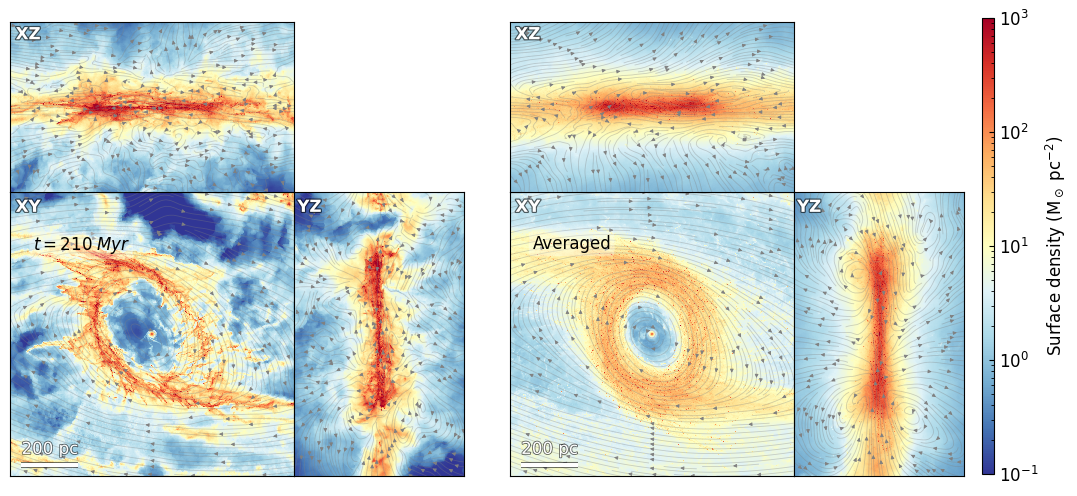}
\centering
\caption{Same as Fig.~\ref{fig:streamline_SN.png}, but for the SNRad simulation.}
\label{fig:streamline_SNRad.png}
\end{figure*}

Figures~\ref{fig:streamline_SN.png} and \ref{fig:streamline_SNRad.png} describe instantaneous and time-averaged gas streamlines overlaid on face-on and edge-on surface-density maps. In the bar, motions are strongly non-circular and follow the  $x_1$ orbit family as discussed in previous work \citep[e.g.][]{athana_92, sel_wil_93,kim_etal_12,li_etal_15, sor_etal_15a}. Inside the CMZ, motions are only mildly non-circular and follow the $x_2$ orbit family. Active star formation in the CMZ drives vertical fountain flows that are visible in edge-on views, especially near the CMZ edges. The ejected gas typically reaches heights of $|z|\sim200\pc$, then drifts mainly outwards before falling back towards regions outside the CMZ. The time-averaged streamlines show ``8''-shaped loops in the $yz$ plane, indicative of continuous cycling.

\begin{figure*}[t!]
\includegraphics[width=\textwidth]{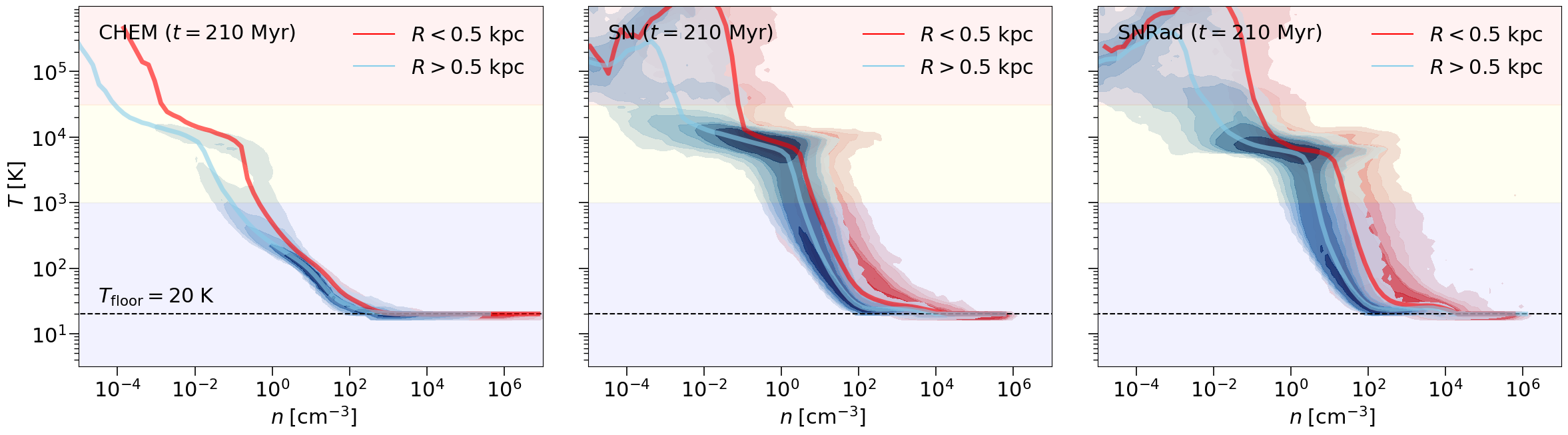}
\centering
\caption{Phase diagrams for CHEM (left), SN (middle), and SNRad (right). Red and blue contours show the distributions for $R<0.5\kpc$ and $R>0.5\kpc$, respectively. The red and blue solid curves show the corresponding median temperature as a function of gas number density. The dashed line marks the temperature floor at $T_{\rm floor} = 20 \;K $. Blue, yellow, red shaded regions highlight the three thermal phases (cold $T < 10^3$ K; warm $10^3 < T < 10^{4.5} \; K$; hot $T > 10^{4.5}\; K$). The warm ($T\simeq 10^4\; K$) and hot ($T\simeq 10^6\; K$) phases are almost absent in the CHEM simulation, while they are present in the SN and SNRad simulations as a result of stellar feedback. }
\label{fig:phase_diagrams}
\end{figure*}

\begin{figure*}[ht!]
\includegraphics[width=\textwidth]{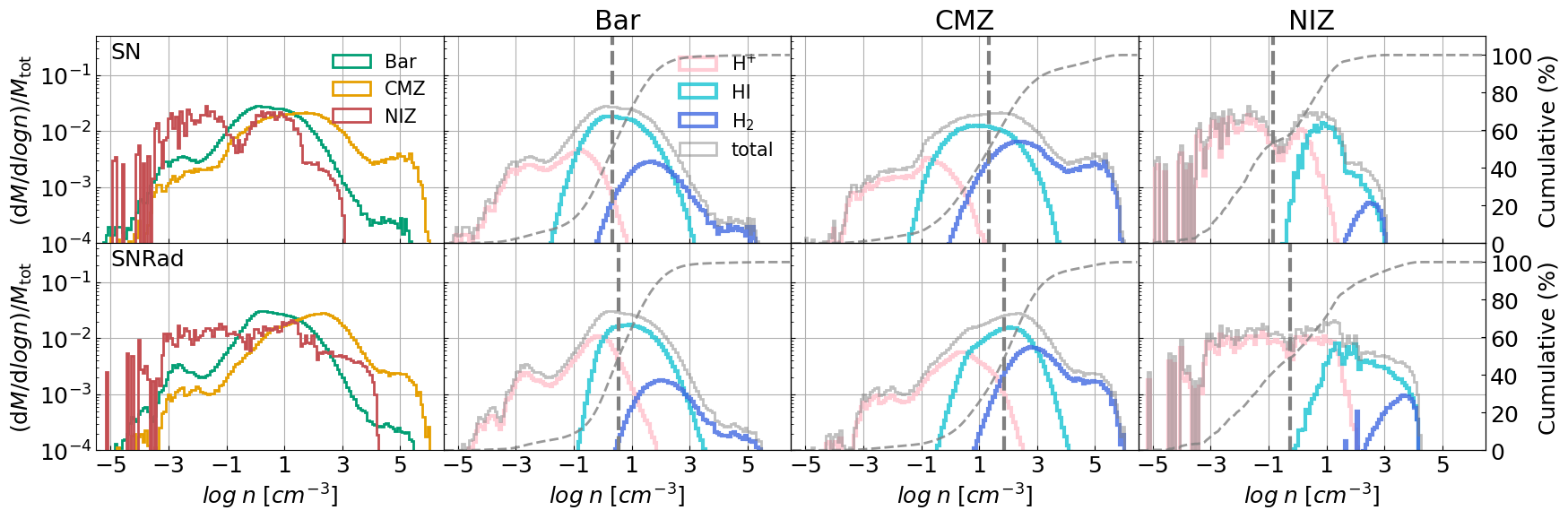}
\centering
\caption{\textit{Left:} Normalised mass distribution as a function of number density in the bar (green), CMZ (orange), and CND (red) at $t=210\Myr$. Top and bottom panels correspond to SN and SNRad simulations, respectively. \textit{Other panels:} Same as the left panel but for chemical tracers. Deep blue, light blue and pink shading indicates $\rm H_2$, $\rm H$, and $\rm H^{+}$ fractions, respectively. Columns show bar, CMZ, and CND (left to right).}
\label{fig:chemistry_f(n)} 
\end{figure*}

\begin{figure*}[ht!]
\includegraphics[width=\textwidth]{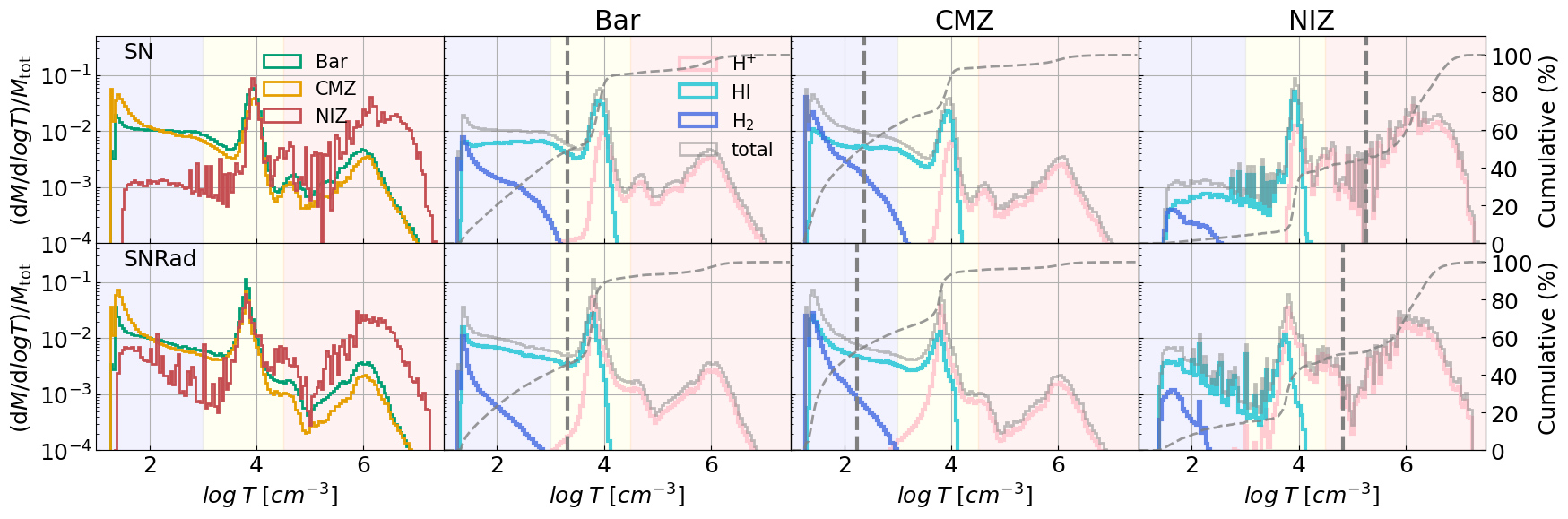}
\centering
\caption{Same as Fig.~\ref{fig:chemistry_f(n)}, but as a function of gas temperature. The vertical shaded regions use the same color coding as Fig.~\ref{fig:phase_diagrams}.}
\label{fig:chemistry_f(T)}
\end{figure*}

\begin{table}
\centering
\begin{tabular}{l l cc}
\toprule
Region & Quantity & SN & SNRad \\
\midrule
\multirow{6}{*}{CMZ}
 & $T_{\rm cold}\;(K)$        & $10^{2.13}$ & $10^{2.06}$ \\
 & $T_{\rm warm}\;(K)$        & $10^{3.86}$ & $10^{3.81}$ \\
 & $T_{\rm hot}\;(K)$         & $10^{6.07}$ & $10^{6.07}$ \\
 & $n_{\rm cold}\;(cm^{-3})$  & $10^{4.33}$ & $10^{4.39}$ \\
 & $n_{\rm warm}\;(cm^{-3})$  & $10^{0.85}$ & $10^{2.32}$ \\
 & $n_{\rm hot}\;(cm^{-3})$   & $10^{-0.48}$ & $10^{-0.36}$ \\
\midrule
\multirow{6}{*}{Bar}
 & $T_{\rm cold}\;(K)$        & $10^{2.33}$ & $10^{2.28}$ \\
 & $T_{\rm warm}\;(K)$        & $10^{3.85}$ & $10^{3.81}$ \\
 & $T_{\rm hot}\;(K)$         & $10^{6.02}$ & $10^{5.98}$ \\
 & $n_{\rm cold}\;(cm^{-3})$  & $10^{3.08}$ & $10^{3.09}$ \\
 & $n_{\rm warm}\;(cm^{-3})$  & $10^{0.28}$ & $10^{0.90}$ \\
 & $n_{\rm hot}\;(cm^{-3})$   & $10^{-1.40}$ & $10^{-1.40}$ \\
\bottomrule
\end{tabular}
\caption{Comparison of gas properties between the SN and SNRad simulations, time-averaged over $t=162\mhyphen240\,\mathrm{Myr}$. The quantities are mass-weighted averaged temperature and density for the three gas phases (cold, warm, and hot).}
\label{tab:ComparisonTable}
\end{table}

\subsubsection{Multi-phase ISM}

Figures~\ref{fig:phase_diagrams},  ~\ref{fig:chemistry_f(n)} and \ref{fig:chemistry_f(T)} present the phase diagram and mass/temperature histograms in our simulations at $t=210\Myr$. The snapshot is chosen after the ISM reaches a quasi-equilibrium state. Choosing another snapshot near this stage does not significantly affect the measured quantities.

The warm and hot phases are almost absent in the CHEM simulation due to the lack of stellar feedback, while they are present in the SN and SNRad simulations. The SNRad simulation has a higher fraction of warm ionised gas compared to the SN simulation, as the radiation feedback increases the temperature at star-forming sites ($n > 100\cm^{-3}$). As a result, the warm-phase ISM has a higher mass-weighted number density in the SNRad simulation than in the SN simulation, both in the CMZ and in the bar region (see Table~\ref{tab:ComparisonTable}). In both simulations the gas in the CMZ ($R<0.5\kpc$) is denser and has larger molecular fraction than in the bar region ($R>0.5\kpc$). The median temperature does not show a significant difference (Fig.~\ref{fig:chemistry_f(T)}). The chemical composition of the gas is approximately constant as a function of time (see Fig.~\ref{fig:chemistry_f(t)} in Appendix).

The gas in the NIZ is generally hotter and contains a larger fraction of diffuse ionized gas ($n < 10^{-1}\cm^{-3}$) than the gas in the CMZ and bar regions. This difference indicates that radial inflow is not simply a direct inward transport of the ISM from the CMZ or bar region. Instead, the inflow is either dominated by particular gas phases or changes the chemical and thermal state of the ISM during transport.

\subsubsection{CMZ mass and SFR}
\label{sec:CMZmass_SFR}

\begin{figure}[t!]
\includegraphics[width=\columnwidth]{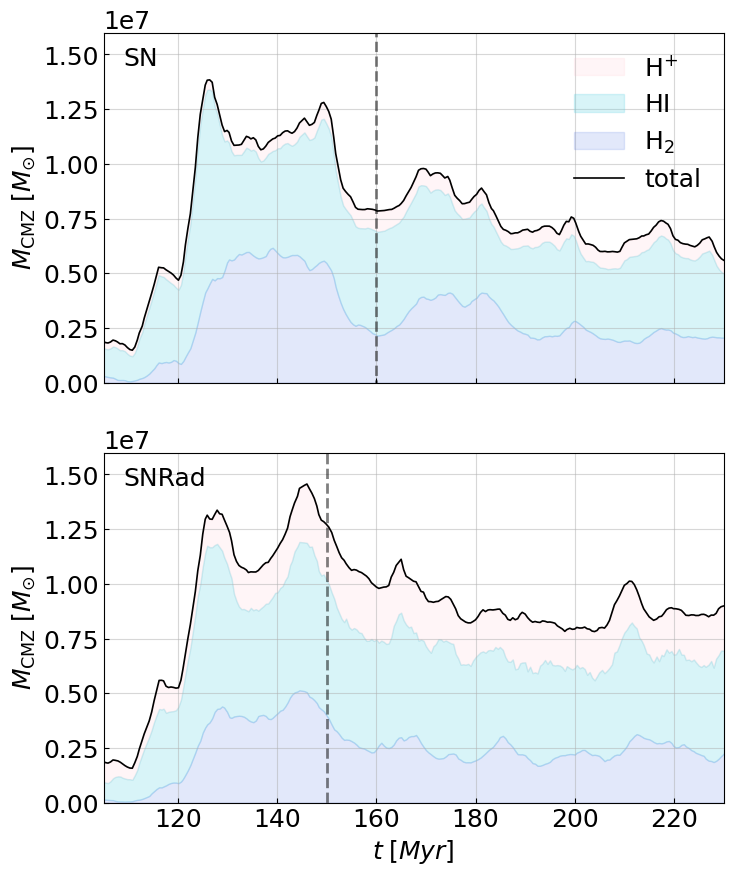}
\centering
\caption{Enclosed gas mass within the CMZ ($R<0.5\kpc$) versus time for SN (top) and SNRad (bottom) simulations. The black line shows total gas mass. The vertical dashed line marks the starting time of Phase II at $t=160\Myr$ for the SN simulation ($t=150\Myr$ for the SNRad). This is the production phase that we use for the inflow analysis (see Sect.~\ref{sec:simulation_phases}). Deep blue, light blue and pink curves indicate $\rm H_2$, $\rm H$, and $\rm H^{+}$ fractions, respectively.} 
\label{fig:Mgcmz.png}
\end{figure}

\begin{figure}[t!]
\includegraphics[width=\columnwidth]{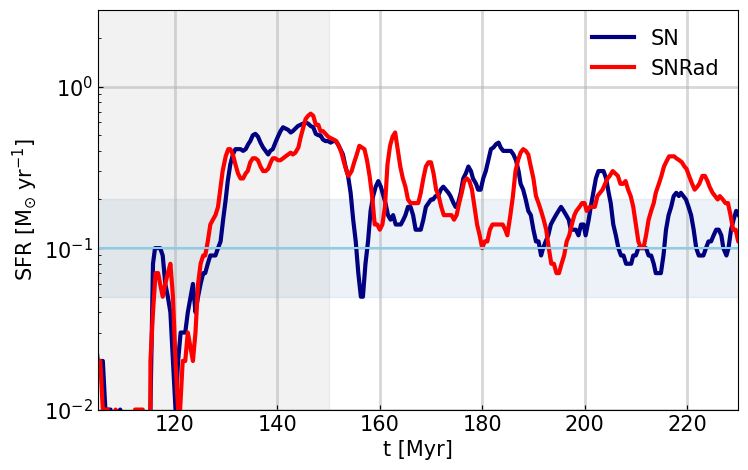}
\centering
\caption{Star formation rate within the CMZ ($R<0.5\kpc$) versus time for SN (blue) and SNRad (red) simulations. The grey shaded region marks the bar turn-on period at $t<150\Myr$ (see Sect.~\ref{sec:simulation_phases}). The horizontal blue band shows the observed range $0.05\mhyphen0.2 \Msun\yr^{-1}$ (see references in Sect.~\ref{sec:CMZmass_SFR}). After the models reach quasi-equilibrium, the SFR lies within the observed range.}
\label{fig:SFR.png}
\end{figure}

Figures ~\ref{fig:Mgcmz.png} and ~\ref{fig:SFR.png} show the time evolution of the CMZ gas mass and star formation rate (SFR) in our simulations. The CMZ mass and the SFR increase during the first $\sim 150 \Myr$ (Phase I, see Sect.~\ref{sec:simulation_phases}) as the bar drives gas inwards along bar lanes. At later times ($t > 150\Myr$), the system approaches a quasi-steady state in which gas inflow from the bar is approximately balanced by gas consumption due to star formation. The CMZ mass stabilises around $M_{\rm CMZ} \lesssim 10^7\Msun$, a value slightly lower than the observational estimate that is primarily based on molecular gas, $M^{\rm obs}_{\rm CMZ}\sim2^{+2}_{-1} \times10^7\Msun$ \citep{dah_etal_98, lon_etal_13, bat_etal_25a}. The observational uncertainties mainly arise from the assumed $X_{\rm CO}$ conversion factor, the subtraction of foreground and background emission, and the assumed dust-to-gas ratio. We refer readers to Sect.~5.3 of \citet{bat_etal_25a} for a more detailed discussion of these uncertainties. For comparison, in the CHEM simulation, in which star formation is turned off, the CMZ mass grows to $M_{\rm CMZ}\sim 5\times10^7\Msun$ at $t \simeq 150 \Myr$. Atomic hydrogen dominates the mass budget in both SN and SNRad simulations. Ionised gas makes only a small contribution in the SN simulation, but a somewhat larger one in SNRad, owing to the influence of the radiation feedback.

The SFR stabilises at ${\rm SFR}\sim 0.2\Msun\yr^{-1}$, consistent with observational estimates of ${\rm SFR} = 0.05\mhyphen0.2 \Msun\yr^{-1}$ (horizontal blue band in Fig.~\ref{fig:SFR.png}), with temporal fluctuations of approximately a factor of $\sim 2$. We refer readers to Sect.~3.1 in \citet{hen_etal_23} and Sect.~4.1 in \citet{sor_etal_20} for detailed review of the observationally determined SFR through difference sources. This level of variability is comparable to that reported in the simulations of \citet{sor_etal_20} and \citet{moo_etal_22}, and contrasts with the results of \citet{ari_etal_19}, who found SFR variations of $\sim 1.5$ dex despite a relatively constant CMZ gas mass. It has been proposed \citep[e.g. Section 3.3 of ][]{hen_etal_23} that the discrepancy between the results of \citet{ari_etal_19} and those of \citet{sor_etal_20} and \citet{moo_etal_22} may arise either because the latter include supernova feedback but lack early stellar feedback, or because insufficient numerical resolution of $2000\Msun$ in the simulations of \citet{ari_etal_19} leads to unphysical behaviour. Since our SNRad simulations include early feedback processes, our results point towards insufficient numerical resolution in \citet{ari_etal_19} as the more likely of the two explanations for their reported large-amplitude SFR variability.

\subsection{Inflows and outflows} \label{sec:inflow}
\subsubsection{Net inflow as a function of time and radius} \label{sec:netinflow}

We start investigating the inflow by considering net mass flows. Mass conservation in any (stationary) volume $V$ implies: 

\begin{equation}
    \label{eq:mass_conservation_2}
    \dot{M} = \mathrm{SFR} + \dot{M}_{\rm gas} \,, 
\end{equation}

where $\dot{M}$ is the net mass inflow into the volume $V$, $\mathrm{SFR}$ is the star formation rate within $V$, and $\dot{M}_{\rm gas}$ is the rate of change in the total gas mass inside $V$. Note that we assumed that the accretion onto the SMBH sink is negligible.

To calculate $\dot{M}$ we proceed as follows. We accurately record the birth position and masses of star particles (this is done at the timestep level, with a time granularity of $\lesssim 100\yr$, rather than at the much coarser output snapshot level with a spacing of $0.5\Myr$). We then divide the total stellar mass $\Delta M_{\rm star}^{\rm born}$ formed within the considered volume between two consecutive snapshots by the time separation $\Delta t$ between the two snapshots to obtain the term $\mathrm{SFR} = \Delta M_{\rm star}^{\rm born}/\Delta t$ on the right-hand-side of Eq.~\eqref{eq:mass_conservation_2}. The term $\dot{M}_{\rm gas}$ is straightforward to calculate by dividing the difference in the total gas mass $M_{\rm gas}$ within the considered volume between two consecutive snapshots by $\Delta t$. We then sum the two to obtain $\dot{M} = {\rm SFR} + \dot{M}_{\rm gas}$.

\begin{figure*}[ht!]
\includegraphics[width=\textwidth]{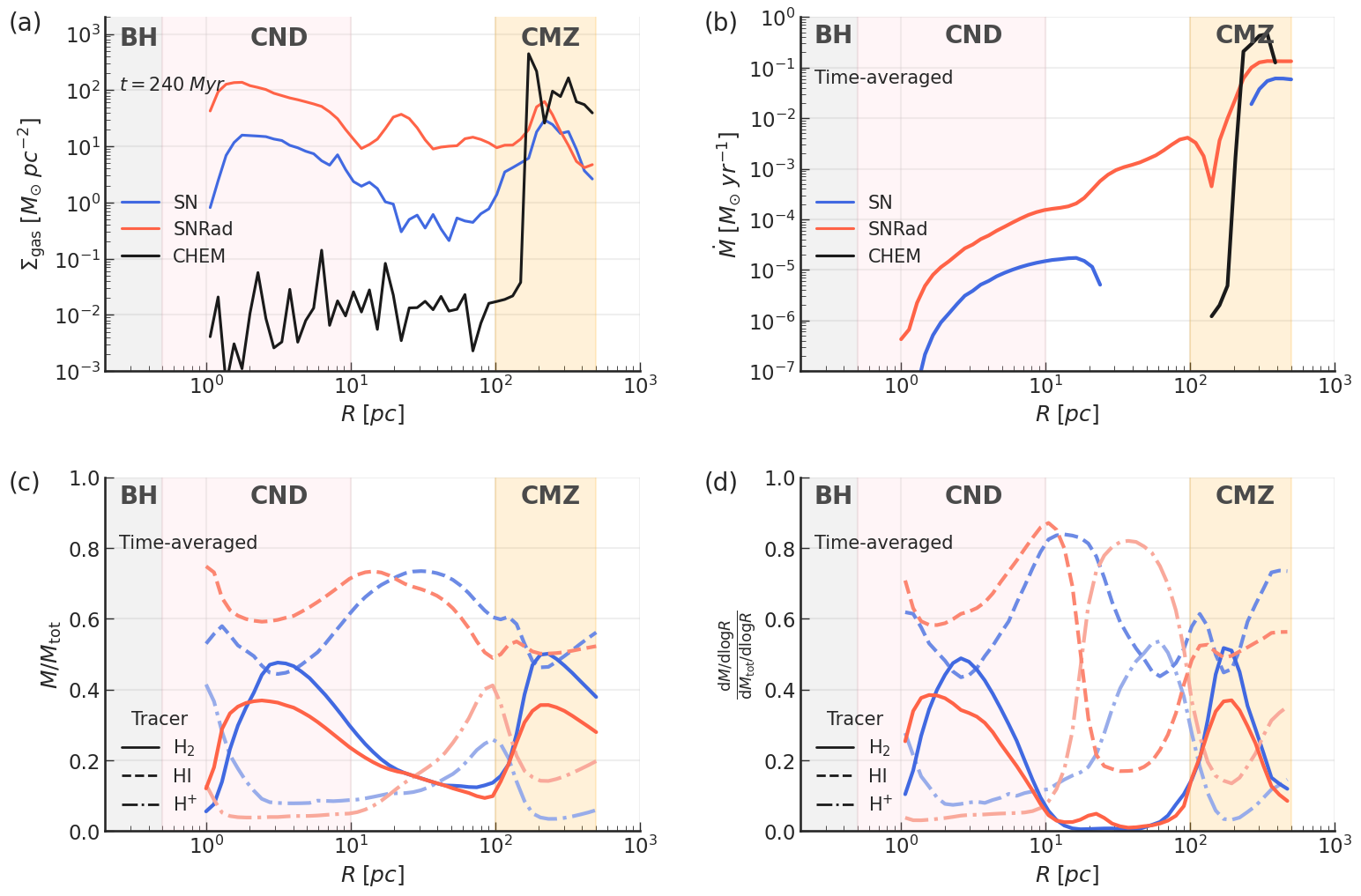}
\centering
\caption{\textit{Panel a:} Gas surface density as a function of cylindrical Galactocentric radius for CHEM, SN, and SNRad at $t=240\Myr$. Colors are the same as in Fig.~\ref{fig:net_inflow_t}. \textit{Panel b:} Time-averaged net inflow rate over $t=162\mhyphen240\Myr$ as a function of radius. SNRad generally produces higher inflow rates than SN. In CHEM, the black curve appears mainly near the CMZ because no mechanism drives gas inflow from the CMZ to smaller radii. \textit{Panel c:} Time-averaged cumulative mass fractions of $\Htwo$ (solid), $\HI$ (dashed), and $\Hp$ (dot-dashed) as a function of radius. \textit{Panel d:} Same as panel c, but showing the mass fractions computed in each logarithmic radial bin.}
\label{fig:net_inflow_R}
\end{figure*}

Panel b of Fig.~\ref{fig:net_inflow_R} shows the time-averaged inflow rate as a function of radius. At large radii ($R=500\pc$) the inflow is dominated by the bar potential and is similar for all three simulations. The inflow decreases quickly with radius for all simulations. Inside the NIZ ($R<100\pc$), the inflow is virtually zero in the CHEM simulation (black line), present but very small in the SN simulation (blue line), and much larger in the SNRad simulation (red line). At $R\simeq 10\pc$, the time averaged net inflow rate in SNRad simulation ($\Mdot\sim2\times10^{-4} \Aunit$) is about an order of magnitude higher than in the SN run ($\Mdot\sim3\times10^{-5} \Aunit$). As we will see below (Sect.~\ref{sec:episodic_radial_inflow}) this difference is mainly due to episodic radial inflow events that are more frequent in the SNRad run than in the SN run. These events temporarily increase the inflow rate to $\Mdot\sim10^{-3} \Aunit$.

\begin{figure*}[t!]
\includegraphics[width=0.9\textwidth]{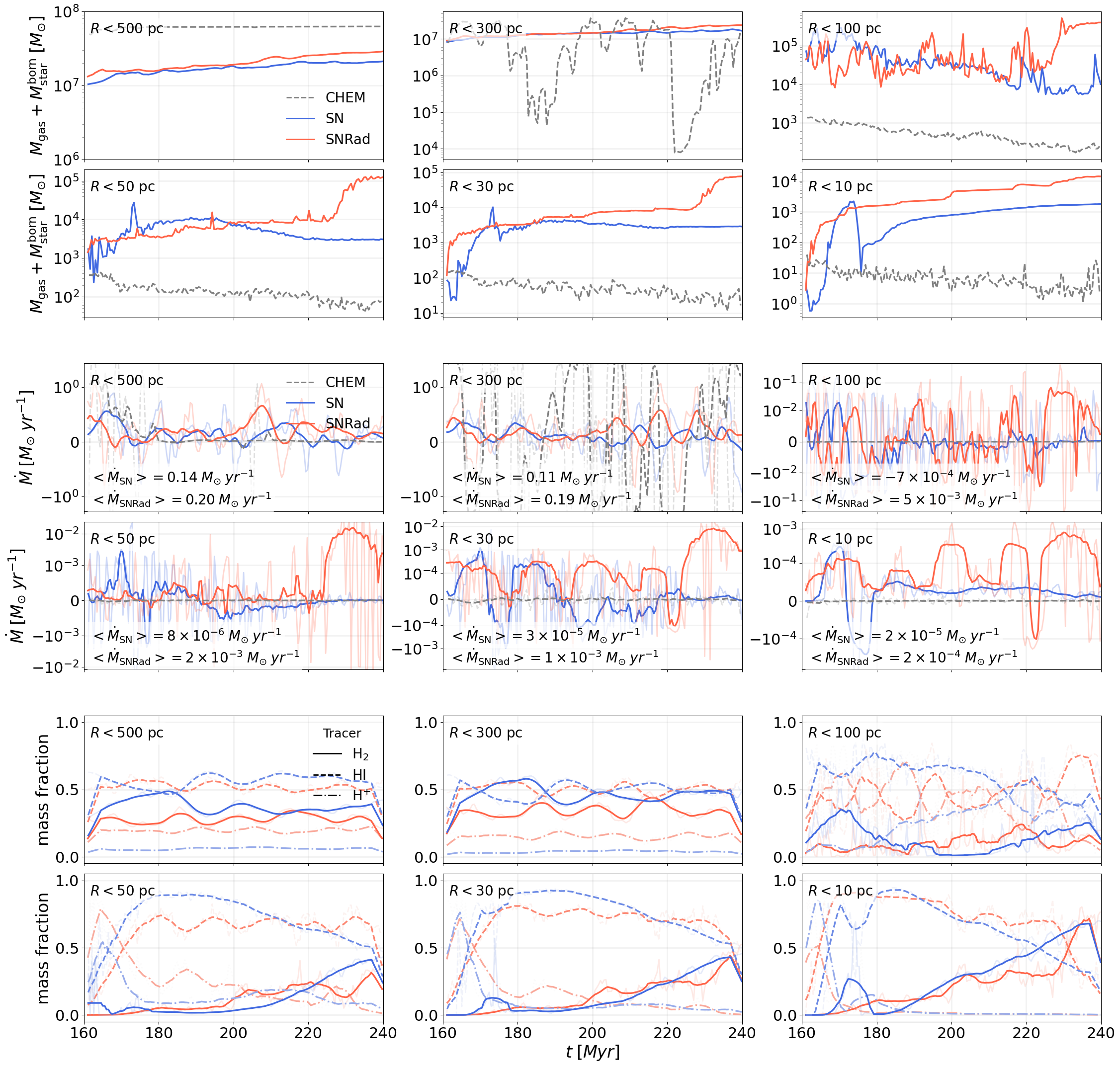}
\centering
\caption{\textit{Top rows:} Sum of the total mass of gas within a given cylindrical volume enclosing $|z| = R/2$ plus total mass of stars born within the same cylindrical volume as a function of time for the CHEM (black dashed), SN (blue), and SNRad (red) simulations. Note that $M^{\rm born}_{\rm gas}$ is \emph{not} the total mass of stars currently within the volume, but the total mass of stars born within the volume (that may have now moved to different volumes). \textit{Middle rows:} Time evolution of the net inflow rate (Eq.~\ref{eq:mass_conservation_2}). The smoothed curves are boxcar averages over a time window of $7\Myr$. \textit{Bottom rows:} Time evolution of the cumulative mass fraction of $\Htwo$ (solid curve), $\HI$ (dashed curve), and $\Hp$ (dot-dashed curve).}
\label{fig:net_inflow_t}
\end{figure*}

\begin{figure*}[ht!]
\includegraphics[width=0.9\textwidth]{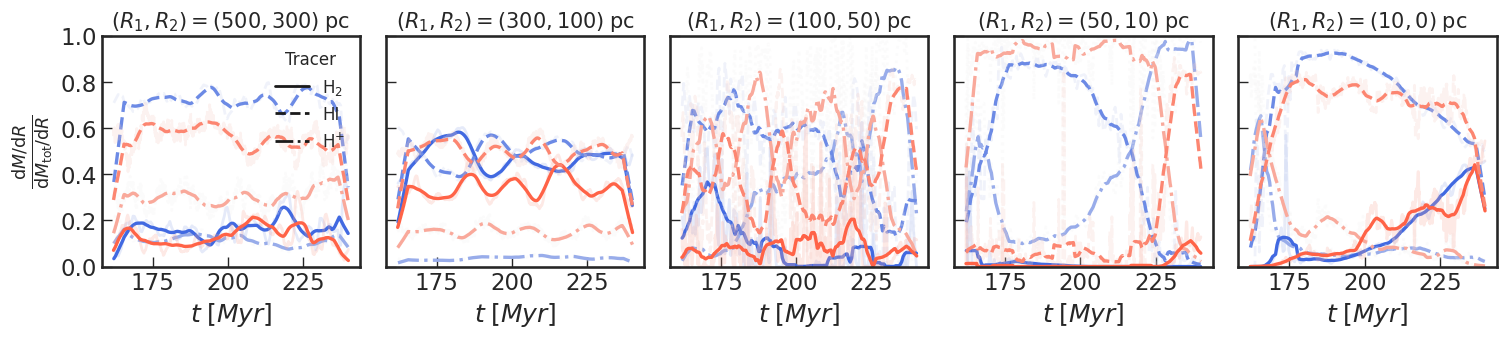}
\centering
\caption{Same as the bottom row of Fig.~\ref{fig:net_inflow_t}, but showing the mass fractions computed in shells between two nested cylinders with radius $R_1$ and $R_2$. Each cylinder of radius $R$ has a vertical extent of $|z|<R/2$.}
\label{fig:mass_fraction_t}
\end{figure*}

Panel a of Fig.~\ref{fig:net_inflow_R} shows the time- and azimuthally-averaged surface mass density profile. The density peaks in the SN and SNRad simulations at $R\simeq 3\pc$ correspond to a CND that spontaneously forms in the simulation as a consequence of the inflow. The SN simulation forms a low-mass CND with surface density $\Sigma_{\rm gas} \simeq 10 \Msun \pc^{-2}$, while the SNRad has larger inflow rates and forms a higher-mass CND with surface density $\Sigma_{\rm gas} \simeq 100 \Msun \pc^{-2}$. The CHEM simulation does not form a CND as there is no inflow in the NIZ. Near the end of the simulation, $t=240\Myr$, the total CND mass ($R<10\pc$) in the SNRad simulations reaches $M_{\rm CND}\sim1.5\times10^4\Msun$, a value similar to the observational estimates of $10^4\mhyphen10^6\Msun$ from dust and gas measurements \citep{mez_etal_89, mez_etal_96, den_etal_93, chr_etal_05, mon_etal_09}.

Panel c of Fig.~\ref{fig:net_inflow_R} shows that neutral gas ($\rm HI$) dominates the cumulative mass within the NIZ in both the SN and SNRad simulations. Panel d of Fig.~\ref{fig:net_inflow_R} further indicates that the radial inflow in the NIZ is dominated by neutral gas ($\rm HI$) in the SN simulation, but by $\Hp$ in the SNRad simulation. In contrast, the $\rm H_2$ fraction is higher in the CND and CMZ than in the region between them, reaching $\sim 40\%$ and $\sim50\%$ in the SN and SNRad runs, respectively. The higher $\rm H_2$ fraction in the SNRad run is potentially due to its larger gas density and more efficient cooling.

So far we have looked at the time-averaged inflow. Let us now analyse the net inflow as a function of time. The top rows in Fig.~\ref{fig:net_inflow_t} show the total gas $M_{\rm gas}$ plus the total mass of stars $M_{\rm star}^{\rm born}$ born within the volume for various cylindrical volumes centred on the Galactic centre with radius $R$ and height $|z| = R/2$. This quantity represents the time-integrated net inflow experienced by these volumes as can be seen integrating Eq.~\eqref{eq:mass_conservation_2}:
\begin{equation}
    \int \dot{M}\, \mathrm{d} t = M_{\rm star}^{\rm born} + M_{\rm gas} \,.
\end{equation}
The curves show the following general features inside the NIZ (cylinders at $R<100\pc$):
\begin{itemize}
    \item The CHEM simulation (black dashed curves) is flat (no inflow).
    \item The SN simulation (blue line) alternates periods of moderate inflow with periods of moderate outflows. The latter are caused by SNe bursts within the central 200 pc over $t = 180 \mhyphen 220 \Myr$ which produce vertical outflows at $R\sim30\mhyphen 200 \pc$ (see Sect.~\ref{sec:outflows}).
    \item The SNRad simulation shows a clear upward average trend in all curves, indicating inflow. On top of the smooth upwards trend, there are transient spikes that correspond to gas clumps that temporarily enter the volume but quickly leave, and step-like sudden increase in mass that correspond to episodic inflow events (see Sect.~\ref{sec:episodic_radial_inflow}).
\end{itemize}

The middle rows of Fig.~\ref{fig:net_inflow_t} show that the net inflow rates fluctuate strongly with time at all radii. The bottom row of Fig.~\ref{fig:net_inflow_t} breaks the enclosed gas mass into molecular, atomic, and ionised components, confirming that the cumulative mass is dominated by neutral gas at all radii.

To investigate which gas phase dominates the inflow, Fig.~\ref{fig:mass_fraction_t} shows the mass fractions of different chemical phases as a function of time in logarithmic radial bins. In the SN simulation, the inflow is dominated by neutral gas at most radii and times, with $\HI$ providing the main contribution. In the SNRad simulation, however, $\Hp$ dominates the inflow within the NIZ, especially during episodic inflow events originating from the CMZ (e.g. $t=170\Myr$, $t=185\Myr$, and $t=200\Myr$). Neutral gas remains important during more secular inflow phases, as well as during episodic events fed by clouds outside the CMZ (e.g. $t=218\Myr$ and $t=230\mhyphen 240\Myr$). The different origins of radial inflow are discussed in more detail in Sect.~\ref{sec:episodic_radial_inflow}. Here we show that the dominant inflowing gas phase depends on both the feedback model and the origin of the inflowing material.

\begin{figure}[ht!]
\includegraphics[width=0.9\columnwidth]{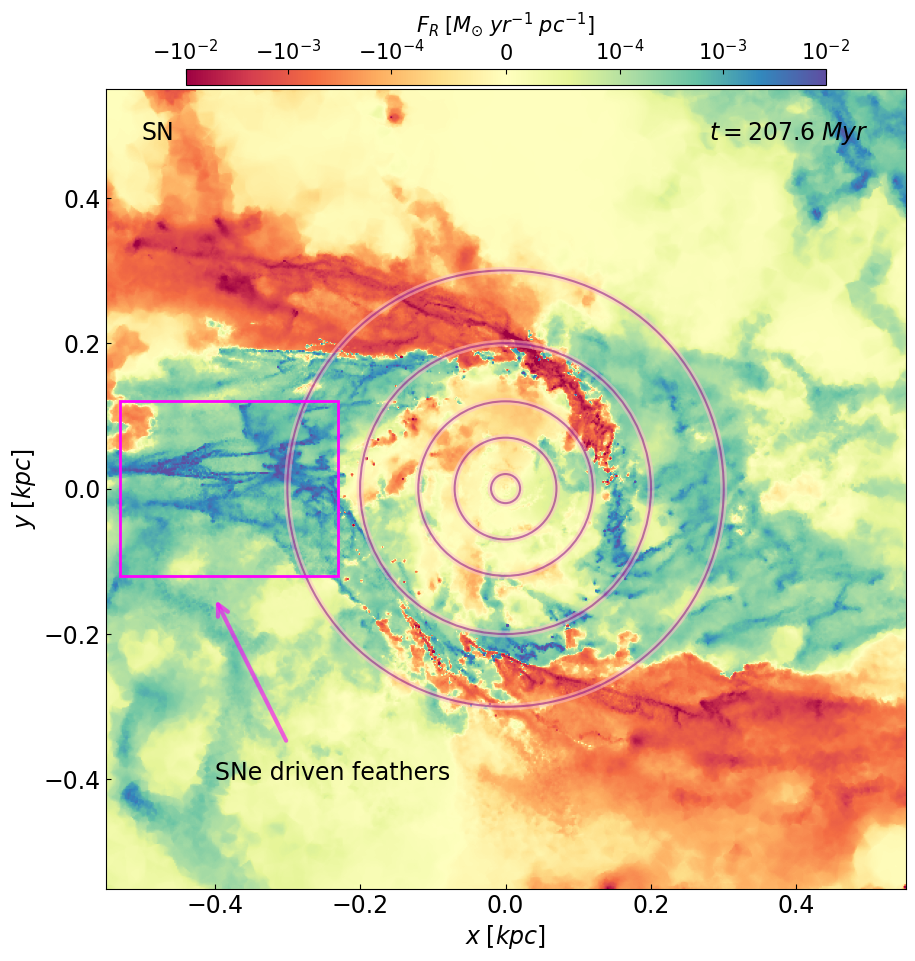}
\centering
\caption{Instantaneous radial mass flux (Eq.~\ref{eq:FR}) in the SN simulation at $t=207.6\Myr$. Red/blue indicate inward/outward motion. Inflows concentrate along the bar lanes. Solid circles correspond to 300, 200, 120, 70, and 20$\pc$. SNe driven turbulence in the CMZ produces feather-like outflows (see boxed region), and inflow into the central 120$\pc$.}
\label{fig:SN_flux_map}
\end{figure}

\begin{figure}[ht!]
\includegraphics[width=0.9\columnwidth]{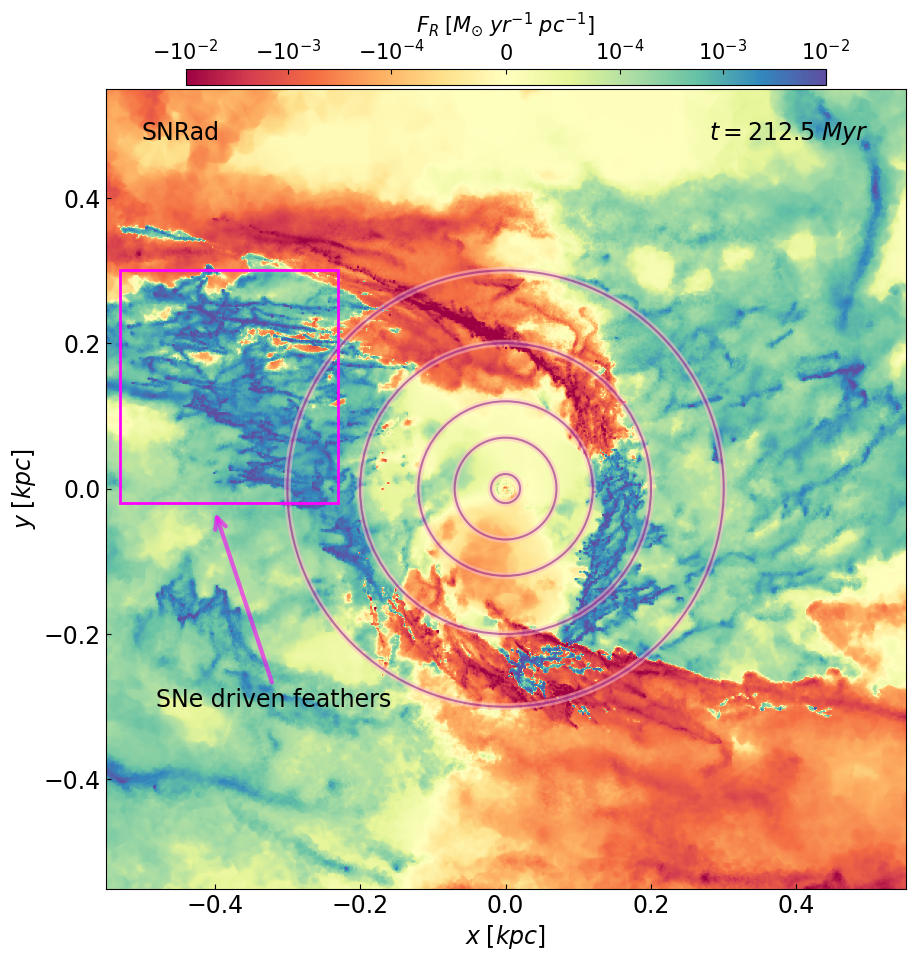}
\centering
\caption{Same as Fig.~\ref{fig:SN_flux_map} but for the SNRad simulation at $t=212.5 \Myr$.} 
\label{fig:SNRad_flux_map}
\end{figure}

\begin{figure*}[ht!]
\includegraphics[width=0.9\textwidth]{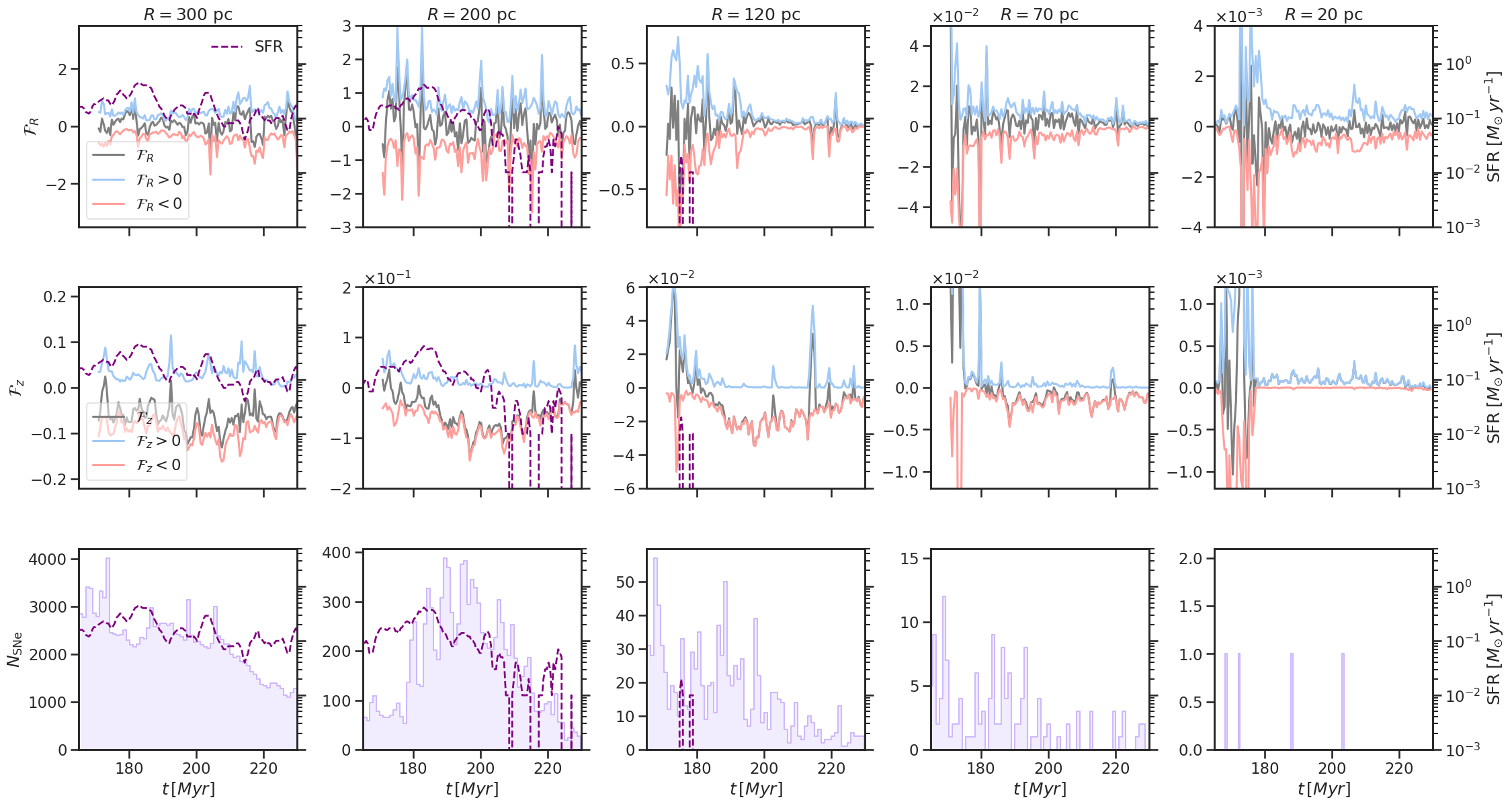}
\centering
\caption{\textit{Top}: Mass flux as a function of time through the lateral surface (top panels) of a cylindrical control volume with height $|z| = R/2$ in the SN model. Panels from left to right correspond to cylinders of radius $R = \{ 300,200, 100,70,20\} \pc$ (see dashed circles in Figs.~\ref{fig:SN_flux_map} and \ref{fig:SNRad_flux_map}). Sign convention is that positive (negative) $\dot{M}$ means inflow (outflow). The dashed purple curve indicates the SFR within the corresponding cylinder. 
\textit{Middle}: Same as the top panel but for the upper and lower surfaces (middle panels) of the cylindrical control volume. \textit{Bottom}: The number of supernova explosion events within cylinders as a function of time, with the star formation rate overplotted as a dashed purple curve.}
\label{fig:SFR_flux_SN_5bin}
\end{figure*}

\begin{figure*}[ht!]
\includegraphics[width=0.9\textwidth]{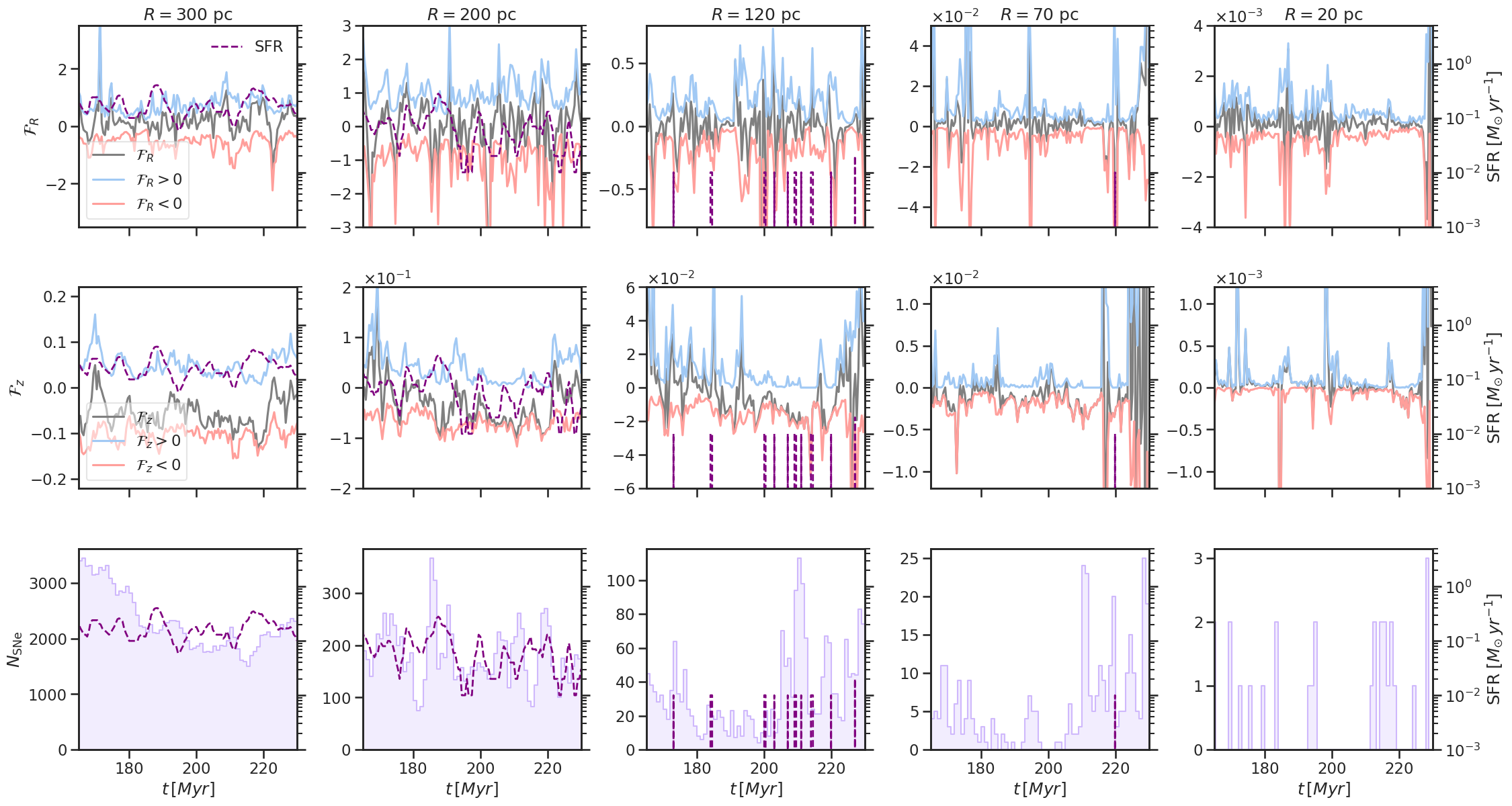}
\centering
\caption{Same as Fig.~\ref{fig:SFR_flux_SN_5bin} but for the SNRad model.}
\label{fig:SFR_flux_SNRad_5bin}
\end{figure*}

\begin{figure}[ht!]
\includegraphics[width=\columnwidth]{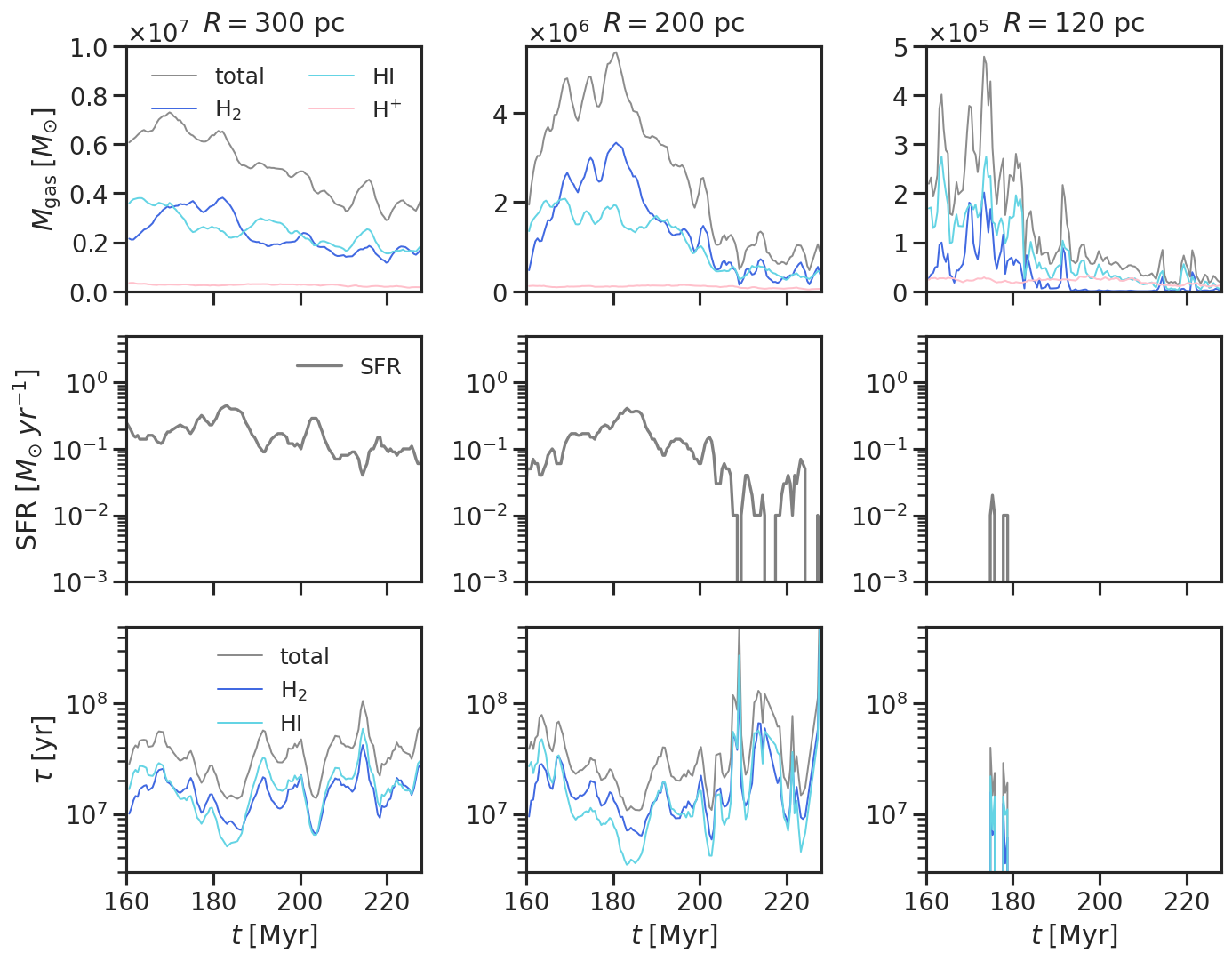}
\centering
\caption{\textit{Top:} Enclosed mass of different chemical tracers for $R=300\pc$ (left), $R=100\pc$ (middle), and $R=120\pc$ (right) in the SN simulation. Deep blue, faint blue, and pink curves represent the enclosed mass of $\rm H_2$, $\rm HI$, and $\rm H^{+}$, respectively. \textit{Middle:} Star formation rate measured within cylinders of different cylindrical radius. \textit{Bottom:} Depletion time as a function of time for different tracers. The color coding is the same as in the top panel.}
\label{fig:SN_depletion_time}
\end{figure}

\begin{figure}[ht!]
\includegraphics[width=\columnwidth]{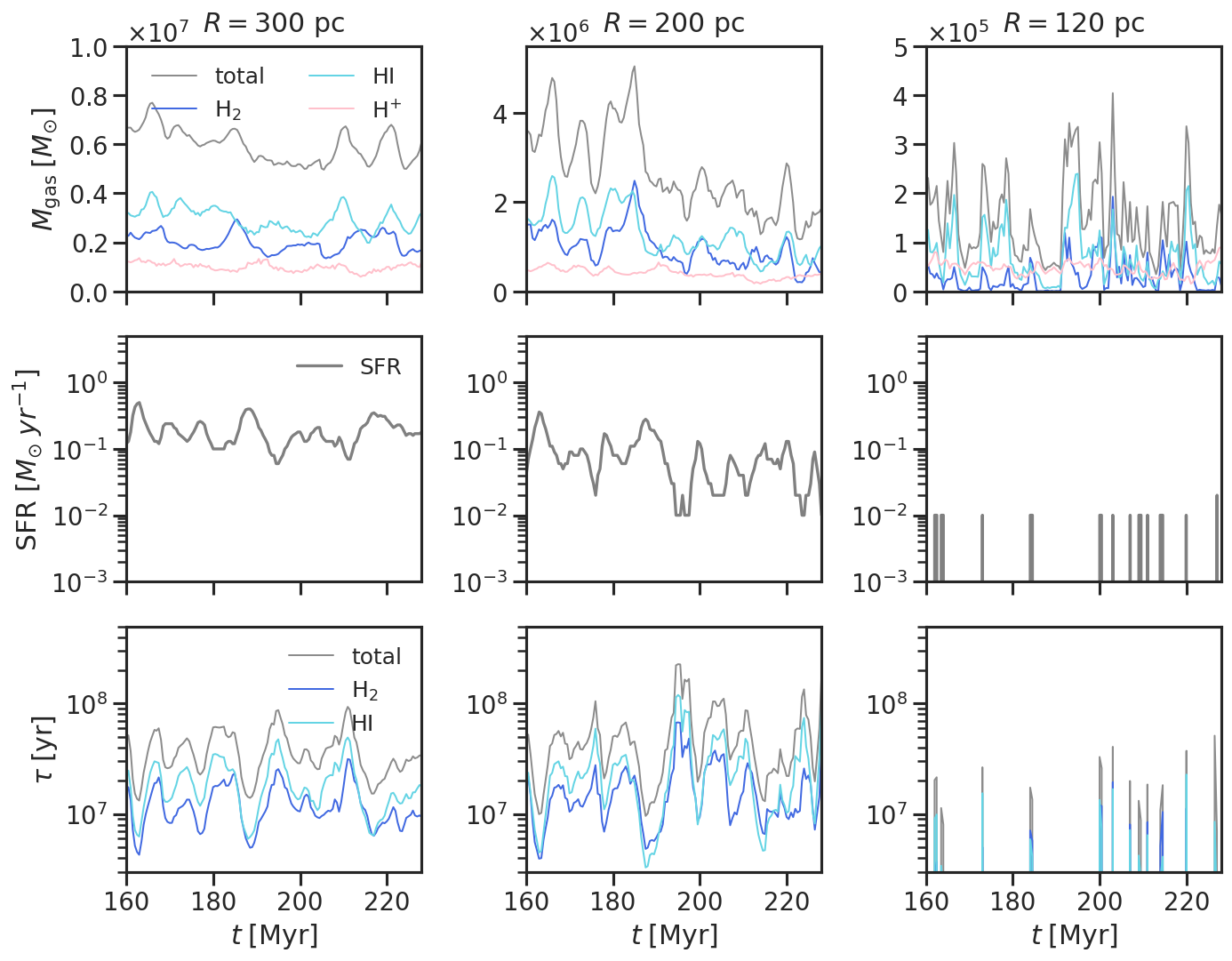}
\centering
\caption{Same as in Fig.~\ref{fig:SN_depletion_time} but for the SNRad model.}
\label{fig:SNRad_depletion_time}
\end{figure}

\begin{figure*}[h!]
\includegraphics[width=\textwidth]{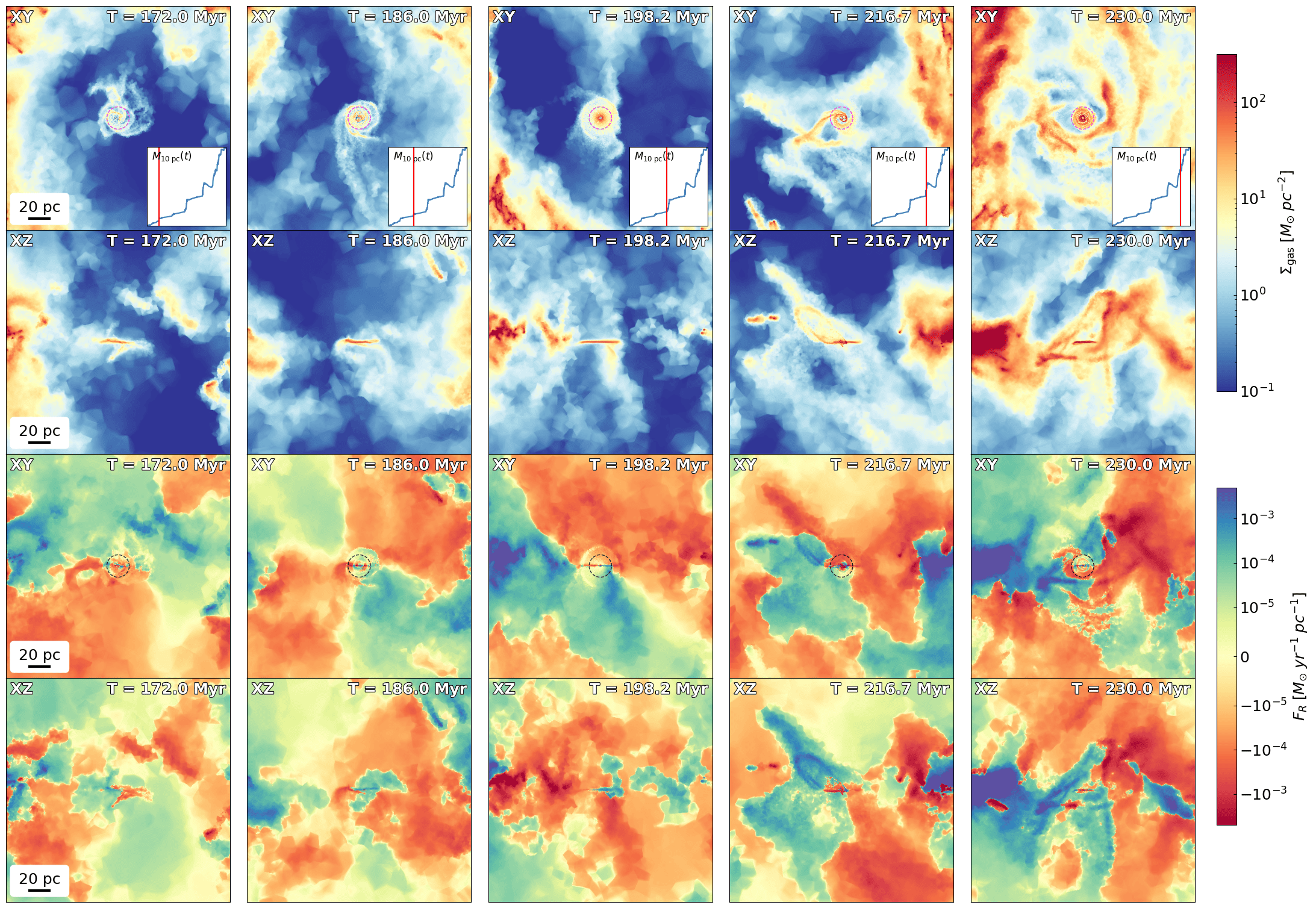}
\centering
\caption{Episodic radial inflow events within $\rm NIZ$ in the SNRad simulation. \textit{Upper rows}: Overall gas morphology (surface density) in the face-on (first row) and edge-on view (second row), with scale bars positioned on the lower left corner. The inset shows the enclosed gas mass within $R=10\pc$ and $|z|<R/2$ as a function of time over $t=160\mhyphen 240\Myr$. Vertical dashed lines mark the timing of the corresponding episodic inflow events. Dashed circles mark $R<10\pc$. \textit{Lower rows:} Same as the upper rows but for the instantaneous radial mass flux. The colormap scale is chosen to highlight the lower-amplitude radial mass flux within the NIZ, and therefore differs from those used in Figs.~\ref{fig:SN_flux_map} and \ref{fig:SNRad_flux_map}.}
\label{fig:edisodic_radial_inflow}
\end{figure*}

\begin{figure}[h!]
\includegraphics[width=\columnwidth]{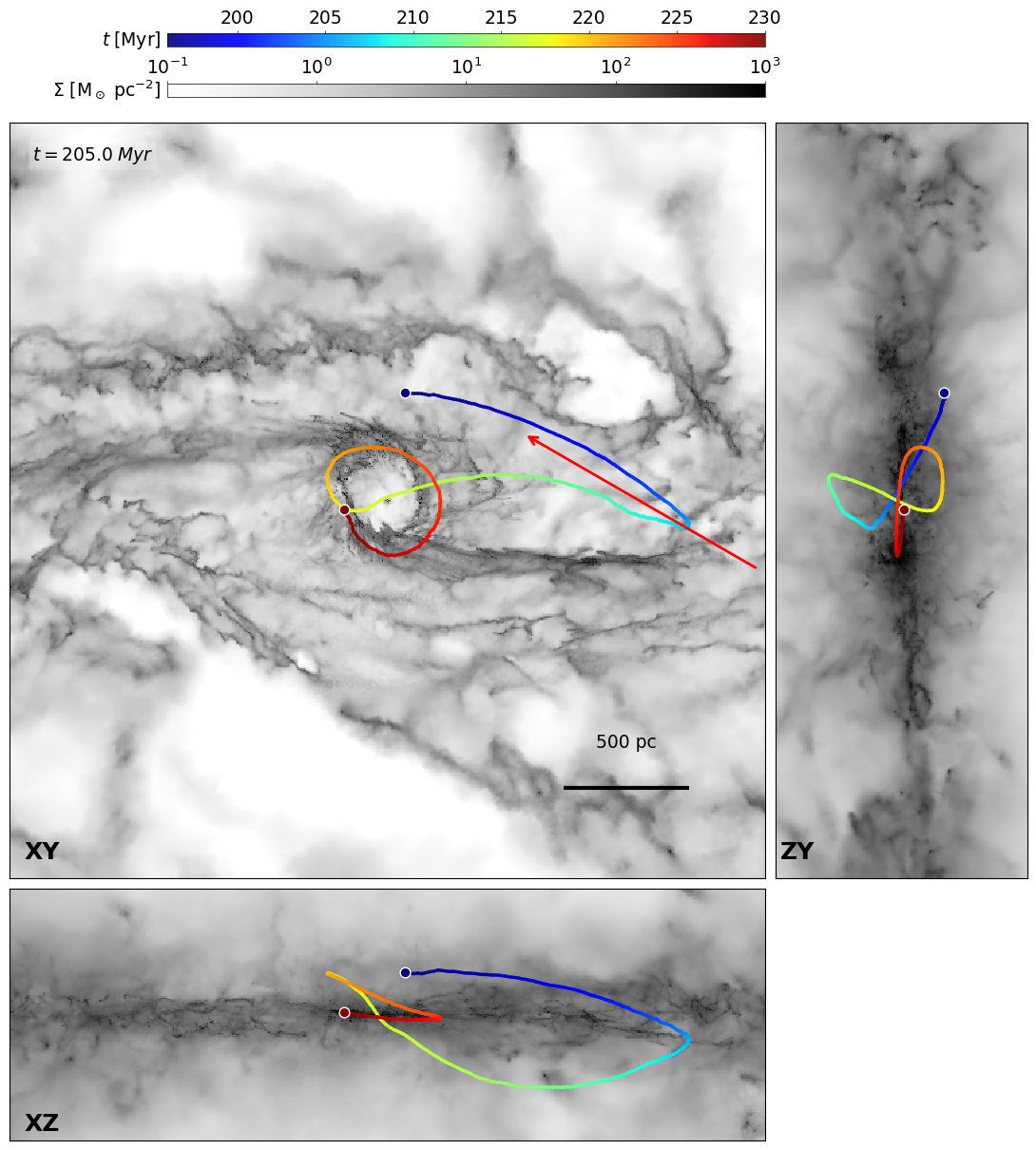}
\centering
\caption{Orbit of the gas cloud that produces the inclined radial inflow at $t=216.7\Myr$ in Fig.~\ref{fig:edisodic_radial_inflow}, overplotted on the gas surface density map at $t=205\Myr$. The line colour indicates the simulation time. The cloud feeds the CND as it crosses the Galactic centre, before eventually settling into the CMZ. The red arrow marks the pseudo-slit covering the region where the cloud interacts with supernova-driven shocks from the dust lane.}
\label{fig:orbit_tracer}
\end{figure}

\begin{figure}[h!]
\includegraphics[width=\columnwidth]{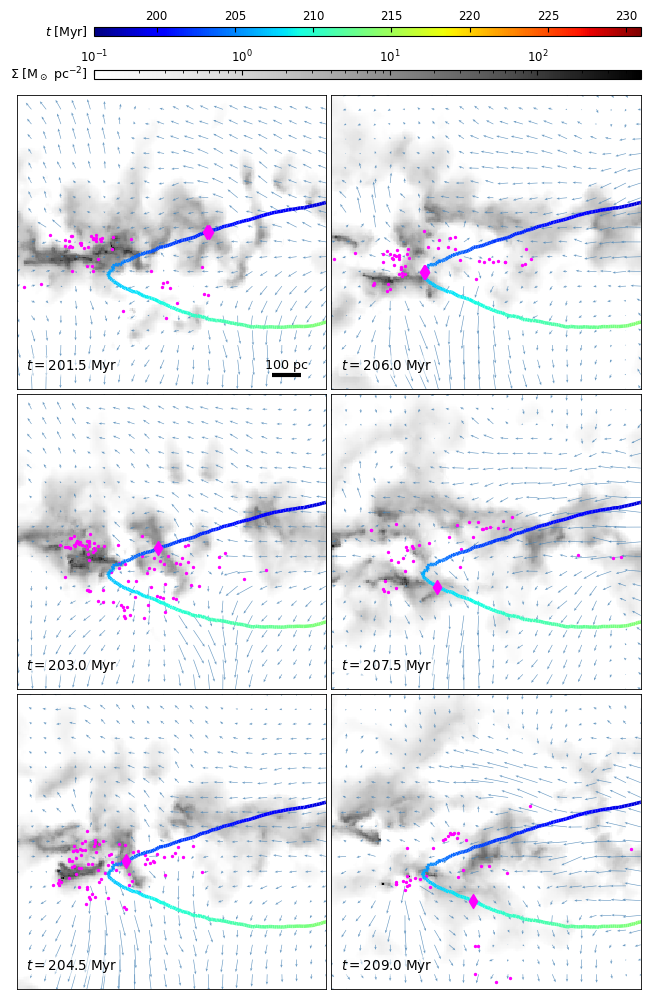}
\centering
\caption{Mass-weighted velocity vectors superimposed on the gas surface density in the plane defined by the pseudo-slit shown in Fig.~\ref{fig:orbit_tracer}. The simulation time is labelled in the bottom-left corner. The pseudo-slit has a half-width of $75\pc$. The $x$- and $y$-axes represent the distance along the pseudo-slit and the vertical direction, respectively. Highlighted dots mark the positions of SNe occurring within $|t-\Delta t|<0.5\Myr$. The orbit is the same as that shown in Fig.~\ref{fig:orbit_tracer}, and the instantaneous position of the cloud is highlighted by the purple marker.}
\label{fig:orbit_tracer_slice}
\end{figure}

\subsubsection{Radial and vertical flows} \label{sec:outflows}

Figures~\ref{fig:SN_flux_map} and ~\ref{fig:SNRad_flux_map} present maps of the instantaneous radial mass flux, defined as:
\begin{equation} \label{eq:FR}
    F_R(x,y) = \int_{-\infty}^{\infty} \rho v_R \, \mathrm{d} z \,.
\end{equation}

These maps confirm that the bar-driven inflow onto the CMZ occurs mainly via the bar lanes, similar to previous work \citep[e.g.][]{tre_etal_20}. Visible in the figures are SNe-driven ``feathers'' (see boxed region) at $R=200 \mhyphen 300\pc$ that correspond to strong radial outflows.

To investigate in more detail radial and vertical flows we consider mass fluxes through the lateral and top/bottom surfaces of cylinders with radius $R$ and height $|z| = R/2$:
\begin{align} \label{eq:FRZ}
\mathcal{F}_R & = - \int_{0}^{2\pi} \int_{-R/2}^{R/2}   \rho v_R\, R   \mathrm{d} \phi \mathrm{d} z \\ 
\mathcal{F}_z & = - \int_{0}^{2\pi} \int_{0}^{R}   \rho v_z\, R   \mathrm{d} \phi \mathrm{d} R
\end{align}

The minus sign is chosen so that $\mathcal{F}>0$ means inflow. Note that the sum of the two gives the net mass inflow rate
\begin{equation} \label{eq:Mdot2}
    \dot{M} = \mathcal{F}_R + \mathcal{F}_z \,.
\end{equation}

We have checked that calculating the inflow rate using Eqs.~\eqref{eq:FRZ} and \eqref{eq:Mdot2} gives the same result as that calculated using the method described in Sect.~\ref{sec:netinflow} (based on Eq.~\ref{eq:mass_conservation_2}) provided that the fluxes are calculated using sufficiently fine snapshots. With our fiducial snapshot spacing $\Delta t_{\rm output}= 0.5\Myr$, the flux method under-samples short inflow/outflow episodes between outputs, particularly within central $50\pc$ where the dynamical time are shorter than the snapshot spacing $\Delta t_{\rm output}= 0.5\Myr$.

We also split the fluxes into positive and negative contributions:
\begin{align} \label{eq:FRZ2}
 \mathcal{F}_R & = \mathcal{F}^{+}_R + \mathcal{F}^{-}_R \\ 
 \mathcal{F}_z & = \mathcal{F}^{+}_z + \mathcal{F}^{-}_z
 \end{align}
These are obtained by integrating Eq.~\eqref{eq:FRZ} only over positive/negative velocities:
 \begin{align} \label{eq:FRZ3}
 \mathcal{F}^{\pm}_R & = - \int_{0}^{2\pi} \int_{-R/2}^{R/2}  H(\pm v_R) \rho v_R\, R   \mathrm{d} \phi \mathrm{d} z \\
 \mathcal{F}^{\pm}_z & = - \int_{0}^{2\pi} \int_{0}^{R} H(\pm v_R)  \rho v_z\, R   \mathrm{d} \phi \mathrm{d} R
 \end{align}
 
 where 
 \begin{equation} H(x) = \begin{cases} 1, & x \geq 0 \\ 0, & x < 0 \end{cases} \end{equation}
 is the Heaviside step function.
 
The fluxes through cylinders with radii $R=\{300, 200,120,70,20\} \pc$ are shown in Figs.~\ref{fig:SFR_flux_SN_5bin} and ~\ref{fig:SFR_flux_SNRad_5bin} (see solid circles in Figs.~\ref{fig:SN_flux_map} and \ref{fig:SNRad_flux_map}). These illustrate the following:
\begin{itemize}
    \item The radial flux ($\mathcal{F}_R$) is typically an order of magnitude larger than the vertical flux ($\mathcal{F}_z$). At small radii ($R=20\pc$), the net radial and vertical inflows become comparable, supplying the CND at a rate of $\sim 10^{-4}\Msun\yr^{-1}$.
    \item The time-averaged net radial flux $\langle \mathcal{F}_R \rangle$ is generally positive (indicating inflow), while the net vertical flux $\langle \mathcal{F}_z \rangle$ is negative (indicating outflow).
    \item The radial flux $\mathcal{F}_R$ is roughly two orders of magnitude higher at $R=300\pc$ than at $R=120\pc$. This is a consequence of the different mechanisms that drive the inflow at these two radii. The former ($R=300\pc$) radius is outside of the CMZ ring and registers the inflow \emph{onto} the CMZ, which is driven by the gravitational torques of the Galactic bar. The latter ($R=120\pc$) is inside the CMZ, and the mass flux, driven by turbulent Reynolds stresses that are present because of stellar feedback, is much smaller. 
    \item Periodic radial inflow episodes supply the CMZ and lead to recurrent enhancements in the SFR (purple dashed curves). The subsequent SNe events (purple shaded regions on the bottom row) then drive outflows.
    \item Despite the net radial inflow inside the CMZ ($R=120\pc$ and $R=70\pc$) being positive for most of the time, the total enclosed gas mass can decrease as vertical outflows remove a substantial amount of gas, particularly during episodes of concentrated SNe activity (e.g., $t=190\mhyphen210\Myr$). 
\end{itemize}

Figs~\ref{fig:SN_depletion_time} and ~\ref{fig:SNRad_depletion_time} show the temporal evolution of the enclosed gas mass for different chemical tracers (top), the star formation rate (middle), and the gas depletion timescale (bottom) within cylinders of $R=300\pc$, $R=200\pc$, and $R=120\pc$. In both the SN and SNRad models, the depletion time lies in the range of $10^7 \mhyphen 10^8\yr$ and closely follows the variations in $\rm SFR$, reaching local minima when the SFR peaks. This correlation indicates that star formation efficiency increases during phases when dense gas clouds enter and collide within the CMZ, leading to enhanced gas compression, and decreases during more quiescent periods when such interactions are weaker. In our simulations, star formation is predominantly concentrated on the CMZ at $R=200\mhyphen 300\pc$, while only a few young stars form within $R\leq120\pc$. 

\subsubsection{Episodic inflow events}
\label{sec:episodic_radial_inflow}

Fig.~\ref{fig:edisodic_radial_inflow} presents five representative radial inflow events within the NIZ in the SNRad simulation at  $t=172\Myr$, $t=186\Myr$, $t=198\Myr$, $t=217\Myr$, and $t=230\Myr$. The timings of these events are marked by vertical dashed lines in the inset panels. The SNRad simulation exhibits a total of five episodic inflow events in the time range $t=160 \mhyphen 240\Myr$, while the SN simulation exhibits one event only in the same time frame.

The first three columns illustrate episodic radial inflow events originating from the CMZ, with streams that remain roughly parallel to the disk midplane. In higher-resolution tests within the CND region, such inflow events efficiently generate turbulence in the CND, reduce the Toomre $Q$, and promote gas cloud fragmentation. The $3rd$ column corresponds to a particularly strong radial inflow that increases the CND mass by more than $2\times10^3\Msun$ (see Fig.~\ref{fig:net_inflow_t}). In contrast, the $2nd$ column shows a case in which the inflow reaches the edge of the CND along nearly circularized orbits, resulting in a substantially lower inflow rate ($\Mdot\sim3\times10^{-5}\Aunit$) compared to the other events. Note that we do not detect similar prominent radial inflow that remains roughly parallel to the disk midplane in the SN simulation.

Gas clouds ejected from star-forming regions in the CMZ can present filamentary structures during their transport toward the Galactic Centre, as shown in Fig.~\ref{fig:edisodic_radial_inflow}. In these episodic inflow events, the gas from CMZ subsequently feed the CND as coherent streams instead of isotropic spherical clouds. This morphology differs from the idealized cloud assumptions often adopted in CND simulations \citep[e.g. ][]{map_tra_16, tra_etal_18}, highlighting the importance to study the formation of CND within a realistic Galactic Centre environment.

The $4th$ and $5th$ columns show cases in which episodic inflows interact with the CND at inclinations over $40^{\circ}$. These interactions perturb the CND and produce tilts of $10^{\circ}\mhyphen30^{\circ}$ that persist for approximately $1\mhyphen2\Myr$. Our test simulations including tracer particles show that these inclined inflow streams originate from ISM clouds in the bar region. Over $t=160\mhyphen240\Myr$, we identify two episodic inflow events in the SNRad simulation that increase the inclination difference between the CND and the CMZ to more than $30^\circ$. By contrast, only one such event occurs in the SN simulation. For most of the time, the CND remains coplanar with the CMZ, follows nearly circular motion, and shows continuous inward gas transport.

Figures~\ref{fig:orbit_tracer} and~\ref{fig:orbit_tracer_slice} trace the orbit of a gas cloud that initially overshoots from the CMZ into the bar region and later produces the inclined radial inflow at $t=217\Myr$ shown in Fig.~\ref{fig:edisodic_radial_inflow}. The cloud first moves inwards toward the CMZ along the dust lane on the left. It then overshoots the CMZ and moves toward the dust lane on the right, where supernova explosions drive shocks that push the surrounding ISM outward. In contrast to the typical scenario in which a cloud loses angular momentum through bar-driven shocks and flows inward along the dust lane, interaction with the SNe-driven shock front changes the cloud trajectory and lifts it below the midplane, reaching $z\sim-260\pc$. After the collision, the cloud loses angular momentum and follows a nearly radial orbit toward the Galactic centre, with an inclination greater than $40\degree$. The cloud feeds the CND while crossing the Galactic centre and eventually settles into the CMZ.

The bottom rows of Fig.~\ref{fig:edisodic_radial_inflow} show the instantaneous radial mass flux in face-on and edge-on views. These panels trace where gas is moving radially inward or outward. Gas within the CND mainly follows nearly circular orbits, therefore the radial flux there is weak compared to the NIZ. In the face-on views, extended streams in the NIZ present opposite signs of radial mass flux on either side of the centre, indicating gas motion along highly eccentric, nearly radial orbits. In the first three edge-on views, the inflows are roughly parallel to the disk midplane and produce overall coherent flux patterns. In the last two edge-on views, the inflowing gas crosses the midplane from below and perturbs the local velocity field, producing opposite flux signs between the tilted stream and the surrounding material.

\begin{figure}[ht!]
\includegraphics[width=\columnwidth]{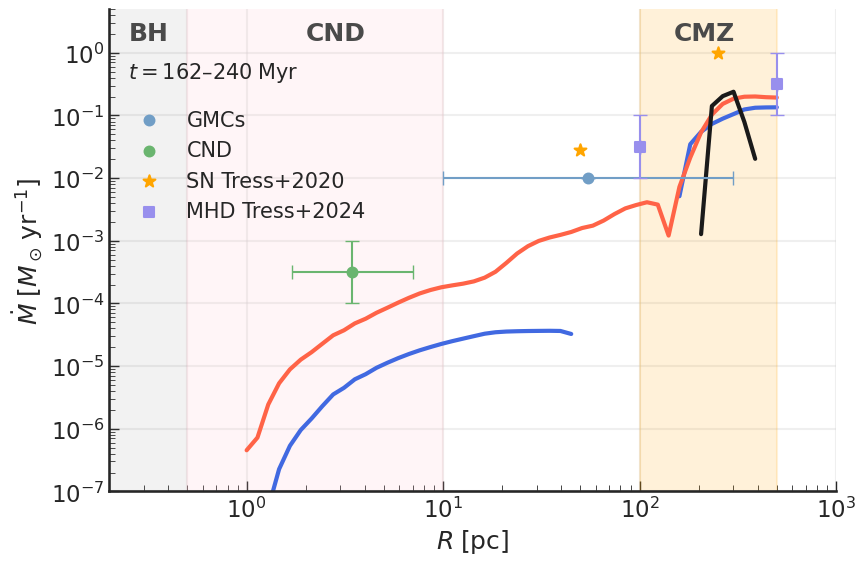}
\centering
\caption{Same as the middle panel of Fig.~\ref{fig:net_inflow_R}, but with observational estimates overlaid: GMC-based dynamical estimates \citep[blue dot;][]{VonLinden1993} and dynamical estimates based on CND CS(2-1) emission \citep[green dot;][]{Vollmer2001, Vollmer2002, Hsieh2017}. Estimates from the Milky Way simulation by \citet{tre_etal_20} including star formation and supernova feedback (orange stars) and the MHD simulation by \citet{tre_etal_24} (purple rectangles) are also marked.}
\label{fig:inflow_comparison}
\end{figure}

\section{Discussion}
\label{sec:discussion}

Figure~\ref{fig:inflow_comparison} shows the mass inflow rate $\Mdot(R)$ from our simulations compared with observational estimates and previous numerical results. For observational estimates, between the CMZ and the CND ($R \sim 10\mhyphen200\pc$), $\Mdot \sim 10^{-2}\Msunyr$ is inferred from dynamical models of giant molecular cloud (GMC) kinematics \citep{VonLinden1993}. Within the CND ($R \sim 2\mhyphen5\pc$), $\Mdot \sim 10^{-3}\mhyphen10^{-4}\Msunyr$ is derived from viscous disk models of cloud-cloud collisions \citep{Vollmer2001, Vollmer2002} and molecular streamer observations \citep{Hsieh2017}. 

\subsection{Comparison with previous simulations of CMZ gas inflow}
\label{sec:comp_tress}

\citet{tre_etal_20a} simulated the CMZ using the moving-mesh code \textsc{arepo}, including gas self-gravity, star formation, supernova feedback and a Milky Way external rotating barred gravitational potential. These simulations are similar to those reported here, but as mentioned in the introduction, they had a less accurate gravitational potential (especially at $R<300\pc$, where they clumped together NSD and NSC into a single power-law component), lacked early feedback, and had lower resolution. They reported a bar-driven inflow of $\sim 1\Msunyr$ from the Galactic disk to the CMZ ($R \simeq 200\pc$), and a supernova-feedback-driven inflow of $\sim 0.03\Msunyr$ from the CMZ inward to the CND, measured as a time-averaged net flux across $R = 50\pc$. \citet{tre_etal_24} further explored MHD simulations of the Milky Way. Their bar driven inflow rates of $\sim 0.3 \Msunyr$ at $R\sim500\pc$ are consistent with our results, but the magnetorotational instability (MRI) driven inflow rates of $\sim0.03 \Msunyr$ are higher than our estimates at $R=100\pc$ (Fig.~\ref{fig:inflow_comparison}). 

Our simulations produces $\Mdot$ lower than both works at comparable $R$. For the \citet{tre_etal_20a}, where the inflow-driving mechanism is stellar feedback as in the present paper, this is attributable to two factors. First, we achieve higher spatial resolution within the central $\sim 100\pc$. Our convergence tests (see Sect.~\ref{sec:resolution_tests}) demonstrate that $\Mdot$ decreases with increasing resolution, as finer grids reduce numerical angular momentum diffusion and better resolve the multiphase ISM. Second, the CMZ mass in our simulations, $M_{\rm CMZ}\sim10^7\Msun$, is around five times lower than in \citet{tre_etal_20}. If the inflow rate scales approximately with the available gas reservoir, this lower CMZ mass could partly explain the lower $\Mdot$ values in our simulations.\footnote{We do not apply any CMZ-mass rescaling to the inflow rates shown in Fig.~\ref{fig:inflow_comparison}. The figure shows the inflow rates measured directly from each simulation.} The different gravitational potential could also have an effect, although it is unlikely to be the dominant factor. The \citet{tre_etal_24} simulation is not directly comparable, as their inflow-driving mechanism is completely different (MRI-driven inflow vs stellar feedback). However, we note that their convergence tests also suggest that their values are upper limits. Despite these differences, all three simulations show the pattern that $\Mdot$ decreases by factors over 10 as clouds move inwards from the CMZ to the innermost $100\pc$.

These comparisons suggest that multiple mechanisms, including stellar feedback and MRI-driven stresses, can contribute to gas inflow toward the central $\sim100\pc$, without a clear dominant mechanism. In the real CMZ, these mechanisms likely operate simultaneously, making it difficult to predict their combined effect or to isolate their individual contributions using observations alone. A systematic comparison of simulations with controlled physical ingredients is therefore needed to clarify their separate and combined roles in driving gas inflow from the CMZ to the innermost $\sim100\pc$.

\subsection{Comparison with recent barred-galaxy simulations of outside-CMZ fueling toward the CND}

\citet{kob_etal_26} proposed a ``vaulting'' mechanism that transports gas from outside the CMZ toward the CND. In their simulations, a gas cloud acquires vertical momentum when it crosses the dust lanes near the nuclear ring. It then moves above or below the nuclear ring and loses angular momentum when interacting with the dust lane on the opposite side, allowing it to flow inward from below the nuclear ring and eventually reach the CND. This mechanism provides a pathway for outside-CMZ gas to bypass the nuclear ring and feed the central $50\pc$.

The event shown in Figs~\ref{fig:orbit_tracer} and~\ref{fig:orbit_tracer_slice} follows a qualitatively similar orbital path, but our simulations identify a different physical driver. In both cases, gas initially located outside the CMZ is lifted away from the disk midplane near the dust lanes and later approaches the CND from off the midplane. However, in \citet{kob_etal_26}, the vertical motion appears to result from purely gas-dynamical interactions with dust lanes near the nuclear ring. In our simulations, the cloud trajectory is instead altered primarily by SNe-driven shocks along the dust lanes.

The zoom-in resolution in the innermost $100\pc$ allows us to follow the cloud trajectories and their interactions with the CND/NIZ in greater detail. We find that outside-CMZ fueling in our simulations is episodic and less massive than in \citet{kob_etal_26}. In the SNRad simulation, the strongest event occurs over $t=225\mhyphen240\Myr$, when $\HI$ clouds with a total mass of $\sim10^5\Msun$ reach the central $50\pc$, giving an average inflow rate of $\dot{M}_{\rm 50pc}\sim7\times10^{-3}\Msunyr$. A smaller event at $t\sim216\Myr$ delivers $\sim5\times10^3\Msun$ to the CND over $\sim5\Myr$. In the SN simulation, only $\sim100\Msun$ remains in the CND after a $\sim10^3\Msun$ cloud passes through it. Thus, individual events in our simulations deliver $\sim10^2\mhyphen10^5\Msun$, with characteristic inflow rates of $\sim10^{-4}\mhyphen10^{-2}\Msunyr$ at $r=50\pc$. By contrast, \citet{kob_etal_26} find that this mechanism delivers more than $10^6\Msun$ of gas to the central $50\pc$ within $\sim30\Myr$, producing a time-averaged inflow rate of $\dot{M}_{\rm 50pc}\sim3\times10^{-2}\Msunyr$.

\subsection{Comparison with multiscale simulations of out-of-equilibrium galaxies}
\label{sec:grav_torques}

\citet{hop_qua_10, Hopkins2011} constructed multiscale simulations of out-of-equilibrium galaxies and found that gravitational torques from non-axisymmetric stellar structures (bars at $\gtrsim 100\pc$ and lopsided $m=1$ modes at $\lesssim 10\pc$) dominate angular momentum transport from kpc to sub-pc scales. Hydrodynamic pressure forces are negligible in comparison, and the resulting $\Mdot$ reaches $0.1\mhyphen10\Msunyr$ at $R<0.1\pc$ and $10\mhyphen100\Msunyr$ at $R\sim1\mhyphen 100\pc$ depending on gas fraction and bulge-to-total mass fraction.  \citet{hop_etal_24} further included magnetic field and found that it plays an important role in enhancing the inflow rates within scales of $0.1\pc$ and feeding the central SMBH. 

Our simulations use a fixed external potential, excluding gravitational torques from a live stellar component by design. This isolates the contribution of stellar feedback to the inflow. We note that the $m=1$ lopsided disk modes mentioned by \citet{hop_qua_10} to drive inflow at $\lesssim 10\pc$ could arise in gas-rich mergers or strongly unstable nuclear disks. The nuclear stellar disk (NSD) in the Milky Way is a kinematically cool, axisymmetric structure dominated by old stars \citep[][]{Nogueras-Lara2020,Sanders2024}. There is no evidence for a significant $m=1$ stellar asymmetry in the current NSD \citep{sch_etal_25}, so the use of a smooth external potential is a reasonable approximation for the present-day Galactic Centre. 

\subsection{More frequent episodic inflow events in the SNRad than in the SN}
\label{sec:SNRad_vs_SN}

Episodic inflow events from the CMZ occur more frequently in the SNRad simulation. Among the five episodic inflow events identified in the SNRad simulation (see Sect.~\ref{sec:episodic_radial_inflow}), three originate from the CMZ and flow towards the CND while remaining roughly parallel to the disk midplane. The other two originate from `over-shooting' clouds in the bar region. These clouds change their orbits when they interact with SNe-driven bubbles from the dust lanes, producing inclined streams that cross the CND region (see Figs.~\ref{fig:orbit_tracer} and~\ref{fig:orbit_tracer_slice}). Here we focus on the CMZ-origin inflows because we do not identify comparably strong events of this type in the SN simulation.

Supernova explosions occur with delays of $\sim2\mhyphen10\Myr$ after the formation of young stars. As young stars drift away from their parent molecular clouds in the CMZ (see Sect.~5.1 of \citealt{Tress2025} for more details), some SNe explode close to the NIZ. Their high-pressure bubbles then expand into the surrounding gas. Gas on the outer side of the bubbles (e.g. molecular gas in the CMZ) is pushed outward, while gas on the inner side (dominated by ionised gas) is accelerated inward. This produces ionised radial inflows that sweep through the NIZ and feed the CND.

This mechanism is less effective in the SN simulation, potentially because ionised gas is scarce within the NIZ. By contrast, the larger ionised gas reservoir in the SNRad simulation may allow SNe near the NIZ to drive stronger ionised inflows toward the CND. This effect can be more prominent during periods of frequent SN explosions near the NIZ.

\begin{figure}[ht!]
\includegraphics[width=\columnwidth]{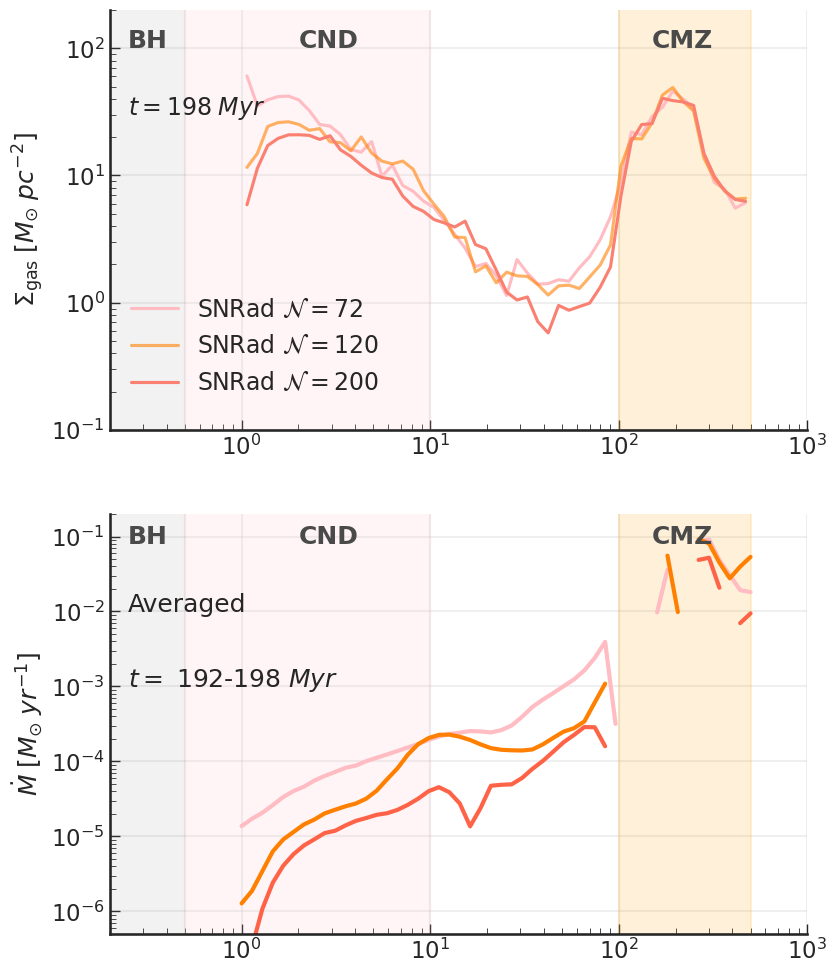}
\centering
\caption{Same as Fig.~\ref{fig:net_inflow_R} but for the SNRad simulation with different spatial resolutions. The inflow rate decreases as spatial resolution improves. }
\label{fig:net_inflow_R_resolution}
\end{figure}

\subsection{Effects of spatial resolutions on inflow rates} \label{sec:resolution_tests} 
We perform a resolution test to study how the inflow rate depends on the spatial resolution. We re-run the SNRad simulation from $t=191\Myr$ to $t=198\Myr$ with \textit{lower} resolution using different $\mathcal{N}$ factors (see the definition in Sect.~\ref{sec:resolution}) of 72 and 120. Fig.~\ref{fig:net_inflow_R_resolution} shows that the inflow rate decreases as the spatial resolution increases. It is beyond the scope of this work to investigate the necessary resolution for convergent inflow rates. 

Therefore, the net inflow rates presented in this paper should be taken as upper limits.

\section{Conclusion}
\label{sec:conclusion}

In this work we present high-resolution simulations of gas inflow in the innermost $5\kpc$ of the Milky Way to investigate the stellar-feedback-regulated transport from the central molecular zone (CMZ; $R\sim200\pc$) to the circum-nuclear disk (CND; $R\sim2\mhyphen5\pc$). The simulations include a realistic Milky Way's externally imposed multi-component gravitational potential, non-equilibrium ISM chemistry and cooling, star formation, supernova feedback, radiation feedback, and radial dependent resolution to zoom-in within the nuclear inflow zone (NIZ; $R<100\pc$). The simulations are designed to follow stellar-feedback-regulated transport of gas in the NIZ fully self-consistently from larger-scale gas flows.

Our main conclusions are as follows:

\begin{enumerate}
    \item Stellar feedback drives a radial inflow that decreases monotonically with Galactocentric radius. The total time-averaged net inflow rate in our fiducial SNRad simulation declines from $\langle \dot{M}\rangle \sim5\times10^{-3}\Msun\yr^{-1}$ at $R=100 \pc$, to $\langle \dot{M}\rangle \sim2\times10^{-4}\Msun\yr^{-1}$ at $R=10 \pc$, to $\langle \dot{M}\rangle \sim 10^{-6}\Msun\yr^{-1}$ at $R=1 \pc$ (see Fig.~\ref{fig:net_inflow_R} and Sect.~\ref{sec:inflow}).
    
    \item The stellar-feedback-driven inflow can be broken down into two components driven by two distinct mechanisms. First, feedback-driven turbulence redistributes the angular momentum of gas clouds and produces a smooth (secular) transport of mass inward, similar to a Shakura-Sunyaev viscous accretion disk (Sect.~\ref{sec:netinflow}). Second, episodic inflow events occur that can increase the inflow rate by several orders of magnitude for short periods of time (Sect.~\ref{sec:episodic_radial_inflow}).
    
    \item The secular component of the inflow rate is similar in the SNRad simulation (that includes both supernova and radiation feedback) and in the SN simulation (that includes only supernova feedback). The time-averaged inflow rate of the secular component declines from $\dot{M}\sim5\times10^{-4}\Msun\yr^{-1}$ at $R\sim 100\pc$ to $\dot{M}\sim 10^{-7}\Msun\yr^{-1}$ at $R\sim1\pc$ (Sect.~\ref{sec:inflow}). 
    
    \item Episodic radial inflow events are more frequent in the SNRad simulation, that includes radiation feedback, than in the SN simulation, that only includes supernova feedback (Sect.~\ref{sec:episodic_radial_inflow}). These events can produce inflow rates as high as $\dot{M}\sim 10^{-3}\Msun\yr^{-1}$ lasting for $\Delta t \sim 3\mhyphen5\Myr$ at $R=10\pc$. Near the end of the SNRad simulation, a massive CND with $M_{\rm CND} \sim 1.5\times10^4\Msun$ forms after several episodic inflow events, consistent with the observational estimate of $\sim2\times10^4\Msun$. We speculate that the higher frequency of episodic inflow in the SNRad simulation is linked to the larger fraction of ionised gas (Sect.~\ref{sec:SNRad_vs_SN}).
   
    \item The total time-averaged inflow rate, including both the smooth and episodic contributions, decreases monotonically with Galactocentric radius (Fig.~\ref{fig:net_inflow_R}). The time-averaged inflow rates at $R=10\pc$ are $\langle \dot{M}\rangle \sim2\times10^{-5}\Msun\yr^{-1}$ and $\langle\dot{M}\rangle\sim2\times10^{-4}\Msun\yr^{-1}$ for the SN and SNRad simulation, respectively.
    
    \item Gas clouds can exhibit filamentary morphologies when ejected from star-forming regions in the CMZ. They enter the CND as filamentary streams rather than isotropic spherical clouds, in contrast to what is often assumed in CND simulations. This difference highlights the importance of studying ISM dynamics and evolution within a realistic Galactic Centre environment (see Fig.~\ref{fig:edisodic_radial_inflow} and Sect.~\ref{sec:SNRad_vs_SN}).

    \item Gas clouds `overshooting' from the CMZ into the bar region can acquire more radial orbits when they interact with supernova-driven bubbles from the dust lanes. These clouds can drift farther from the Galactic plane, flow inward below the CMZ, and enter the CND. This process provides an additional channel for feeding the CND (Sect.~\ref{sec:episodic_radial_inflow}).
    
    \item The dominant inflowing gas phase depends on both the feedback model and the origin of the inflowing material. Neutral gas dominates the radial inflow in the SN simulation at nearly all radii. By contrast, ionised gas dominates the radial inflow within the NIZ in the SNRad simulation, particularly during CMZ-origin episodic inflow events. Neutral gas remains important when the inflowing gas is originated from `overshooting' clouds outside the CMZ (Sect.~\ref{sec:netinflow}).

    \item Finally, the resolution tests show that the inflow rate decreases with increasing numerical resolution, suggesting that our simulations have not reached convergence. Therefore, all the inflow rates reported in this paper should be considered upper limits (Sect.~\ref{sec:resolution_tests}).
    
\end{enumerate}

\begin{acknowledgements}
    ZF, MCS and KF acknowledge financial support from the European Research Council under the ERC Starting Grant “GalFlow” (grant 101116226). MCS further acknowledges financial support from the Fondazione Cariplo under the grant ERC attrattivit\`{a} n. 2023-3014. RSK acknowledges financial support from the ERC via Synergy Grant ``ECOGAL'' (project ID 855130) and from the German Excellence Strategy via the Heidelberg Cluster ``STRUCTURES'' (EXC 2181 - 390900948). In addition RSK is grateful for funding from the German Ministry for Economy and Energy (BMWE) in project ``MAINN'' (funding ID 50OO2206), and from DFG and ANR for project ``STARCLUSTERS'' (funding ID KL 1358/22-1). JP and MH acknowledge funding from the Swiss National Science Foundation (SNF) via a PRIMA grant PR00P2 193577 `From cosmic dawn to high noon: the role of BHs for young galaxies'. CB gratefully  acknowledges  funding  from  National  Science  Foundation  under  Award  Nos. 2108938, 2206510, 2414862, and CAREER 2145689, as well as from the National Aeronautics and Space Administration through the Astrophysics Data Analysis Program under Award ``3-D MC: Mapping Circumnuclear Molecular Clouds from X-ray to Radio,” Grant No. 80NSSC22K1125 as well as participation in the PRIMA project under Grant No. 80NSSC25K7944. DL gratefully acknowledges funding from the National Science Foundation under Award Nos. 1816715, 2108938, and CAREER 2145689; and NASA FINESST Award No: 80NSSC24K1474. All authors are grateful for computing resources provided by the Ministry of Science, Research and the Arts (MWK) of the State of Baden-W\"{u}rttemberg through bwHPC and the German Science Foundation (DFG) through grants INST 35/1134-1 FUGG and 35/1597-1 FUGG, and also for data storage at SDS@hd funded through grants INST 35/1314-1 FUGG and INST 35/1503-1 FUGG. We thank Zhi Li for insightful discussions on multi-phase Milky Way simulations.  

\end{acknowledgements}

\bibliography{reference}{}

@ARTICLE{Jaura2018,
       author = {{Jaura}, O. and {Glover}, S.~C.~O. and {Klessen}, R.~S. and {Paardekooper}, J.-P.},
        title = "{SPRAI: coupling of radiative feedback and primordial chemistry in moving mesh hydrodynamics}",
      journal = {\mnras},
         year = 2018,
        month = apr,
       volume = {475},
       number = {2},
        pages = {2822-2834},
          doi = {10.1093/mnras/stx3356},
archivePrefix = {arXiv},
       eprint = {1711.02542},
 primaryClass = {astro-ph.IM},
       adsurl = {https://ui.adsabs.harvard.edu/abs/2018MNRAS.475.2822J}
}

@ARTICLE{Jaura2020,
       author = {{Jaura}, Ondrej and {Magg}, Mattis and {Glover}, Simon C.~O. and {Klessen}, Ralf S.},
        title = "{SPRAI-II: multifrequency radiative transfer for variable gas densities}",
      journal = {\mnras},
         year = 2020,
        month = dec,
       volume = {499},
       number = {3},
        pages = {3594-3609},
          doi = {10.1093/mnras/staa3054},
       adsurl = {https://ui.adsabs.harvard.edu/abs/2020MNRAS.499.3594J}
}

@ARTICLE{Lau2013,
       author = {{Lau}, R.~M. and {Herter}, T.~L. and {Morris}, M.~R. and {Becklin}, E.~E. and {Adams}, J.~D.},
        title = "{SOFIA/FORCAST Imaging of the Circumnuclear Ring at the Galactic Center}",
      journal = {\apj},
         year = 2013,
        month = sep,
       volume = {775},
       number = {1},
          eid = {37},
        pages = {37},
          doi = {10.1088/0004-637X/775/1/37},
archivePrefix = {arXiv},
       eprint = {1307.8443},
 primaryClass = {astro-ph.GA},
       adsurl = {https://ui.adsabs.harvard.edu/abs/2013ApJ...775...37L}
}

@ARTICLE{Su2024,
       author = {{Su}, Yang and {Zhang}, Shiyu and {Sun}, Yan and {Yang}, Ji and {Yan}, Qing-Zeng and {Zhang}, Shaobo and {Chen}, Zhiwei and {Chen}, Xuepeng and {Zhou}, Xin and {Yuan}, Lixia},
        title = "{Revealing Gas Inflows Toward the Galactic Central Molecular Zone}",
      journal = {\apjl},
         year = 2024,
        month = aug,
       volume = {971},
       number = {1},
          eid = {L6},
        pages = {L6},
          doi = {10.3847/2041-8213/ad656d},
archivePrefix = {arXiv},
       eprint = {2407.10857},
 primaryClass = {astro-ph.GA},
       adsurl = {https://ui.adsabs.harvard.edu/abs/2024ApJ...971L...6S}
}

@ARTICLE{Erwin2024,
       author = {{Erwin}, Peter},
        title = "{The frequency and sizes of inner bars and nuclear rings in barred galaxies and their dependence on galaxy properties}",
      journal = {\mnras},
         year = 2024,
        month = feb,
       volume = {528},
       number = {2},
        pages = {3613-3628},
          doi = {10.1093/mnras/stad3944},
archivePrefix = {arXiv},
       eprint = {2312.12893},
 primaryClass = {astro-ph.GA},
       adsurl = {https://ui.adsabs.harvard.edu/abs/2024MNRAS.528.3613E}
}

@ARTICLE{Gramze2023,
       author = {{Gramze}, Savannah R. and {Ginsburg}, Adam and {Meier}, David S. and {Ott}, Juergen and {Shirley}, Yancy and {Sormani}, Mattia C. and {Svoboda}, Brian E.},
        title = "{Evidence of a Cloud{\textendash}Cloud Collision from Overshooting Gas in the Galactic Center}",
      journal = {\apj},
         year = 2023,
        month = dec,
       volume = {959},
       number = {2},
          eid = {93},
        pages = {93},
          doi = {10.3847/1538-4357/ad01be},
archivePrefix = {arXiv},
       eprint = {2309.16403},
 primaryClass = {astro-ph.GA},
       adsurl = {https://ui.adsabs.harvard.edu/abs/2023ApJ...959...93G}
}

@ARTICLE{Ryu2013,
       author = {{Ryu}, Syukyo Gando and {Nobukawa}, Masayoshi and {Nakashima}, Shinya and {Tsuru}, Takeshi Go and {Koyama}, Katsuji and {Uchiyama}, Hideki},
        title = "{X-Ray Echo from the Sagittarius C Complex and 500-year Activity History of Sagittarius A*}",
      journal = {\pasj},
         year = 2013,
        month = apr,
       volume = {65},
       number = {2},
          eid = {33},
        pages = {33},
          doi = {10.1093/pasj/65.2.33},
archivePrefix = {arXiv},
       eprint = {1211.4529},
 primaryClass = {astro-ph.GA},
       adsurl = {https://ui.adsabs.harvard.edu/abs/2013PASJ...65...33R}
}

@ARTICLE{Chuard2018,
       author = {{Chuard}, D. and {Terrier}, R. and {Goldwurm}, A. and {Clavel}, M. and {Soldi}, S. and {Morris}, M.~R. and {Ponti}, G. and {Walls}, M. and {Chernyakova}, M.},
        title = "{Glimpses of the past activity of Sgr A$^{★}$ inferred from X-ray echoes in Sgr C}",
      journal = {\aap},
         year = 2018,
        month = feb,
       volume = {610},
          eid = {A34},
        pages = {A34},
          doi = {10.1051/0004-6361/201731864},
archivePrefix = {arXiv},
       eprint = {1712.02678},
 primaryClass = {astro-ph.HE},
       adsurl = {https://ui.adsabs.harvard.edu/abs/2018A&A...610A..34C}
}

@ARTICLE{Zhang2024,
       author = {{Zhang}, Hengyue and {Bureau}, Martin and {Smith}, Mark D. and {Cappellari}, Michele and {Davis}, Timothy A. and {Dominiak}, Pandora and {Elford}, Jacob S. and {Liang}, Fu-Heng and {Ruffa}, Ilaria and {Williams}, Thomas G.},
        title = "{WISDOM Project - XIX. Figures of merit for supermassive black hole mass measurements using molecular gas and/or megamaser kinematics}",
      journal = {\mnras},
         year = 2024,
        month = may,
       volume = {530},
       number = {3},
        pages = {3240-3251},
          doi = {10.1093/mnras/stae1106},
archivePrefix = {arXiv},
       eprint = {2404.16345},
 primaryClass = {astro-ph.GA},
       adsurl = {https://ui.adsabs.harvard.edu/abs/2024MNRAS.530.3240Z}
}

@ARTICLE{Storchi-Bergmann2019,
       author = {{Storchi-Bergmann}, Thaisa and {Schnorr-M{\"u}ller}, Allan},
        title = "{Observational constraints on the feeding of supermassive black holes}",
      journal = {Nature Astronomy},
         year = 2019,
        month = jan,
       volume = {3},
        pages = {48-61},
          doi = {10.1038/s41550-018-0611-0},
archivePrefix = {arXiv},
       eprint = {1904.03338},
 primaryClass = {astro-ph.GA},
       adsurl = {https://ui.adsabs.harvard.edu/abs/2019NatAs...3...48S}
}

@ARTICLE{Oppenheimer2020,
       author = {{Oppenheimer}, Benjamin D. and {Davies}, Jonathan J. and {Crain}, Robert A. and {Wijers}, Nastasha A. and {Schaye}, Joop and {Werk}, Jessica K. and {Burchett}, Joseph N. and {Trayford}, James W. and {Horton}, Ryan},
        title = "{Feedback from supermassive black holes transforms centrals into passive galaxies by ejecting circumgalactic gas}",
      journal = {\mnras},
         year = 2020,
        month = jan,
       volume = {491},
       number = {2},
        pages = {2939-2952},
          doi = {10.1093/mnras/stz3124},
archivePrefix = {arXiv},
       eprint = {1904.05904},
 primaryClass = {astro-ph.GA},
       adsurl = {https://ui.adsabs.harvard.edu/abs/2020MNRAS.491.2939O}
}

@ARTICLE{Tumlinson2017,
       author = {{Tumlinson}, Jason and {Peeples}, Molly S. and {Werk}, Jessica K.},
        title = "{The Circumgalactic Medium}",
      journal = {\araa},
         year = 2017,
        month = aug,
       volume = {55},
       number = {1},
        pages = {389-432},
          doi = {10.1146/annurev-astro-091916-055240},
archivePrefix = {arXiv},
       eprint = {1709.09180},
 primaryClass = {astro-ph.GA},
       adsurl = {https://ui.adsabs.harvard.edu/abs/2017ARA&A..55..389T}
}

@ARTICLE{Heywood2019,
       author = {{Heywood}, I. and {Camilo}, F. and {Cotton}, W.~D. and {Yusef-Zadeh}, F. and {Abbott}, T.~D. and {Adam}, R.~M. and {Aldera}, M.~A. and {Bauermeister}, E.~F. and {Booth}, R.~S. and {Botha}, A.~G. and {Botha}, D.~H. and {Brederode}, L.~R.~S. and {Brits}, Z.~B. and {Buchner}, S.~J. and {Burger}, J.~P. and {Chalmers}, J.~M. and {Cheetham}, T. and {de Villiers}, D. and {Dikgale-Mahlakoana}, M.~A. and {du Toit}, L.~J. and {Esterhuyse}, S.~W.~P. and {Fanaroff}, B.~L. and {Foley}, A.~R. and {Fourie}, D.~J. and {Gamatham}, R.~R.~G. and {Goedhart}, S. and {Gounden}, S. and {Hlakola}, M.~J. and {Hoek}, C.~J. and {Hokwana}, A. and {Horn}, D.~M. and {Horrell}, J.~M.~G. and {Hugo}, B. and {Isaacson}, A.~R. and {Jonas}, J.~L. and {Jordaan}, J.~D.~B.~L. and {Joubert}, A.~F. and {J{\'o}zsa}, G.~I.~G. and {Julie}, R.~P.~M. and {Kapp}, F.~B. and {Kenyon}, J.~S. and {Kotz{\'e}}, P.~P.~A. and {Kriel}, H. and {Kusel}, T.~W. and {Lehmensiek}, R. and {Liebenberg}, D. and {Loots}, A. and {Lord}, R.~T. and {Lunsky}, B.~M. and {Macfarlane}, P.~S. and {Magnus}, L.~G. and {Magozore}, C.~M. and {Mahgoub}, O. and {Main}, J.~P.~L. and {Malan}, J.~A. and {Malgas}, R.~D. and {Manley}, J.~R. and {Maree}, M.~D.~J. and {Merry}, B. and {Millenaar}, R. and {Mnyandu}, N. and {Moeng}, I.~P.~T. and {Monama}, T.~E. and {Mphego}, M.~C. and {New}, W.~S. and {Ngcebetsha}, B. and {Oozeer}, N. and {Otto}, A.~J. and {Passmoor}, S.~S. and {Patel}, A.~A. and {Peens-Hough}, A. and {Perkins}, S.~J. and {Ratcliffe}, S.~M. and {Renil}, R. and {Rust}, A. and {Salie}, S. and {Schwardt}, L.~C. and {Serylak}, M. and {Siebrits}, R. and {Sirothia}, S.~K. and {Smirnov}, O.~M. and {Sofeya}, L. and {Swart}, P.~S. and {Tasse}, C. and {Taylor}, D.~T. and {Theron}, I.~P. and {Thorat}, K. and {Tiplady}, A.~J. and {Tshongweni}, S. and {van Balla}, T.~J. and {van der Byl}, A. and {van der Merwe}, C. and {van Dyk}, C.~L. and {Van Rooyen}, R. and {Van Tonder}, V. and {Van Wyk}, R. and {Wallace}, B.~H. and {Welz}, M.~G. and {Williams}, L.~P.},
        title = "{Inflation of 430-parsec bipolar radio bubbles in the Galactic Centre by an energetic event}",
      journal = {\nat},
         year = 2019,
        month = sep,
       volume = {573},
       number = {7773},
        pages = {235-237},
          doi = {10.1038/s41586-019-1532-5},
archivePrefix = {arXiv},
       eprint = {1909.05534},
 primaryClass = {astro-ph.GA},
       adsurl = {https://ui.adsabs.harvard.edu/abs/2019Natur.573..235H}
}

@ARTICLE{Su2010,
       author = {{Su}, Meng and {Slatyer}, Tracy R. and {Finkbeiner}, Douglas P.},
        title = "{Giant Gamma-ray Bubbles from Fermi-LAT: Active Galactic Nucleus Activity or Bipolar Galactic Wind?}",
      journal = {\apj},
         year = 2010,
        month = dec,
       volume = {724},
       number = {2},
        pages = {1044-1082},
          doi = {10.1088/0004-637X/724/2/1044},
archivePrefix = {arXiv},
       eprint = {1005.5480},
 primaryClass = {astro-ph.HE},
       adsurl = {https://ui.adsabs.harvard.edu/abs/2010ApJ...724.1044S}
}

@ARTICLE{Ponti2019,
       author = {{Ponti}, G. and {Hofmann}, F. and {Churazov}, E. and {Morris}, M.~R. and {Haberl}, F. and {Nandra}, K. and {Terrier}, R. and {Clavel}, M. and {Goldwurm}, A.},
        title = "{An X-ray chimney extending hundreds of parsecs above and below the Galactic Centre}",
      journal = {\nat},
         year = 2019,
        month = mar,
       volume = {567},
       number = {7748},
        pages = {347-350},
          doi = {10.1038/s41586-019-1009-6},
archivePrefix = {arXiv},
       eprint = {1904.05969},
 primaryClass = {astro-ph.HE},
       adsurl = {https://ui.adsabs.harvard.edu/abs/2019Natur.567..347P}
}

@ARTICLE{Predehl2020,
       author = {{Predehl}, P. and {Sunyaev}, R.~A. and {Becker}, W. and {Brunner}, H. and {Burenin}, R. and {Bykov}, A. and {Cherepashchuk}, A. and {Chugai}, N. and {Churazov}, E. and {Doroshenko}, V. and {Eismont}, N. and {Freyberg}, M. and {Gilfanov}, M. and {Haberl}, F. and {Khabibullin}, I. and {Krivonos}, R. and {Maitra}, C. and {Medvedev}, P. and {Merloni}, A. and {Nandra}, K. and {Nazarov}, V. and {Pavlinsky}, M. and {Ponti}, G. and {Sanders}, J.~S. and {Sasaki}, M. and {Sazonov}, S. and {Strong}, A.~W. and {Wilms}, J.},
        title = "{Detection of large-scale X-ray bubbles in the Milky Way halo}",
      journal = {\nat},
         year = 2020,
        month = dec,
       volume = {588},
       number = {7837},
        pages = {227-231},
          doi = {10.1038/s41586-020-2979-0},
archivePrefix = {arXiv},
       eprint = {2012.05840},
 primaryClass = {astro-ph.GA},
       adsurl = {https://ui.adsabs.harvard.edu/abs/2020Natur.588..227P}
}

@ARTICLE{Weinberger2018,
       author = {{Weinberger}, Rainer and {Springel}, Volker and {Pakmor}, R{\"u}diger and {Nelson}, Dylan and {Genel}, Shy and {Pillepich}, Annalisa and {Vogelsberger}, Mark and {Marinacci}, Federico and {Naiman}, Jill and {Torrey}, Paul and {Hernquist}, Lars},
        title = "{Supermassive black holes and their feedback effects in the IllustrisTNG simulation}",
      journal = {\mnras},
         year = 2018,
        month = sep,
       volume = {479},
       number = {3},
        pages = {4056-4072},
          doi = {10.1093/mnras/sty1733},
archivePrefix = {arXiv},
       eprint = {1710.04659},
 primaryClass = {astro-ph.GA},
       adsurl = {https://ui.adsabs.harvard.edu/abs/2018MNRAS.479.4056W}
}

@ARTICLE{kor_ho_13,
       author = {{Kormendy}, John and {Ho}, Luis C.},
        title = "{Coevolution (Or Not) of Supermassive Black Holes and Host Galaxies}",
      journal = {\araa},
         year = 2013,
        month = aug,
       volume = {51},
       number = {1},
        pages = {511-653},
          doi = {10.1146/annurev-astro-082708-101811},
archivePrefix = {arXiv},
       eprint = {1304.7762},
 primaryClass = {astro-ph.CO},
       adsurl = {https://ui.adsabs.harvard.edu/abs/2013ARA&A..51..511K}
}

@ARTICLE{kor_etal_11,
       author = {{Kormendy}, John and {Bender}, R. and {Cornell}, M.~E.},
        title = "{Supermassive black holes do not correlate with galaxy disks or pseudobulges}",
      journal = {\nat},
         year = 2011,
        month = jan,
       volume = {469},
       number = {7330},
        pages = {374-376},
          doi = {10.1038/nature09694},
archivePrefix = {arXiv},
       eprint = {1101.3781},
 primaryClass = {astro-ph.GA},
       adsurl = {https://ui.adsabs.harvard.edu/abs/2011Natur.469..374K}
}

@ARTICLE{hop_qua_10,
       author = {{Hopkins}, Philip F. and {Quataert}, Eliot},
        title = "{How do massive black holes get their gas?}",
      journal = {\mnras},
         year = 2010,
        month = sep,
       volume = {407},
       number = {3},
        pages = {1529-1564},
          doi = {10.1111/j.1365-2966.2010.17064.x},
archivePrefix = {arXiv},
       eprint = {0912.3257},
 primaryClass = {astro-ph.CO},
       adsurl = {https://ui.adsabs.harvard.edu/abs/2010MNRAS.407.1529H}
}

@ARTICLE{hop_etal_07,
       author = {{Hopkins}, Philip F. and {Hernquist}, Lars and {Cox}, Thomas J. and {Robertson}, Brant and {Krause}, Elisabeth},
        title = "{A Theoretical Interpretation of the Black Hole Fundamental Plane}",
      journal = {\apj},
         year = 2007,
        month = nov,
       volume = {669},
       number = {1},
        pages = {45-66},
          doi = {10.1086/521590},
archivePrefix = {arXiv},
       eprint = {astro-ph/0701351},
 primaryClass = {astro-ph},
       adsurl = {https://ui.adsabs.harvard.edu/abs/2007ApJ...669...45H}
}

@ARTICLE{hop_etal_07a,
       author = {{Hopkins}, Philip F. and {Hernquist}, Lars and {Cox}, Thomas J. and {Robertson}, Brant and {Krause}, Elisabeth},
        title = "{An Observed Fundamental Plane Relation for Supermassive Black Holes}",
      journal = {\apj},
         year = 2007,
        month = nov,
       volume = {669},
       number = {1},
        pages = {67-73},
          doi = {10.1086/521601},
archivePrefix = {arXiv},
       eprint = {0707.4005},
 primaryClass = {astro-ph},
       adsurl = {https://ui.adsabs.harvard.edu/abs/2007ApJ...669...67H}
}

@ARTICLE{all_ric_07,
       author = {{Aller}, M.~C. and {Richstone}, D.~O.},
        title = "{Host Galaxy Bulge Predictors of Supermassive Black Hole Mass}",
      journal = {\apj},
         year = 2007,
        month = aug,
       volume = {665},
       number = {1},
        pages = {120-156},
          doi = {10.1086/519298},
       adsurl = {https://ui.adsabs.harvard.edu/abs/2007ApJ...665..120A}
}

@ARTICLE{geb_etal_00,
       author = {{Gebhardt}, Karl and {Kormendy}, John and {Ho}, Luis C. and {Bender}, Ralf and {Bower}, Gary and {Dressler}, Alan and {Faber}, S.~M. and {Filippenko}, Alexei V. and {Green}, Richard and {Grillmair}, Carl and {Lauer}, Tod R. and {Magorrian}, John and {Pinkney}, Jason and {Richstone}, Douglas and {Tremaine}, Scott},
        title = "{Black Hole Mass Estimates from Reverberation Mapping and from Spatially Resolved Kinematics}",
      journal = {\apjl},
         year = 2000,
        month = nov,
       volume = {543},
       number = {1},
        pages = {L5-L8},
          doi = {10.1086/318174},
archivePrefix = {arXiv},
       eprint = {astro-ph/0007123},
 primaryClass = {astro-ph},
       adsurl = {https://ui.adsabs.harvard.edu/abs/2000ApJ...543L...5G}
}

@ARTICLE{fer_mer_00,
       author = {{Ferrarese}, Laura and {Merritt}, David},
        title = "{A Fundamental Relation between Supermassive Black Holes and Their Host Galaxies}",
      journal = {\apjl},
         year = 2000,
        month = aug,
       volume = {539},
       number = {1},
        pages = {L9-L12},
          doi = {10.1086/312838},
archivePrefix = {arXiv},
       eprint = {astro-ph/0006053},
 primaryClass = {astro-ph},
       adsurl = {https://ui.adsabs.harvard.edu/abs/2000ApJ...539L...9F}
}

@ARTICLE{mag_etal_98,
       author = {{Magorrian}, John and {Tremaine}, Scott and {Richstone}, Douglas and {Bender}, Ralf and {Bower}, Gary and {Dressler}, Alan and {Faber}, S.~M. and {Gebhardt}, Karl and {Green}, Richard and {Grillmair}, Carl and {Kormendy}, John and {Lauer}, Tod},
        title = "{The Demography of Massive Dark Objects in Galaxy Centers}",
      journal = {\aj},
         year = 1998,
        month = jun,
       volume = {115},
       number = {6},
        pages = {2285-2305},
          doi = {10.1086/300353},
archivePrefix = {arXiv},
       eprint = {astro-ph/9708072},
 primaryClass = {astro-ph},
       adsurl = {https://ui.adsabs.harvard.edu/abs/1998AJ....115.2285M}
}

@ARTICLE{shl_etal_89,
       author = {{Shlosman}, Isaac and {Frank}, Juhan and {Begelman}, Mitchell C.},
        title = "{Bars within bars: a mechanism for fuelling active galactic nuclei}",
      journal = {\nat},
         year = 1989,
        month = mar,
       volume = {338},
       number = {6210},
        pages = {45-47},
          doi = {10.1038/338045a0},
       adsurl = {https://ui.adsabs.harvard.edu/abs/1989Natur.338...45S}
}

@ARTICLE{ari_etal_19,
       author = {{Armillotta}, Lucia and {Krumholz}, Mark R. and {Di Teodoro}, Enrico M. and {McClure-Griffiths}, N.~M.},
        title = "{The life cycle of the Central Molecular Zone - I. Inflow, star formation, and winds}",
      journal = {\mnras},
         year = 2019,
        month = dec,
       volume = {490},
       number = {3},
        pages = {4401-4418},
          doi = {10.1093/mnras/stz2880},
archivePrefix = {arXiv},
       eprint = {1905.01309},
 primaryClass = {astro-ph.GA},
       adsurl = {https://ui.adsabs.harvard.edu/abs/2019MNRAS.490.4401A}
}

@ARTICLE{eks_etal_12,
       author = {{Ekstr{\"o}m}, S. and {Georgy}, C. and {Eggenberger}, P. and {Meynet}, G. and {Mowlavi}, N. and {Wyttenbach}, A. and {Granada}, A. and {Decressin}, T. and {Hirschi}, R. and {Frischknecht}, U. and {Charbonnel}, C. and {Maeder}, A.},
        title = "{Grids of stellar models with rotation. I. Models from 0.8 to 120 M$_{{\ensuremath{\odot}}}$ at solar metallicity (Z = 0.014)}",
      journal = {\aap},
         year = 2012,
        month = jan,
       volume = {537},
          eid = {A146},
        pages = {A146},
          doi = {10.1051/0004-6361/201117751},
archivePrefix = {arXiv},
       eprint = {1110.5049},
 primaryClass = {astro-ph.SR},
       adsurl = {https://ui.adsabs.harvard.edu/abs/2012A&A...537A.146E}
}

@ARTICLE{hu_etal_17,
       author = {{Hu}, Chia-Yu and {Naab}, Thorsten and {Glover}, Simon C.~O. and {Walch}, Stefanie and {Clark}, Paul C.},
        title = "{Variable interstellar radiation fields in simulated dwarf galaxies: supernovae versus photoelectric heating}",
      journal = {\mnras},
         year = 2017,
        month = oct,
       volume = {471},
       number = {2},
        pages = {2151-2173},
          doi = {10.1093/mnras/stx1773},
archivePrefix = {arXiv},
       eprint = {1701.08779},
 primaryClass = {astro-ph.GA},
       adsurl = {https://ui.adsabs.harvard.edu/abs/2017MNRAS.471.2151H}
}

@article{athana_92,
  adsurl = {http://ukads.nottingham.ac.uk/abs/1992MNRAS.259..345A},
  author = {{Athanassoula}, E.},
  journal = {\mnras},
  month = {nov},
  pages = {345-364},
  title = {{The existence and shapes of dust lanes in galactic bars}},
  volume = {259},
  year = {1992}
}

@ARTICLE{bal_haw_98,
  author = {{Balbus}, Steven A. and {Hawley}, John F.},
  title = {{Instability, turbulence, and enhanced transport in accretion disks}},
  journal = {Reviews of Modern Physics},
  year = {1998},
  month = {jan},
  volume = {70},
  number = {1},
  pages = {1-53},
  doi = {10.1103/RevModPhys.70.1},
  adsurl = {https://ui.adsabs.harvard.edu/abs/1998RvMP...70....1B}
}

@ARTICLE{cau_etal_20,
  author = {{Cautun}, Marius and {Ben{\'\i}tez-Llambay}, Alejandro and {Deason}, Alis J. and {Frenk}, Carlos S. and {Fattahi}, Azadeh and {G{\'o}mez}, Facundo A. and {Grand}, Robert J.~J. and {Oman}, Kyle A. and {Navarro}, Julio F. and {Simpson}, Christine M.},
  title = {{The milky way total mass profile as inferred from Gaia DR2}},
  journal = {\mnras},
  year = {2020},
  month = {may},
  volume = {494},
  number = {3},
  pages = {4291-4313},
  doi = {10.1093/mnras/staa1017},
  archiveprefix = {arXiv},
  eprint = {1911.04557},
  primaryclass = {astro-ph.GA},
  adsurl = {https://ui.adsabs.harvard.edu/abs/2020MNRAS.494.4291C}
}

@ARTICLE{chr_etal_05,
  author = {{Christopher}, M.~H. and {Scoville}, N.~Z. and {Stolovy}, S.~R. and {Yun}, Min S.},
  title = {{HCN and HCO$^{+}$ Observations of the Galactic Circumnuclear Disk}},
  journal = {\apj},
  year = {2005},
  month = {mar},
  volume = {622},
  number = {1},
  pages = {346-365},
  doi = {10.1086/427911},
  archiveprefix = {arXiv},
  eprint = {astro-ph/0502532},
  primaryclass = {astro-ph},
  adsurl = {https://ui.adsabs.harvard.edu/abs/2005ApJ...622..346C}
}

@ARTICLE{cla_etal_12,
  author = {{Clark}, Paul C. and {Glover}, Simon C.~O. and {Klessen}, Ralf S.},
  title = {{TreeCol: a novel approach to estimating column densities in astrophysical simulations}},
  journal = {\mnras},
  year = {2012},
  month = {feb},
  volume = {420},
  number = {1},
  pages = {745-756},
  doi = {10.1111/j.1365-2966.2011.20087.x},
  archiveprefix = {arXiv},
  eprint = {1109.3861},
  primaryclass = {astro-ph.GA},
  adsurl = {https://ui.adsabs.harvard.edu/abs/2012MNRAS.420..745C}
}

@ARTICLE{cla_etal_13,
  author = {{Clark}, Paul C. and {Glover}, Simon C.~O. and {Ragan}, Sarah E. and {Shetty}, Rahul and {Klessen}, Ralf S.},
  title = {{On the Temperature Structure of the Galactic Center Cloud G0.253+0.016}},
  journal = {\apjl},
  year = {2013},
  month = {may},
  volume = {768},
  number = {2},
  eid = {L34},
  pages = {L34},
  doi = {10.1088/2041-8205/768/2/L34},
  archiveprefix = {arXiv},
  eprint = {1303.1978},
  primaryclass = {astro-ph.GA},
  adsurl = {https://ui.adsabs.harvard.edu/abs/2013ApJ...768L..34C}
}

@ARTICLE{kob_etal_26,
       author = {{Kobayashi}, Kotaro and {Yutani}, Naomichi and {Saitoh}, Takayuki R. and {Baba}, Junichi and {Wada}, Keiichi},
        title = "{Vaulting the barrier: An intrinsic mechanism to fuel the gas beyond the nuclear ring into the central region of barred galaxies}",
      journal = {arXiv e-prints},
         year = 2026,
        month = apr,
          eid = {arXiv:2604.17955},
        pages = {arXiv:2604.17955},
          doi = {10.48550/arXiv.2604.17955},
archivePrefix = {arXiv},
       eprint = {2604.17955},
 primaryClass = {astro-ph.GA},
       adsurl = {https://ui.adsabs.harvard.edu/abs/2026arXiv260417955K}
}

@ARTICLE{gra_etal_21,
       author = {{GRAVITY Collaboration} and {Abuter}, R. and {Amorim}, A. and {Baub{\"o}ck}, M. and {Berger}, J.~P. and {Bonnet}, H. and {Brandner}, W. and {Cl{\'e}net}, Y. and {Davies}, R. and {de Zeeuw}, P.~T. and {Dexter}, J. and {Dallilar}, Y. and {Drescher}, A. and {Eckart}, A. and {Eisenhauer}, F. and {F{\"o}rster Schreiber}, N.~M. and {Garcia}, P. and {Gao}, F. and {Gendron}, E. and {Genzel}, R. and {Gillessen}, S. and {Habibi}, M. and {Haubois}, X. and {Hei{\ss}el}, G. and {Henning}, T. and {Hippler}, S. and {Horrobin}, M. and {Jim{\'e}nez-Rosales}, A. and {Jochum}, L. and {Jocou}, L. and {Kaufer}, A. and {Kervella}, P. and {Lacour}, S. and {Lapeyr{\`e}re}, V. and {Le Bouquin}, J.-B. and {L{\'e}na}, P. and {Lutz}, D. and {Nowak}, M. and {Ott}, T. and {Paumard}, T. and {Perraut}, K. and {Perrin}, G. and {Pfuhl}, O. and {Rabien}, S. and {Rodr{\'\i}guez-Coira}, G. and {Shangguan}, J. and {Shimizu}, T. and {Scheithauer}, S. and {Stadler}, J. and {Straub}, O. and {Straubmeier}, C. and {Sturm}, E. and {Tacconi}, L.~J. and {Vincent}, F. and {von Fellenberg}, S. and {Waisberg}, I. and {Widmann}, F. and {Wieprecht}, E. and {Wiezorrek}, E. and {Woillez}, J. and {Yazici}, S. and {Young}, A. and {Zins}, G.},
        title = "{Improved GRAVITY astrometric accuracy from modeling optical aberrations}",
      journal = {\aap},
         year = 2021,
        month = mar,
       volume = {647},
          eid = {A59},
        pages = {A59},
          doi = {10.1051/0004-6361/202040208},
archivePrefix = {arXiv},
       eprint = {2101.12098},
 primaryClass = {astro-ph.GA},
       adsurl = {https://ui.adsabs.harvard.edu/abs/2021A&A...647A..59G}
}

@ARTICLE{fit_etal_26,
       author = {{Fiteni}, Karl and {Li}, Xingchen and {Sormani}, Mattia C. and {Debattista}, Victor P. and {Vasini}, Arianna and {Nogueras-Lara}, Francisco and {Sanders}, Jason L. and {Deg}, Nathan and {Schultheis}, Mathias and {Donati}, Marco and {Feng}, Zi-Xuan},
        title = "{Kinematic diagnostics for non-axisymmetry in the Milky Way's nuclear stellar disc}",
      journal = {arXiv e-prints},
         year = 2026,
        month = mar,
          eid = {arXiv:2603.18738},
        pages = {arXiv:2603.18738},
          doi = {10.48550/arXiv.2603.18738},
archivePrefix = {arXiv},
       eprint = {2603.18738},
 primaryClass = {astro-ph.GA},
       adsurl = {https://ui.adsabs.harvard.edu/abs/2026arXiv260318738F}
}

@ARTICLE{gra_etal_19,
  author = {{GRAVITY Collaboration} and {Abuter}, R. and {Amorim}, A. and {Baub{\"o}ck}, M. and {Berger}, J.~P. and {Bonnet}, H. and {Brandner}, W. and {Cl{\'e}net}, Y. and {Coud{\'e} Du Foresto}, V. and {de Zeeuw}, P.~T. and {Dexter}, J. and {Duvert}, G. and {Eckart}, A. and {Eisenhauer}, F. and {F{\"o}rster Schreiber}, N.~M. and {Garcia}, P. and {Gao}, F. and {Gendron}, E. and {Genzel}, R. and {Gerhard}, O. and {Gillessen}, S. and {Habibi}, M. and {Haubois}, X. and {Henning}, T. and {Hippler}, S. and {Horrobin}, M. and {Jim{\'e}nez-Rosales}, A. and {Jocou}, L. and {Kervella}, P. and {Lacour}, S. and {Lapeyr{\`e}re}, V. and {Le Bouquin}, J. -B. and {L{\'e}na}, P. and {Ott}, T. and {Paumard}, T. and {Perraut}, K. and {Perrin}, G. and {Pfuhl}, O. and {Rabien}, S. and {Rodriguez Coira}, G. and {Rousset}, G. and {Scheithauer}, S. and {Sternberg}, A. and {Straub}, O. and {Straubmeier}, C. and {Sturm}, E. and {Tacconi}, L.~J. and {Vincent}, F. and {von Fellenberg}, S. and {Waisberg}, I. and {Widmann}, F. and {Wieprecht}, E. and {Wiezorrek}, E. and {Woillez}, J. and {Yazici}, S.},
  title = {{A geometric distance measurement to the Galactic center black hole with 0.3\% uncertainty}},
  journal = {\aap},
  year = {2019},
  month = {may},
  volume = {625},
  eid = {L10},
  pages = {L10},
  doi = {10.1051/0004-6361/201935656},
  archiveprefix = {arXiv},
  eprint = {1904.05721},
  primaryclass = {astro-ph.GA},
  adsurl = {https://ui.adsabs.harvard.edu/abs/2019A&A...625L..10G}
}

@ARTICLE{cor_vas_22,
  author = {{Correa Magnus}, Lilia and {Vasiliev}, Eugene},
  title = {{Measuring the Milky Way mass distribution in the presence of the LMC}},
  journal = {\mnras},
  year = {2022},
  month = {apr},
  volume = {511},
  number = {2},
  pages = {2610-2630},
  doi = {10.1093/mnras/stab3726},
  archiveprefix = {arXiv},
  eprint = {2110.00018},
  primaryclass = {astro-ph.GA},
  adsurl = {https://ui.adsabs.harvard.edu/abs/2022MNRAS.511.2610C}
}

@ARTICLE{dah_etal_98,
  author = {{Dahmen}, G. and {Huttemeister}, S. and {Wilson}, T.~L. and {Mauersberger}, R.},
  title = {{Molecular gas in the Galactic center region. II. Gas mass and N\_, = H\_2/I\_\^(12)CO conversion based on a C\^(18)O(J = 1 -> 0) survey}},
  journal = {\aap},
  year = {1998},
  month = {mar},
  volume = {331},
  pages = {959-976},
  doi = {10.48550/arXiv.astro-ph/9711117},
  archiveprefix = {arXiv},
  eprint = {astro-ph/9711117},
  primaryclass = {astro-ph},
  adsurl = {https://ui.adsabs.harvard.edu/abs/1998A&A...331..959D}
}

@ARTICLE{den_etal_93,
  author = {{Dent}, W.~R.~F. and {Matthews}, H.~E. and {Wade}, R. and {Duncan}, W.~D.},
  title = {{Submillimeter Continuum Emission from the Galactic Center Region}},
  journal = {\apj},
  year = {1993},
  month = {jun},
  volume = {410},
  pages = {650},
  doi = {10.1086/172781},
  adsurl = {https://ui.adsabs.harvard.edu/abs/1993ApJ...410..650D}
}

@ARTICLE{draine_78,
  author = {{Draine}, B.~T.},
  title = {{Photoelectric heating of interstellar gas.}},
  journal = {\apjs},
  year = {1978},
  month = {apr},
  volume = {36},
  pages = {595-619},
  doi = {10.1086/190513},
  adsurl = {https://ui.adsabs.harvard.edu/abs/1978ApJS...36..595D}
}

@ARTICLE{Sanders2024,
       author = {{Sanders}, Jason L. and {Kawata}, Daisuke and {Matsunaga}, Noriyuki and {Sormani}, Mattia C. and {Smith}, Leigh C. and {Minniti}, Dante and {Gerhard}, Ortwin},
        title = "{The epoch of the Milky Way's bar formation: dynamical modelling of Mira variables in the nuclear stellar disc}",
      journal = {\mnras},
         year = 2024,
        month = may,
       volume = {530},
       number = {3},
        pages = {2972-2993},
          doi = {10.1093/mnras/stae711},
archivePrefix = {arXiv},
       eprint = {2311.00035},
 primaryClass = {astro-ph.GA},
       adsurl = {https://ui.adsabs.harvard.edu/abs/2024MNRAS.530.2972S}
}

@ARTICLE{einast_69,
  author = {{Einasto}, J.},
  title = {{On Galactic Descriptive Functions}},
  journal = {AN},
  year = {1969},
  month = {feb},
  volume = {291},
  number = {3},
  pages = {97},
  doi = {10.1002/asna.19682910303},
  adsurl = {https://ui.adsabs.harvard.edu/abs/1969AN....291...97E}
}

@ARTICLE{tru_etal_97,
       author = {{Truelove}, J. Kelly and {Klein}, Richard I. and {McKee}, Christopher F. and {Holliman}, II, John H. and {Howell}, Louis H. and {Greenough}, Jeffrey A.},
        title = "{The Jeans Condition: A New Constraint on Spatial Resolution in Simulations of Isothermal Self-gravitational Hydrodynamics}",
      journal = {\apjl},
         year = 1997,
        month = nov,
       volume = {489},
       number = {2},
        pages = {L179-L183},
          doi = {10.1086/310975},
       adsurl = {https://ui.adsabs.harvard.edu/abs/1997ApJ...489L.179T}
}

@ARTICLE{gen_etal_85,
  author = {{Genzel}, R. and {Watson}, D.~M. and {Crawford}, M.~K. and {Townes}, C.~H.},
  title = {{The neutral-gas disk around the galactic center.}},
  journal = {\apj},
  year = {1985},
  month = {oct},
  volume = {297},
  pages = {766-786},
  doi = {10.1086/163574},
  adsurl = {https://ui.adsabs.harvard.edu/abs/1985ApJ...297..766G}
}

@ARTICLE{pet_etal_25,
  author = {{Petersson}, Jonathan and {Hirschmann}, Michaela and {Tress}, Robin G. and {Farcy}, Marion and {Glover}, Simon C.~O. and {Klessen}, Ralf S. and {Naab}, Thorsten and {Partmann}, Christian and {Whitworth}, David J.},
  title = {{The Noctua Suite of Simulations -- The Difficulty of Growing Massive Black Holes in Low-Mass Dwarf Galaxies}},
  journal = {arXiv e-prints},
  year = {2025},
  month = {apr},
  eid = {arXiv:2504.08035},
  pages = {arXiv:2504.08035},
  doi = {10.48550/arXiv.2504.08035},
  archivePrefix = {arXiv},
  eprint = {2504.08035},
  primaryClass = {astro-ph.GA},
  adsurl = {https://ui.adsabs.harvard.edu/abs/2025arXiv250408035P}
}

@ARTICLE{Tress2025,
       author = {{Tress}, R.~G. and {Brucy}, N. and {Girichidis}, P. and {Glover}, S.~C.~O. and {Goeller}, J. and {Hirschmann}, M. and {Klessen}, R. and {Peter}, T. and {Petersson}, J. and {Sormani}, M.~C. and {Armillotta}, L. and {Battersby}, C.~D. and {Donati}, M. and {Feng}, Z.~X. and {Henshaw}, J.~D. and {Lipman}, D.~R. and {Longmore}, S.~N. and {Nogueras-Lara}, F. and {Pelkonen}, V.~M. and {Peschken}, N. and {Petkova}, M.~A. and {Plat}, A. and {Reissl}, S. and {Smith}, R. and {Soler}, J.~D.},
        title = "{Rhea-RT: Dynamical impact of Central Molecular Zone conditions on ISM properties and stellar feedback coupling}",
      journal = {arXiv e-prints},
         year = 2025,
        month = dec,
          eid = {arXiv:2512.09981},
        pages = {arXiv:2512.09981},
          doi = {10.48550/arXiv.2512.09981},
archivePrefix = {arXiv},
       eprint = {2512.09981},
 primaryClass = {astro-ph.GA},
       adsurl = {https://ui.adsabs.harvard.edu/abs/2025arXiv251209981T}
}

@ARTICLE{fri_etal_16,
  author = {{Fritz}, T.~K. and {Chatzopoulos}, S. and {Gerhard}, O. and {Gillessen}, S. and {Genzel}, R. and {Pfuhl}, O. and {Tacchella}, S. and {Eisenhauer}, F. and {Ott}, T.},
  title = {{The Nuclear Cluster of the Milky Way: Total Mass and Luminosity}},
  journal = {\apj},
  year = {2016},
  month = {apr},
  volume = {821},
  number = {1},
  eid = {44},
  pages = {44},
  doi = {10.3847/0004-637X/821/1/44},
  archivePrefix = {arXiv},
  eprint = {1406.7568},
  primaryClass = {astro-ph.GA},
  adsurl = {https://ui.adsabs.harvard.edu/abs/2016ApJ...821...44F}
}

@ARTICLE{map_tra_16,
  author = {{Mapelli}, Michela and {Trani}, Alessandro A.},
  title = {{Modelling the formation of the circumnuclear ring in the Galactic centre}},
  journal = {\aap},
  year = {2016},
  month = {jan},
  volume = {585},
  eid = {A161},
  pages = {A161},
  doi = {10.1051/0004-6361/201527195},
  archivePrefix = {arXiv},
  eprint = {1510.05259},
  primaryClass = {astro-ph.GA},
  adsurl = {https://ui.adsabs.harvard.edu/abs/2016A&A...585A.161M}
}

@ARTICLE{cha_etal_15,
  author = {{Chatzopoulos}, S. and {Fritz}, T.~K. and {Gerhard}, O. and {Gillessen}, S. and {Wegg}, C. and {Genzel}, R. and {Pfuhl}, O.},
  title = {{The old nuclear star cluster in the Milky Way: dynamics, mass, statistical parallax, and black hole mass}},
  journal = {\mnras},
  year = {2015},
  month = {feb},
  volume = {447},
  number = {1},
  pages = {948-968},
  doi = {10.1093/mnras/stu2452},
  archivePrefix = {arXiv},
  eprint = {1403.5266},
  primaryClass = {astro-ph.GA},
  adsurl = {https://ui.adsabs.harvard.edu/abs/2015MNRAS.447..948C}
}

@ARTICLE{fed_etal_10,
  author = {{Federrath}, Christoph and {Banerjee}, Robi and {Clark}, Paul C. and {Klessen}, Ralf S.},
  title = {{Modeling Collapse and Accretion in Turbulent Gas Clouds: Implementation and Comparison of Sink Particles in AMR and SPH}},
  journal = {\apj},
  year = {2010},
  month = {apr},
  volume = {713},
  number = {1},
  pages = {269-290},
  doi = {10.1088/0004-637X/713/1/269},
  archivePrefix = {arXiv},
  eprint = {1001.4456},
  primaryClass = {astro-ph.SR},
  adsurl = {https://ui.adsabs.harvard.edu/abs/2010ApJ...713..269F}
}

@ARTICLE{kat_etal_96,
  author = {{Katz}, Neal and {Weinberg}, David H. and {Hernquist}, Lars},
  title = {{Cosmological Simulations with TreeSPH}},
  journal = {\apjs},
  year = {1996},
  month = {jul},
  volume = {105},
  pages = {19},
  doi = {10.1086/192305},
  archivePrefix = {arXiv},
  eprint = {astro-ph/9509107},
  primaryClass = {astro-ph},
  adsurl = {https://ui.adsabs.harvard.edu/abs/1996ApJS..105...19K}
}

@ARTICLE{katz_92,
  author = {{Katz}, Neal},
  title = {{Dissipational Galaxy Formation. II. Effects of Star Formation}},
  journal = {\apj},
  year = {1992},
  month = {jun},
  volume = {391},
  pages = {502},
  doi = {10.1086/171366},
  adsurl = {https://ui.adsabs.harvard.edu/abs/1992ApJ...391..502K}
}

@ARTICLE{glo_cla_12,
  author = {{Glover}, Simon C.~O. and {Clark}, Paul C.},
  title = {{Approximations for modelling CO chemistry in giant molecular clouds: a comparison of approaches}},
  journal = {\mnras},
  year = {2012},
  month = {mar},
  volume = {421},
  number = {1},
  pages = {116-131},
  doi = {10.1111/j.1365-2966.2011.20260.x},
  archiveprefix = {arXiv},
  eprint = {1102.0670},
  primaryclass = {astro-ph.GA},
  adsurl = {https://ui.adsabs.harvard.edu/abs/2012MNRAS.421..116G}
}

@ARTICLE{glo_mac_07,
  author = {{Glover}, Simon C.~O. and {Mac Low}, Mordecai-Mark},
  title = {{Simulating the Formation of Molecular Clouds. I. Slow Formation by Gravitational Collapse from Static Initial Conditions}},
  journal = {\apjs},
  year = {2007},
  month = {apr},
  volume = {169},
  number = {2},
  pages = {239-268},
  doi = {10.1086/512238},
  archiveprefix = {arXiv},
  eprint = {astro-ph/0605120},
  primaryclass = {astro-ph},
  adsurl = {https://ui.adsabs.harvard.edu/abs/2007ApJS..169..239G}
}

@ARTICLE{glo_mac_07a,
  author = {{Glover}, Simon C.~O. and {Mac Low}, Mordecai-Mark},
  title = {{Simulating the Formation of Molecular Clouds. II. Rapid Formation from Turbulent Initial Conditions}},
  journal = {\apj},
  year = {2007},
  month = {apr},
  volume = {659},
  number = {2},
  pages = {1317-1337},
  doi = {10.1086/512227},
  archiveprefix = {arXiv},
  eprint = {astro-ph/0605121},
  primaryclass = {astro-ph},
  adsurl = {https://ui.adsabs.harvard.edu/abs/2007ApJ...659.1317G}
}

@ARTICLE{gol_etal_25,
  author = {{G{\"o}ller}, Junia and {Girichidis}, Philipp and {Brucy}, No{\'e} and {Hunter}, Glen and {Kjellgren}, Karin and {Tress}, Robin and {Klessen}, Ralf S. and {Glover}, Simon C.~O. and {Hennebelle}, Patrick and {Molinari}, Sergio and {Smith}, Rowan and {Soler}, Juan D. and {Sormani}, Mattia C. and {Testi}, Leonardo},
  title = {{Introducing the Rhea simulations of Milky-Way-like galaxies I: Effect of gravitational potential on morphology and star formation}},
  journal = {arXiv e-prints},
  year = {2025},
  month = {feb},
  eid = {arXiv:2502.02646},
  pages = {arXiv:2502.02646},
  doi = {10.48550/arXiv.2502.02646},
  archiveprefix = {arXiv},
  eprint = {2502.02646},
  primaryclass = {astro-ph.GA},
  adsurl = {https://ui.adsabs.harvard.edu/abs/2025arXiv250202646G}
}

@ARTICLE{gol_lan_78,
  author = {{Goldsmith}, P.~F. and {Langer}, W.~D.},
  title = {{Molecular cooling and thermal balance of dense interstellar clouds}},
  journal = {\apj},
  year = {1978},
  month = {jun},
  volume = {222},
  pages = {881-895},
  doi = {10.1086/156206},
  adsurl = {http://adsabs.harvard.edu/abs/1978ApJ...222..881G}
}

@ARTICLE{hat_etal_21,
  author = {{Hatchfield}, H. Perry and {Sormani}, Mattia C. and {Tress}, Robin G. and {Battersby}, Cara and {Smith}, Rowan J. and {Glover}, Simon C.~O. and {Klessen}, Ralf S.},
  title = {{Dynamically Driven Inflow onto the Galactic Center and its Effect upon Molecular Clouds}},
  journal = {\apj},
  year = {2021},
  month = {nov},
  volume = {922},
  number = {1},
  eid = {79},
  pages = {79},
  doi = {10.3847/1538-4357/ac1e89},
  archiveprefix = {arXiv},
  eprint = {2106.08461},
  primaryclass = {astro-ph.GA},
  adsurl = {https://ui.adsabs.harvard.edu/abs/2021ApJ...922...79H}
}

@INPROCEEDINGS{hen_etal_23,
  author = {{Henshaw}, J.~D. and {Barnes}, A.~T. and {Battersby}, C. and {Ginsburg}, A. and {Sormani}, M.~C. and {Walker}, D.~L.},
  title = {{Star Formation in the Central Molecular Zone of the Milky Way}},
  booktitle = {Protostars and Planets VII},
  year = {2023},
  editor = {{Inutsuka}, S. and {Aikawa}, Y. and {Muto}, T. and {Tomida}, K. and {Tamura}, M.},
  series = {Astronomical Society of the Pacific Conference Series},
  volume = {534},
  month = {jul},
  pages = {83},
  doi = {10.48550/arXiv.2203.11223},
  archiveprefix = {arXiv},
  eprint = {2203.11223},
  primaryclass = {astro-ph.GA},
  adsurl = {https://ui.adsabs.harvard.edu/abs/2023ASPC..534...83H}
}

@ARTICLE{hop_etal_24,
  author = {{Hopkins}, Philip F. and {Grudic}, Michael Y. and {Kremer}, Kyle and {Offner}, Stella S.~R. and {Guszejnov}, David and {Rosen}, Anna L.},
  title = {{FORGE'd in FIRE III: The IMF in Quasar Accretion Disks from STARFORGE}},
  journal = {The Open Journal of Astrophysics},
  year = {2024},
  month = {aug},
  volume = {7},
  eid = {71},
  pages = {71},
  doi = {10.33232/001c.122857},
  archiveprefix = {arXiv},
  eprint = {2404.08046},
  primaryclass = {astro-ph.GA},
  adsurl = {https://ui.adsabs.harvard.edu/abs/2024OJAp....7E..71H}
}

@ARTICLE{hop_etal_24b,
  author = {{Hopkins}, Philip F. and {Grudic}, Michael Y. and {Su}, Kung-Yi and {Wellons}, Sarah and {Angles-Alcazar}, Daniel and {Steinwandel}, Ulrich P. and {Guszejnov}, David and {Murray}, Norman and {Faucher-Giguere}, Claude-Andre and {Quataert}, Eliot and {Keres}, Dusan},
  title = {{FORGE'd in FIRE: Resolving the End of Star Formation and Structure of AGN Accretion Disks from Cosmological Initial Conditions}},
  journal = {The Open Journal of Astrophysics},
  year = {2024},
  month = {mar},
  volume = {7},
  eid = {18},
  pages = {18},
  doi = {10.21105/astro.2309.13115},
  archiveprefix = {arXiv},
  eprint = {2309.13115},
  primaryclass = {astro-ph.GA},
  adsurl = {https://ui.adsabs.harvard.edu/abs/2024OJAp....7E..18H}
}

@ARTICLE{hun_etal_24,
  author = {{Hunter}, Glen H. and {Sormani}, Mattia C. and {Beckmann}, Jan P. and {Vasiliev}, Eugene and {Glover}, Simon C.~O. and {Klessen}, Ralf S. and {Soler}, Juan D. and {Brucy}, No{\'e} and {Girichidis}, Philipp and {G{\"o}ller}, Junia and {Ohlin}, Loke and {Tress}, Robin and {Molinari}, Sergio and {Gerhard}, Ortwin and {Benedettini}, Milena and {Smith}, Rowan and {Hennebelle}, Patrick and {Testi}, Leonardo},
  title = {{Testing kinematic distances under a realistic Galactic potential}},
  journal = {arXiv e-prints},
  year = {2024},
  month = {mar},
  eid = {arXiv:2403.18000},
  pages = {arXiv:2403.18000},
  doi = {10.48550/arXiv.2403.18000},
  archiveprefix = {arXiv},
  eprint = {2403.18000},
  primaryclass = {astro-ph.GA},
  adsurl = {https://ui.adsabs.harvard.edu/abs/2024arXiv240318000H}
}

@ARTICLE{kim_etal_12,
  author = {{Kim}, Woong-Tae and {Seo}, Woo-Young and {Stone}, James M. and {Yoon}, Doosoo and {Teuben}, Peter J.},
  title = {{Central Regions of Barred Galaxies: Two-dimensional Non-self-gravitating Hydrodynamic Simulations}},
  journal = {\apj},
  year = {2012},
  month = {mar},
  volume = {747},
  number = {1},
  eid = {60},
  pages = {60},
  doi = {10.1088/0004-637X/747/1/60},
  archiveprefix = {arXiv},
  eprint = {1112.6055},
  primaryclass = {astro-ph.GA},
  adsurl = {https://ui.adsabs.harvard.edu/abs/2012ApJ...747...60K}
}

@ARTICLE{kim_ost_15,
  author = {{Kim}, Chang-Goo and {Ostriker}, Eve C.},
  title = {{Momentum Injection by Supernovae in the Interstellar Medium}},
  journal = {\apj},
  year = {2015},
  month = {apr},
  volume = {802},
  number = {2},
  eid = {99},
  pages = {99},
  doi = {10.1088/0004-637X/802/2/99},
  archiveprefix = {arXiv},
  eprint = {1410.1537},
  primaryclass = {astro-ph.GA},
  adsurl = {https://ui.adsabs.harvard.edu/abs/2015ApJ...802...99K}
}

@ARTICLE{kop_etal_23,
  author = {{Koposov}, Sergey E. and {Erkal}, Denis and {Li}, Ting S. and {Da Costa}, Gary S. and {Cullinane}, Lara R. and {Ji}, Alexander P. and {Kuehn}, Kyler and {Lewis}, Geraint F. and {Pace}, Andrew B. and {Shipp}, Nora and {Zucker}, Daniel B. and {Bland-Hawthorn}, Joss and {Lilleengen}, Sophia and {Martell}, Sarah L. and {S5 Collaboration}},
  title = {{S $^{5}$: Probing the Milky Way and Magellanic Clouds potentials with the 6D map of the Orphan-Chenab stream}},
  journal = {\mnras},
  year = {2023},
  month = {jun},
  volume = {521},
  number = {4},
  pages = {4936-4962},
  doi = {10.1093/mnras/stad551},
  archiveprefix = {arXiv},
  eprint = {2211.04495},
  primaryclass = {astro-ph.GA},
  adsurl = {https://ui.adsabs.harvard.edu/abs/2023MNRAS.521.4936K}
}

@ARTICLE{kroupa_01,
  author = {{Kroupa}, Pavel},
  title = {{On the variation of the initial mass function}},
  journal = {\mnras},
  year = {2001},
  month = {apr},
  volume = {322},
  number = {2},
  pages = {231-246},
  doi = {10.1046/j.1365-8711.2001.04022.x},
  archiveprefix = {arXiv},
  eprint = {astro-ph/0009005},
  primaryclass = {astro-ph},
  adsurl = {https://ui.adsabs.harvard.edu/abs/2001MNRAS.322..231K}
}

@ARTICLE{li_etal_15,
  author = {{Li}, Zhi and {Shen}, Juntai and {Kim}, Woong-Tae},
  title = {{Hydrodynamical Simulations of Nuclear Rings in Barred Galaxies}},
  journal = {\apj},
  year = {2015},
  month = {jun},
  volume = {806},
  number = {2},
  eid = {150},
  pages = {150},
  doi = {10.1088/0004-637X/806/2/150},
  archiveprefix = {arXiv},
  eprint = {1503.02594},
  primaryclass = {astro-ph.GA},
  adsurl = {https://ui.adsabs.harvard.edu/abs/2015ApJ...806..150L}
}

@ARTICLE{li_etal_22,
  author = {{Li}, Zhi and {Shen}, Juntai and {Gerhard}, Ortwin and {Clarke}, Jonathan P.},
  title = {{Gas Dynamics in the Galaxy: Total Mass Distribution and the Bar Pattern Speed}},
  journal = {\apj},
  year = {2022},
  month = {jan},
  volume = {925},
  number = {1},
  eid = {71},
  pages = {71},
  doi = {10.3847/1538-4357/ac3823},
  archiveprefix = {arXiv},
  eprint = {2103.10342},
  primaryclass = {astro-ph.GA},
  adsurl = {https://ui.adsabs.harvard.edu/abs/2022ApJ...925...71L}
}

@ARTICLE{lon_etal_13,
  author = {{Longmore}, S.~N. and {Bally}, J. and {Testi}, L. and {Purcell}, C.~R. and {Walsh}, A.~J. and {Bressert}, E. and {Pestalozzi}, M. and {Molinari}, S. and {Ott}, J. and {Cortese}, L. and {Battersby}, C. and {Murray}, N. and {Lee}, E. and {Kruijssen}, J.~M.~D. and {Schisano}, E. and {Elia}, D.},
  title = {{Variations in the Galactic star formation rate and density thresholds for star formation}},
  journal = {\mnras},
  year = {2013},
  month = {feb},
  volume = {429},
  number = {2},
  pages = {987-1000},
  doi = {10.1093/mnras/sts376},
  archiveprefix = {arXiv},
  eprint = {1208.4256},
  primaryclass = {astro-ph.GA},
  adsurl = {https://ui.adsabs.harvard.edu/abs/2013MNRAS.429..987L}
}

@book{maeder_08,
  title = {Physics, formation and evolution of rotating stars},
  author = {Maeder, Andr{\'e}},
  year = {2008},
  publisher = {Springer Science \& Business Media}
}

@ARTICLE{mcmill_17,
  author = {{McMillan}, Paul J.},
  title = {{The mass distribution and gravitational potential of the Milky Way}},
  journal = {\mnras},
  year = {2017},
  month = {feb},
  volume = {465},
  number = {1},
  pages = {76-94},
  doi = {10.1093/mnras/stw2759},
  archiveprefix = {arXiv},
  eprint = {1608.00971},
  primaryclass = {astro-ph.GA},
  adsurl = {https://ui.adsabs.harvard.edu/abs/2017MNRAS.465...76M}
}

@ARTICLE{mez_etal_89,
  author = {{Mezger}, P.~G. and {Zylka}, R. and {Salter}, C.~J. and {Wink}, J.~E. and {Chini}, R. and {Kreysa}, E. and {Tuffs}, R.},
  title = {{Continuum observations of SGR A at mm/submm wavelengths.}},
  journal = {\aap},
  year = {1989},
  month = {jan},
  volume = {209},
  pages = {337-348},
  adsurl = {https://ui.adsabs.harvard.edu/abs/1989A&A...209..337M}
}

@ARTICLE{mez_etal_96,
  author = {{Mezger}, Peter G. and {Duschl}, Wolfgang J. and {Zylka}, Robert},
  title = {{The Galactic Center: a laboratory for AGN?}},
  journal = {\aapr},
  year = {1996},
  month = {jan},
  volume = {7},
  number = {4},
  pages = {289-388},
  doi = {10.1007/s001590050007},
  adsurl = {https://ui.adsabs.harvard.edu/abs/1996A&ARv...7..289M}
}

@ARTICLE{mon_etal_09,
  author = {{Montero-Casta{\~n}o}, Mar{\'\i}a and {Herrnstein}, Robeson M. and {Ho}, Paul T.~P.},
  title = {{Gas Infall Toward Sgr A* from the Clumpy Circumnuclear Disk}},
  journal = {\apj},
  year = {2009},
  month = {apr},
  volume = {695},
  number = {2},
  pages = {1477-1494},
  doi = {10.1088/0004-637X/695/2/1477},
  archiveprefix = {arXiv},
  eprint = {0903.0886},
  primaryclass = {astro-ph.GA},
  adsurl = {https://ui.adsabs.harvard.edu/abs/2009ApJ...695.1477M}
}

@ARTICLE{moo_etal_23,
  author = {{Moon}, Sanghyuk and {Kim}, Woong-Tae and {Kim}, Chang-Goo and {Ostriker}, Eve C.},
  title = {{Effects of Magnetic Fields on Gas Dynamics and Star Formation in Nuclear Rings}},
  journal = {\apj},
  year = {2023},
  month = {apr},
  volume = {946},
  number = {2},
  eid = {114},
  pages = {114},
  doi = {10.3847/1538-4357/acc250},
  archiveprefix = {arXiv},
  eprint = {2303.04206},
  primaryclass = {astro-ph.GA},
  adsurl = {https://ui.adsabs.harvard.edu/abs/2023ApJ...946..114M}
}

@ARTICLE{mor_ser_96,
  author = {{Morris}, Mark and {Serabyn}, Eugene},
  title = {{The Galactic Center Environment}},
  journal = {\araa},
  year = {1996},
  month = {jan},
  volume = {34},
  pages = {645-702},
  doi = {10.1146/annurev.astro.34.1.645},
  adsurl = {https://ui.adsabs.harvard.edu/abs/1996ARA&A..34..645M}
}

@ARTICLE{nel_lan_97,
  author = {{Nelson}, Richard P. and {Langer}, William D.},
  title = {{The Dynamics of Low-Mass Molecular Clouds in External Radiation Fields}},
  journal = {\apj},
  year = {1997},
  month = {jun},
  volume = {482},
  number = {2},
  pages = {796-826},
  doi = {10.1086/304167},
  adsurl = {https://ui.adsabs.harvard.edu/abs/1997ApJ...482..796N}
}

@ARTICLE{oka_etal_19,
  author = {{Oka}, Takeshi and {Geballe}, T.~R. and {Goto}, Miwa and {Usuda}, Tomonori and {Benjamin} and {McCall}, J. and {Indriolo}, Nick},
  title = {{The Central 300 pc of the Galaxy Probed by Infrared Spectra of \{\{\textbackslashrm\{H\}\}\}\_\{3\}\^\{+\} and CO. I. Predominance of Warm and Diffuse Gas and High H$_{2}$ Ionization Rate}},
  journal = {\apj},
  year = {2019},
  month = {sep},
  volume = {883},
  number = {1},
  eid = {54},
  pages = {54},
  doi = {10.3847/1538-4357/ab3647},
  archiveprefix = {arXiv},
  eprint = {1910.04762},
  primaryclass = {astro-ph.HE},
  adsurl = {https://ui.adsabs.harvard.edu/abs/2019ApJ...883...54O}
}

@ARTICLE{pak_etal_16,
  author = {{Pakmor}, R{\"u}diger and {Springel}, Volker and {Bauer}, Andreas and {Mocz}, Philip and {Munoz}, Diego J. and {Ohlmann}, Sebastian T. and {Schaal}, Kevin and {Zhu}, Chenchong},
  title = {{Improving the convergence properties of the moving-mesh code AREPO}},
  journal = {\mnras},
  year = {2016},
  month = {jan},
  volume = {455},
  number = {1},
  pages = {1134-1143},
  doi = {10.1093/mnras/stv2380},
  archiveprefix = {arXiv},
  eprint = {1503.00562},
  primaryclass = {astro-ph.GA},
  adsurl = {https://ui.adsabs.harvard.edu/abs/2016MNRAS.455.1134P}
}

@ARTICLE{guo_etal_24,
       author = {{Guo}, Minghao and {Stone}, James M. and {Quataert}, Eliot and {Kim}, Chang-Goo},
        title = "{Magnetized Accretion onto and Feedback from Supermassive Black Holes in Elliptical Galaxies}",
      journal = {\apj},
         year = 2024,
        month = oct,
       volume = {973},
       number = {2},
          eid = {141},
        pages = {141},
          doi = {10.3847/1538-4357/ad5fe7},
archivePrefix = {arXiv},
       eprint = {2405.11711},
 primaryClass = {astro-ph.HE},
       adsurl = {https://ui.adsabs.harvard.edu/abs/2024ApJ...973..141G}
}

@ARTICLE{bat_etal_25a,
       author = {{Battersby}, Cara and {Walker}, Daniel L. and {Barnes}, Ashley and {Ginsburg}, Adam and {Lipman}, Dani and {Alboslani}, Danya and {Hatchfield}, H. Perry and {Bally}, John and {Glover}, Simon C.~O. and {Henshaw}, Jonathan D. and {Immer}, Katharina and {Klessen}, Ralf S. and {Longmore}, Steven N. and {Mills}, Elisabeth A.~C. and {Molinari}, Sergio and {Smith}, Rowan and {Sormani}, Mattia C. and {Tress}, Robin G. and {Zhang}, Qizhou},
        title = "{3D CMZ. I. Central Molecular Zone Overview}",
      journal = {\apj},
         year = 2025,
        month = may,
       volume = {984},
       number = {2},
          eid = {156},
        pages = {156},
          doi = {10.3847/1538-4357/adb5f0},
archivePrefix = {arXiv},
       eprint = {2410.17334},
 primaryClass = {astro-ph.GA},
       adsurl = {https://ui.adsabs.harvard.edu/abs/2025ApJ...984..156B}
}

@ARTICLE{moo_etal_22,
       author = {{Moon}, Sanghyuk and {Kim}, Woong-Tae and {Kim}, Chang-Goo and {Ostriker}, Eve C.},
        title = "{Effects of Varying Mass Inflows on Star Formation in Nuclear Rings of Barred Galaxies}",
      journal = {\apj},
         year = 2022,
        month = jan,
       volume = {925},
       number = {1},
          eid = {99},
        pages = {99},
          doi = {10.3847/1538-4357/ac3a7b},
archivePrefix = {arXiv},
       eprint = {2110.14882},
 primaryClass = {astro-ph.GA},
       adsurl = {https://ui.adsabs.harvard.edu/abs/2022ApJ...925...99M}
}

@ARTICLE{pet_etal_23,
  author = {{Peter}, Toni and {Klessen}, Ralf S. and {Kanschat}, Guido and {Glover}, Simon C.~O. and {Bastian}, Peter},
  title = {{The sweep method for radiative transfer in AREPO}},
  journal = {\mnras},
  year = {2023},
  month = {mar},
  volume = {519},
  number = {3},
  pages = {4263-4278},
  doi = {10.1093/mnras/stac3034},
  archiveprefix = {arXiv},
  eprint = {2207.12848},
  primaryclass = {astro-ph.IM},
  adsurl = {https://ui.adsabs.harvard.edu/abs/2023MNRAS.519.4263P}
}

@ARTICLE{por_etal_17,
  author = {{Portail}, Matthieu and {Gerhard}, Ortwin and {Wegg}, Christopher and {Ness}, Melissa},
  title = {{Dynamical modelling of the galactic bulge and bar: the Milky Way's pattern speed, stellar and dark matter mass distribution}},
  journal = {\mnras},
  year = {2017},
  month = {feb},
  volume = {465},
  number = {2},
  pages = {1621-1644},
  doi = {10.1093/mnras/stw2819},
  archiveprefix = {arXiv},
  eprint = {1608.07954},
  primaryclass = {astro-ph.GA},
  adsurl = {https://ui.adsabs.harvard.edu/abs/2017MNRAS.465.1621P}
}

@ARTICLE{req_etal_12,
  author = {{Requena-Torres}, M.~A. and {G{\"u}sten}, R. and {Wei{\ss}}, A. and {Harris}, A.~I. and {Mart{\'\i}n-Pintado}, J. and {Stutzki}, J. and {Klein}, B. and {Heyminck}, S. and {Risacher}, C.},
  title = {{GREAT confirms transient nature of the circum-nuclear disk}},
  journal = {\aap},
  year = {2012},
  month = {jun},
  volume = {542},
  eid = {L21},
  pages = {L21},
  doi = {10.1051/0004-6361/201219068},
  archiveprefix = {arXiv},
  eprint = {1203.6687},
  primaryclass = {astro-ph.GA},
  adsurl = {https://ui.adsabs.harvard.edu/abs/2012A&A...542L..21R}
}

@ARTICLE{sch_etal_25,
  author = {{Schultheis}, Mathias and {Sormani}, Mattia C. and {Gadotti}, Dimitri A.},
  title = {{Nuclear Stellar Discs}},
  journal = {arXiv e-prints},
  year = {2025},
  month = {sep},
  eid = {arXiv:2509.04562},
  pages = {arXiv:2509.04562},
  doi = {10.48550/arXiv.2509.04562},
  archiveprefix = {arXiv},
  eprint = {2509.04562},
  primaryclass = {astro-ph.GA},
  adsurl = {https://ui.adsabs.harvard.edu/abs/2025arXiv250904562S}
}

@ARTICLE{sel_wil_93,
  author = {{Sellwood}, J.~A. and {Wilkinson}, A.},
  title = {{Dynamics of barred galaxies}},
  journal = {Reports on Progress in Physics},
  year = {1993},
  month = {feb},
  volume = {56},
  number = {2},
  pages = {173-256},
  doi = {10.1088/0034-4885/56/2/001},
  archiveprefix = {arXiv},
  eprint = {astro-ph/0608665},
  primaryclass = {astro-ph},
  adsurl = {https://ui.adsabs.harvard.edu/abs/1993RPPh...56..173S}
}

@ARTICLE{sem_etal_00,
  author = {{Sembach}, Kenneth R. and {Howk}, J. Christopher and {Ryans}, Robert S.~I. and {Keenan}, Francis P.},
  title = {{Modeling the Warm Ionized Interstellar Medium and Its Impact on Elemental Abundance Studies}},
  journal = {\apj},
  year = {2000},
  month = {jan},
  volume = {528},
  number = {1},
  pages = {310-324},
  doi = {10.1086/308173},
  archiveprefix = {arXiv},
  eprint = {astro-ph/9908051},
  primaryclass = {astro-ph},
  adsurl = {https://ui.adsabs.harvard.edu/abs/2000ApJ...528..310S}
}

@ARTICLE{shi_etal_25,
  author = {{Shin}, Eun-jin and {Sijacki}, Debora and {Smith}, Matthew C. and {Bourne}, Martin A. and {Koudmani}, Sophie},
  title = {{The MandelZoom project I: modelling black hole accretion through an $α$-disc in dwarf galaxies with a resolved interstellar medium}},
  journal = {arXiv e-prints},
  year = {2025},
  month = {apr},
  eid = {arXiv:2504.18384},
  pages = {arXiv:2504.18384},
  doi = {10.48550/arXiv.2504.18384},
  archiveprefix = {arXiv},
  eprint = {2504.18384},
  primaryclass = {astro-ph.GA},
  adsurl = {https://ui.adsabs.harvard.edu/abs/2025arXiv250418384S}
}

@article{smi_etal_21,
  title = {Efficient early stellar feedback can suppress galactic outflows by reducing supernova clustering},
  author = {Smith, Matthew C and Bryan, Greg L and Somerville, Rachel S and Hu, Chia-Yu and Teyssier, Romain and Burkhart, Blakesley and Hernquist, Lars},
  journal = {\mnras},
  volume = {506},
  number = {3},
  pages = {3882--3915},
  year = {2021},
  publisher = {Oxford University Press}
}

@ARTICLE{sor_bar_19,
  author = {{Sormani}, Mattia C. and {Barnes}, Ashley T.},
  title = {{Mass inflow rate into the Central Molecular Zone: observational determination and evidence of episodic accretion}},
  journal = {\mnras},
  year = {2019},
  month = {mar},
  volume = {484},
  number = {1},
  pages = {1213-1219},
  doi = {10.1093/mnras/stz046},
  archiveprefix = {arXiv},
  eprint = {1901.00867},
  primaryclass = {astro-ph.GA},
  adsurl = {https://ui.adsabs.harvard.edu/abs/2019MNRAS.484.1213S}
}

@ARTICLE{sor_etal_15a,
  author = {{Sormani}, Mattia C. and {Binney}, James and {Magorrian}, John},
  title = {{Gas flow in barred potentials}},
  journal = {\mnras},
  year = {2015},
  month = {may},
  volume = {449},
  number = {3},
  pages = {2421-2435},
  doi = {10.1093/mnras/stv441},
  archiveprefix = {arXiv},
  eprint = {1502.02740},
  primaryclass = {astro-ph.GA},
  adsurl = {https://ui.adsabs.harvard.edu/abs/2015MNRAS.449.2421S}
}

@article{sor_etal_17,
  title = {A simple method to convert sink particles into stars},
  author = {Sormani, Mattia C and Tre{\ss}, Robin G and Klessen, Ralf S and Glover, Simon CO},
  journal = {\mnras},
  volume = {466},
  number = {1},
  pages = {407--412},
  year = {2017},
  publisher = {Oxford University Press}
}

@ARTICLE{sor_etal_18,
  author = {{Sormani}, Mattia C. and {Tre{\ss}}, Robin G. and {Ridley}, Matthew and {Glover}, Simon C.~O. and {Klessen}, Ralf S. and {Binney}, James and {Magorrian}, John and {Smith}, Rowan},
  title = {{A theoretical explanation for the Central Molecular Zone asymmetry}},
  journal = {\mnras},
  year = {2018},
  month = {apr},
  volume = {475},
  number = {2},
  pages = {2383-2402},
  doi = {10.1093/mnras/stx3258},
  archiveprefix = {arXiv},
  eprint = {1707.03650},
  primaryclass = {astro-ph.GA},
  adsurl = {https://ui.adsabs.harvard.edu/abs/2018MNRAS.475.2383S}
}

@ARTICLE{sor_etal_20,
  author = {{Sormani}, Mattia C. and {Tress}, Robin G. and {Glover}, Simon C.~O. and {Klessen}, Ralf S. and {Battersby}, Cara D. and {Clark}, Paul C. and {Hatchfield}, H. Perry and {Smith}, Rowan J.},
  title = {{Simulations of the Milky Way's Central Molecular Zone - II. Star formation}},
  journal = {\mnras},
  year = {2020},
  month = {oct},
  volume = {497},
  number = {4},
  pages = {5024-5040},
  doi = {10.1093/mnras/staa1999},
  archiveprefix = {arXiv},
  eprint = {2004.06731},
  primaryclass = {astro-ph.GA},
  adsurl = {https://ui.adsabs.harvard.edu/abs/2020MNRAS.497.5024S}
}

@ARTICLE{sor_etal_20a,
  author = {{Sormani}, Mattia C. and {Magorrian}, John and {Nogueras-Lara}, Francisco and {Neumayer}, Nadine and {Sch{\"o}nrich}, Ralph and {Klessen}, Ralf S. and {Mastrobuono-Battisti}, Alessandra},
  title = {{Jeans modelling of the Milky Way's nuclear stellar disc}},
  journal = {\mnras},
  year = {2020},
  month = {nov},
  volume = {499},
  number = {1},
  pages = {7-24},
  doi = {10.1093/mnras/staa2785},
  archiveprefix = {arXiv},
  eprint = {2007.06577},
  primaryclass = {astro-ph.GA},
  adsurl = {https://ui.adsabs.harvard.edu/abs/2020MNRAS.499....7S}
}

@ARTICLE{sor_etal_22,
  author = {{Sormani}, Mattia C. and {Gerhard}, Ortwin and {Portail}, Matthieu and {Vasiliev}, Eugene and {Clarke}, Jonathan},
  title = {{The stellar mass distribution of the Milky Way's bar: an analytical model}},
  journal = {\mnras},
  year = {2022},
  month = {jul},
  volume = {514},
  number = {1},
  pages = {L1-L5},
  doi = {10.1093/mnrasl/slac046},
  archiveprefix = {arXiv},
  eprint = {2204.13114},
  primaryclass = {astro-ph.GA},
  adsurl = {https://ui.adsabs.harvard.edu/abs/2022MNRAS.514L...1S}
}

@ARTICLE{spring_10,
  author = {{Springel}, Volker},
  title = {{E pur si muove: Galilean-invariant cosmological hydrodynamical simulations on a moving mesh}},
  journal = {\mnras},
  year = {2010},
  month = {jan},
  volume = {401},
  number = {2},
  pages = {791-851},
  doi = {10.1111/j.1365-2966.2009.15715.x},
  archiveprefix = {arXiv},
  eprint = {0901.4107},
  primaryclass = {astro-ph.CO},
  adsurl = {https://ui.adsabs.harvard.edu/abs/2010MNRAS.401..791S}
}

@ARTICLE{sti_etal_06,
  author = {{Stil}, J.~M. and {Taylor}, A.~R. and {Dickey}, J.~M. and {Kavars}, D.~W. and {Martin}, P.~G. and {Rothwell}, T.~A. and {Boothroyd}, A.~I. and {Lockman}, Felix J. and {McClure-Griffiths}, N.~M.},
  title = {{The VLA Galactic Plane Survey}},
  journal = {\aj},
  year = {2006},
  month = {sep},
  volume = {132},
  number = {3},
  pages = {1158-1176},
  doi = {10.1086/505940},
  archiveprefix = {arXiv},
  eprint = {astro-ph/0605422},
  primaryclass = {astro-ph},
  adsurl = {https://ui.adsabs.harvard.edu/abs/2006AJ....132.1158S}
}

@ARTICLE{tra_etal_18,
  author = {{Trani}, Alessandro A. and {Mapelli}, Michela and {Ballone}, Alessandro},
  title = {{Forming Circumnuclear Disks and Rings in Galactic Nuclei: A Competition Between Supermassive Black Hole and Nuclear Star Cluster}},
  journal = {\apj},
  year = {2018},
  month = {sep},
  volume = {864},
  number = {1},
  eid = {17},
  pages = {17},
  doi = {10.3847/1538-4357/aad414},
  archiveprefix = {arXiv},
  eprint = {1807.09780},
  primaryclass = {astro-ph.GA},
  adsurl = {https://ui.adsabs.harvard.edu/abs/2018ApJ...864...17T}
}

@article{tre_etal_20,
  title = {Simulations of the star-forming molecular gas in an interacting M51-like galaxy},
  author = {Tress, Robin G and Smith, Rowan J and Sormani, Mattia C and Glover, Simon CO and Klessen, Ralf S and Mac Low, Mordecai-Mark and Clark, Paul C},
  journal = {\mnras},
  volume = {492},
  number = {2},
  pages = {2973--2995},
  year = {2020},
  publisher = {Oxford University Press}
}

@ARTICLE{tre_etal_20a,
  author = {{Tress}, Robin G. and {Sormani}, Mattia C. and {Glover}, Simon C.~O. and {Klessen}, Ralf S. and {Battersby}, Cara D. and {Clark}, Paul C. and {Hatchfield}, H. Perry and {Smith}, Rowan J.},
  title = {{Simulations of the Milky Way's central molecular zone - I. Gas dynamics}},
  journal = {\mnras},
  year = {2020},
  month = {dec},
  volume = {499},
  number = {3},
  pages = {4455-4478},
  doi = {10.1093/mnras/staa3120},
  archiveprefix = {arXiv},
  eprint = {2004.06724},
  primaryclass = {astro-ph.GA},
  adsurl = {https://ui.adsabs.harvard.edu/abs/2020MNRAS.499.4455T}
}

@ARTICLE{tre_etal_24,
       author = {{Tress}, R.~G. and {Sormani}, M.~C. and {Girichidis}, P. and {Glover}, S.~C.~O. and {Klessen}, R.~S. and {Smith}, R.~J. and {Sobacchi}, E. and {Armillotta}, L. and {Barnes}, A.~T. and {Battersby}, C. and {Bogue}, K.~R.~J. and {Brucy}, N. and {Colzi}, L. and {Federrath}, C. and {Garc{\'\i}a}, P. and {Ginsburg}, A. and {G{\"o}ller}, J. and {Hatchfield}, H.~P. and {Henkel}, C. and {Hennebelle}, P. and {Henshaw}, J.~D. and {Hirschmann}, M. and {Hu}, Y. and {Kauffmann}, J. and {Kruijssen}, J.~M.~D. and {Lazarian}, A. and {Lipman}, D. and {Longmore}, S.~N. and {Morris}, M.~R. and {Nogueras-Lara}, F. and {Petkova}, M.~A. and {Pillai}, T.~G.~S. and {Rivilla}, V.~M. and {S{\'a}nchez-Monge}, {\'A}. and {Soler}, J.~D. and {Whitworth}, D. and {Zhang}, Q.},
        title = "{Magnetic field morphology and evolution in the Central Molecular Zone and its effect on gas dynamics}",
      journal = {\aap},
         year = 2024,
        month = nov,
       volume = {691},
          eid = {A303},
        pages = {A303},
          doi = {10.1051/0004-6361/202450035},
archivePrefix = {arXiv},
       eprint = {2403.13048},
 primaryClass = {astro-ph.GA},
       adsurl = {https://ui.adsabs.harvard.edu/abs/2024A&A...691A.303T}
}

@ARTICLE{vas_etal_21,
  author = {{Vasiliev}, Eugene and {Belokurov}, Vasily and {Erkal}, Denis},
  title = {{Tango for three: Sagittarius, LMC, and the Milky Way}},
  journal = {\mnras},
  year = {2021},
  month = {feb},
  volume = {501},
  number = {2},
  pages = {2279-2304},
  doi = {10.1093/mnras/staa3673},
  archiveprefix = {arXiv},
  eprint = {2009.10726},
  primaryclass = {astro-ph.GA},
  adsurl = {https://ui.adsabs.harvard.edu/abs/2021MNRAS.501.2279V}
}

@ARTICLE{vasili_19,
  author = {{Vasiliev}, Eugene},
  title = {{AGAMA: action-based galaxy modelling architecture}},
  journal = {\mnras},
  year = {2019},
  month = {jan},
  volume = {482},
  number = {2},
  pages = {1525-1544},
  doi = {10.1093/mnras/sty2672},
  archiveprefix = {arXiv},
  eprint = {1802.08239},
  primaryclass = {astro-ph.GA},
  adsurl = {https://ui.adsabs.harvard.edu/abs/2019MNRAS.482.1525V}
}

@ARTICLE{wei_etal_20,
  author = {{Weinberger}, Rainer and {Springel}, Volker and {Pakmor}, R{\"u}diger},
  title = {{The AREPO Public Code Release}},
  journal = {\apjs},
  year = {2020},
  month = {jun},
  volume = {248},
  number = {2},
  eid = {32},
  pages = {32},
  doi = {10.3847/1538-4365/ab908c},
  archiveprefix = {arXiv},
  eprint = {1909.04667},
  primaryclass = {astro-ph.IM},
  adsurl = {https://ui.adsabs.harvard.edu/abs/2020ApJS..248...32W}
}

@ARTICLE{Nogueras-Lara2020,
       author = {{Nogueras-Lara}, Francisco and {Sch{\"o}del}, Rainer and {Gallego-Calvente}, Aurelia Teresa and {Gallego-Cano}, Eulalia and {Shahzamanian}, Banafsheh and {Dong}, Hui and {Neumayer}, Nadine and {Hilker}, Michael and {Najarro}, Francisco and {Nishiyama}, Shogo and {Feldmeier-Krause}, Anja and {Girard}, Julien H.~V. and {Cassisi}, Santi},
        title = "{Early formation and recent starburst activity in the nuclear disk of the Milky Way}",
      journal = {Nature Astronomy},
         year = 2020,
        month = jan,
       volume = {4},
        pages = {377-381},
          doi = {10.1038/s41550-019-0967-9},
archivePrefix = {arXiv},
       eprint = {1910.06968},
 primaryClass = {astro-ph.GA},
       adsurl = {https://ui.adsabs.harvard.edu/abs/2020NatAs...4..377N}
}

@ARTICLE{Hsieh2017,
       author = {{Hsieh}, Pei-Ying and {Koch}, Patrick M. and {Ho}, Paul T.~P. and {Kim}, Woong-Tae and {Tang}, Ya-Wen and {Wang}, Hsiang-Hsu and {Yen}, Hsi-Wei and {Hwang}, Chorng-Yuan},
        title = "{Molecular Gas Feeding the Circumnuclear Disk of the Galactic Center}",
      journal = {\apj},
         year = 2017,
        month = sep,
       volume = {847},
       number = {1},
          eid = {3},
        pages = {3},
          doi = {10.3847/1538-4357/aa8329},
archivePrefix = {arXiv},
       eprint = {1708.08579},
 primaryClass = {astro-ph.GA},
       adsurl = {https://ui.adsabs.harvard.edu/abs/2017ApJ...847....3H}
}

@ARTICLE{Hopkins2011,
       author = {{Hopkins}, Philip F. and {Quataert}, Eliot},
        title = "{An analytic model of angular momentum transport by gravitational torques: from galaxies to massive black holes}",
      journal = {\mnras},
         year = 2011,
        month = aug,
       volume = {415},
       number = {2},
        pages = {1027-1050},
          doi = {10.1111/j.1365-2966.2011.18542.x},
archivePrefix = {arXiv},
       eprint = {1007.2647},
 primaryClass = {astro-ph.CO},
       adsurl = {https://ui.adsabs.harvard.edu/abs/2011MNRAS.415.1027H}
}

@ARTICLE{Vollmer2002,
       author = {{Vollmer}, B. and {Duschl}, W.~J.},
        title = "{The Dynamics of the Circumnuclear Disk and its environment in the Galactic Centre}",
      journal = {\aap},
         year = 2002,
        month = jun,
       volume = {388},
        pages = {128-148},
          doi = {10.1051/0004-6361:20020422},
archivePrefix = {arXiv},
       eprint = {astro-ph/0203428},
 primaryClass = {astro-ph},
       adsurl = {https://ui.adsabs.harvard.edu/abs/2002A&A...388..128V}
}

@ARTICLE{Vollmer2001,
       author = {{Vollmer}, B. and {Duschl}, W.~J.},
        title = "{A cloudy model for the Circumnuclear Disk in the Galactic Centre}",
      journal = {\aap},
         year = 2001,
        month = feb,
       volume = {367},
        pages = {72-85},
          doi = {10.1051/0004-6361:20000425},
archivePrefix = {arXiv},
       eprint = {astro-ph/0012315},
 primaryClass = {astro-ph},
       adsurl = {https://ui.adsabs.harvard.edu/abs/2001A&A...367...72V}
}

@ARTICLE{VonLinden1993,
       author = {{von Linden}, Susanne and {Duschl}, Wolfgang J. and {Biermann}, Peter L.},
        title = "{Molecular clouds as tracers of the dynamics in the central region of the galaxy.}",
      journal = {\aap},
         year = 1993,
        month = mar,
       volume = {269},
        pages = {169-174},
       adsurl = {https://ui.adsabs.harvard.edu/abs/1993A&A...269..169V}
}
\bibliographystyle{aasjournal}

\appendix

\section{Overall properties of ISM and stellar components}
\begin{figure*}[ht!]
\includegraphics[width=0.9\textwidth]{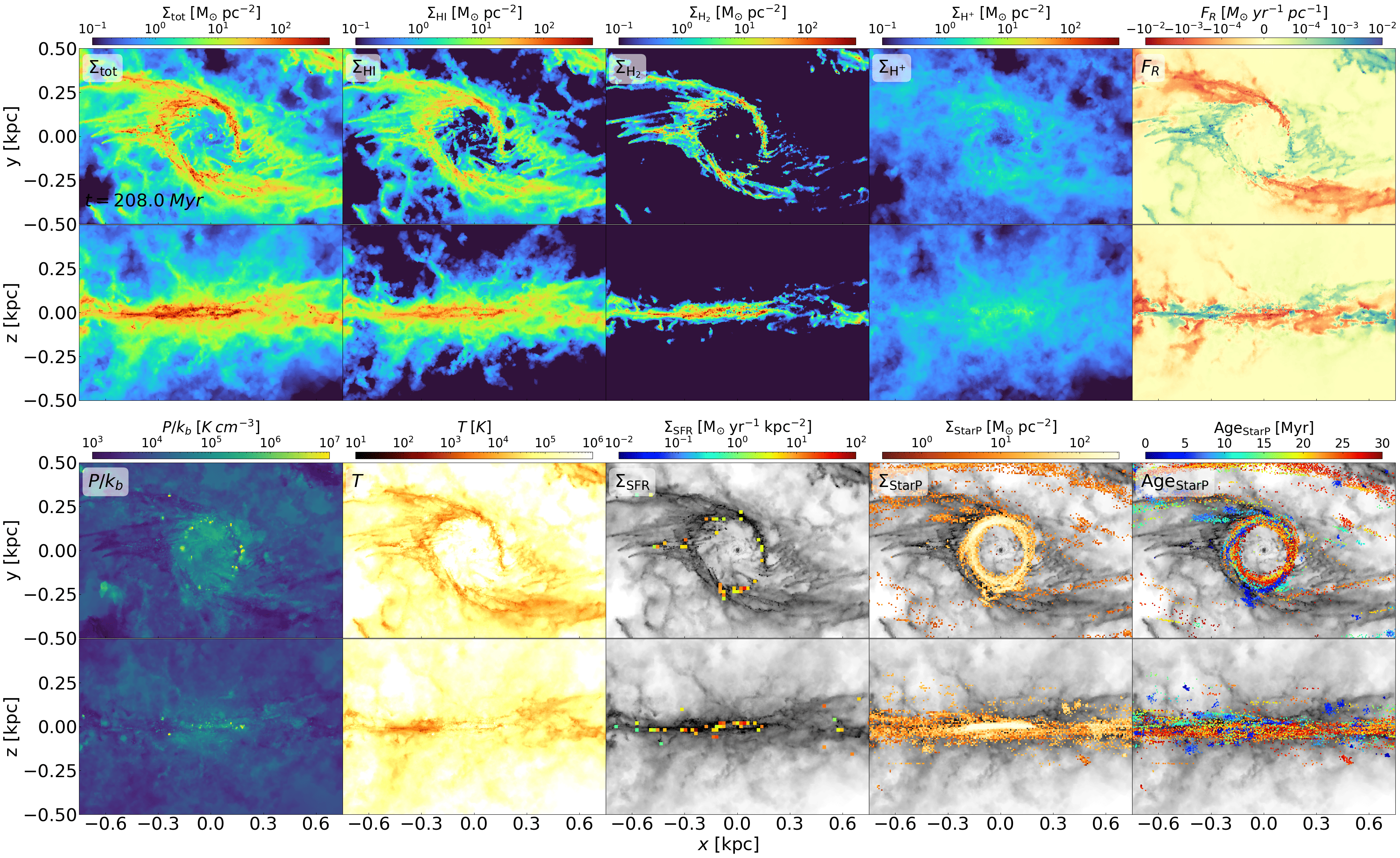}
\centering
\caption{General properties of ISM and stellar components in the SN simulation at $t=208\Myr$.  Both face-on and edge-on views are presented. \textit{top row:} surface densities of the total ISM, $\HI$, $\Htwo$, $\Hp$, and the radial flux distribution. \textit{bottom row:} distributions of ISM pressure, temperature, SFRD, stellar surface density, and stellar age.}
\label{fig:diagnostic_SN.png}
\end{figure*}

\begin{figure*}[ht!]
\includegraphics[width=0.9\textwidth]{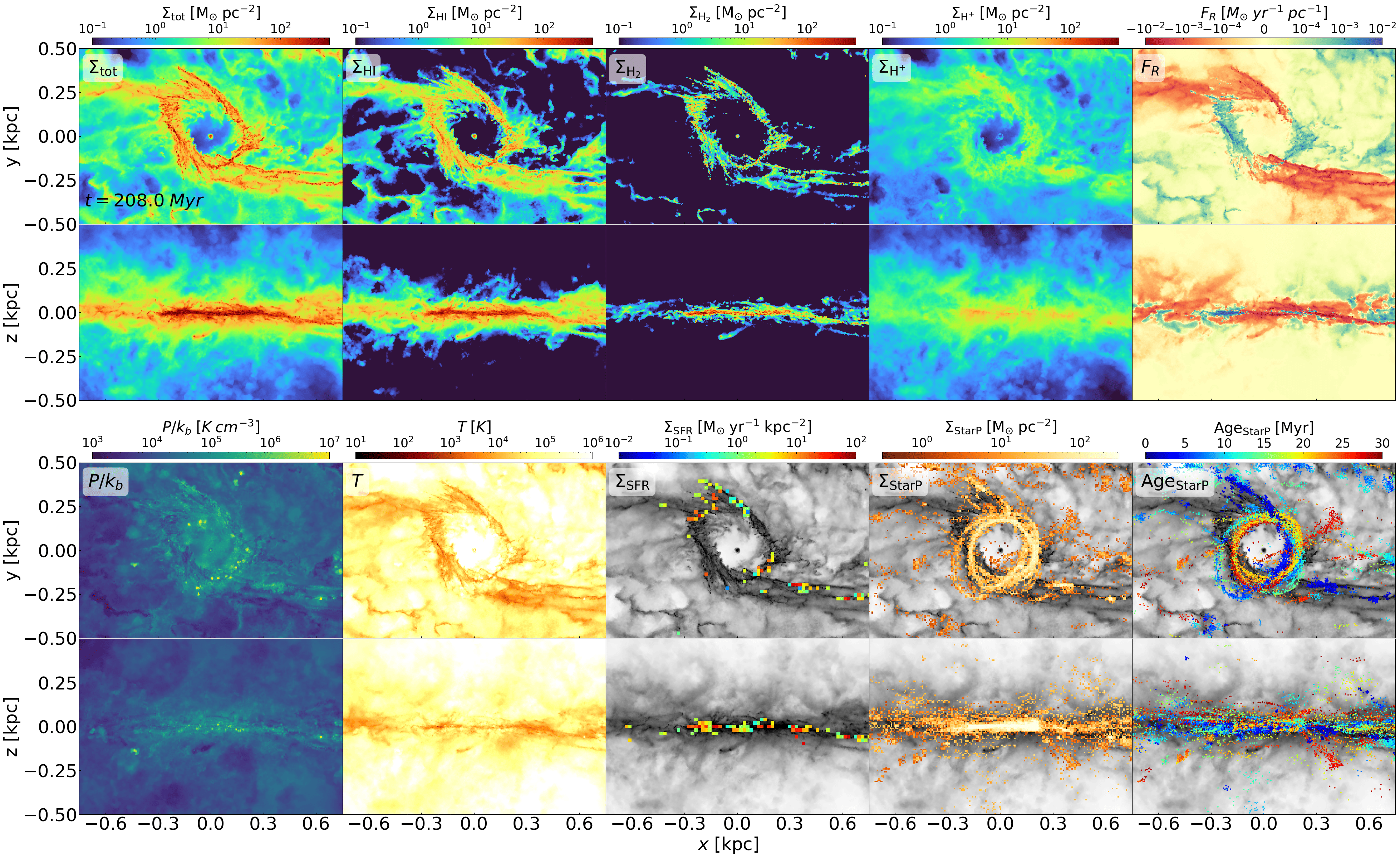}
\centering
\caption{Same as Fig.~\ref{fig:diagnostic_SN.png} but for the SNRad simulation.}
\label{fig:diagnostic_SNRad.png}
\end{figure*}

Figures.~\ref{fig:diagnostic_SN.png} and \ref{fig:diagnostic_SNRad.png} show general properties of the ISM and stellar components. The panels include the surface densities of the total ISM, $\HI$, $\Htwo$, and $\Hp$, together with the instantaneous radial flux, ISM pressure, temperature, star formation rate surface density (SFRD), stellar surface density, and stellar age distribution. Compared to the SN simulation, the SNRad simulation produces larger amounts of $\Hp$. 

Surface density maps are obtained by taking the integral of densities along the line of sight, with vertical heights $|z| < 0.5\kpc$ and $|y| < 0.5\kpc$. Distributions of ISM radial flux, pressure, temperature are obtained by taking the mass-weighted average. The SFRD is computed with newly formed stars younger than $0.5\Myr$.

\section{Time evolution of mass fractions of chemical tracers}

The mass fractions of $\Htwo$, $\HI$, and $\Hp$ remain nearly unchanged with time in the bar and CMZ region. In the NIZ, the $\Hp$ fraction increases with time in the SN simulation, but it remains almost unchanged in the SNRad simulation.

\begin{figure}[ht!]
\includegraphics[width=\columnwidth]{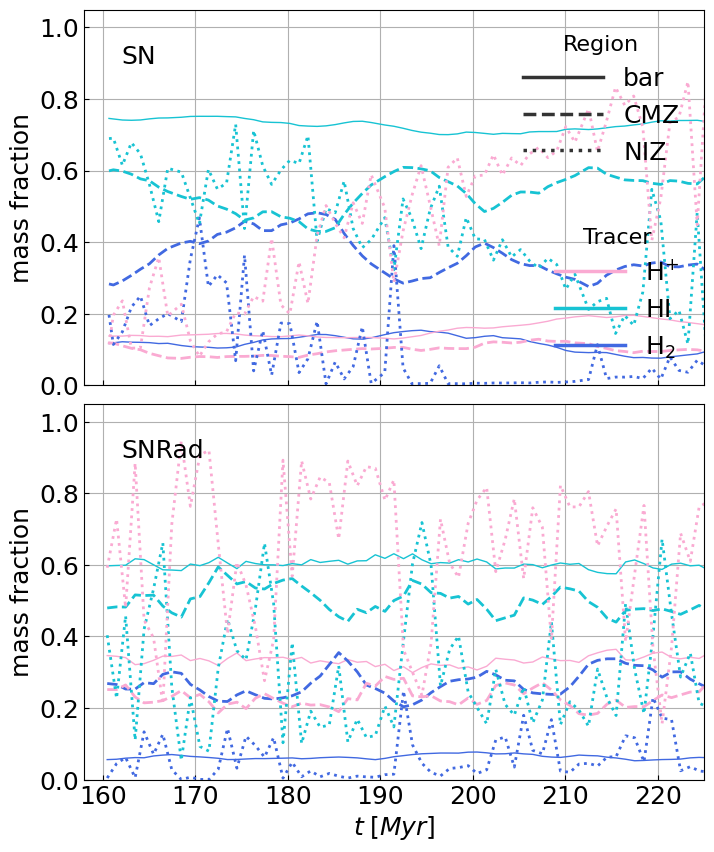}
\centering
\caption{Time evolution of mass fractions of different chemical compositions in SN (top) and SNRad (bottom) simulations. The color code is the same as the right panels of Fig.~\ref{fig:chemistry_f(n)}. Solid, dashed, and dotted curves correspond to the bar, CMZ, and NIZ regions.}
\label{fig:chemistry_f(t)}
\end{figure}

\end{document}